\documentclass[twocolumn]{autart}    

\usepackage{graphicx}          
\usepackage[T1]{fontenc}
\usepackage{arydshln}
\usepackage{tikz}
\usepackage[english]{babel}
\usepackage{amsmath,amssymb,amsfonts}
\usepackage{algorithm}
\usepackage[algo2e,lined,norelsize]{algorithm2e}
\usepackage{dsfont}

\newtheorem{theorem}{Theorem}[section]
\newtheorem{lemma}{Lemma}[section]
\newtheorem{proposition}{Proposition}[section]
\newtheorem{remark}{Remark}[section]

\begin{document}
	
	\begin{frontmatter}
		
		\title{\vspace{-8mm}Network-Realised Model Predictive Control\\ Part I: NRF-Enabled Closed-loop Decomposition\\{\color{white}(Andrei's Version)}\vspace{-12mm}} 
		
		
		
		\author[UPB]{Andrei Speril\u a}\ead{andrei.sperila@upb.ro},    
		\author[L2S]{Alessio Iovine}\ead{alessio.iovine@centralesupelec.fr},               
		\author[L2S]{Sorin Olaru}\ead{sorin.olaru@centralesupelec.fr},  
		\author[L2S]{Patrick Panciatici}\ead{patrick.panciatici@ieee.org}  
		
		\address[UPB]{Department of Automatic Control and Systems Engineering, ``Politehnica'' University of Bucharest, Bucharest, Romania\vspace{-2.5mm}}
		\address[L2S]{Laboratoire des Signaux et Syst\`emes, CentraleSup\'elec, Paris-Saclay University, Gif-sur-Yvette, France\vspace{-10mm}}  

		\begin{keyword}                           
			Distributed control; convex optimisation; model predictive control; networked systems; scalable implementations.   \vspace{-3mm}            
		\end{keyword}                             

		\begin{abstract}
			A two-layer control architecture is proposed to enable scalable implementations for constraint-based decision strategies, such as model predictive controllers. The bottom layer is based upon a distributed feedback-feedforward scheme that directs the controlled network's information flow according to a pre-specified communication infrastructure. Explicit expressions for the resulting closed-loop maps are obtained, and an offline model-matching procedure is proposed for designing the first layer. The obtained control laws are deployed via distributed state-space-based implementations, and the resulting closed-loop models enable predictive control design for the constraint management procedure described in our companion paper.			
			\vspace{-6mm}
		\end{abstract}
		
	\end{frontmatter}
	
	{\color{black}
		
		\section{Introduction}\label{sec:intro}\vspace{-4mm}
		
		\subsection{Context}\vspace{-4mm}
		
		Due to the ever-increasing interconnection and spatial dispersion of modern technical applications, distributed control has become a primary research topic in system-theoretical literature \cite{Luca2}. Consequently, new families of control laws, such as the \emph{System Level Synthesis} (SLS) framework \cite{SLS} and the \emph{Input-Output Parametrisation} (IOP) \cite{Luca1}, have been developed with the express aim of deploying scalable and highly efficient implementations.\vspace{-3mm}
		
		While significant effort has been expended to investigate this topic \cite{Luca3} in the context of networks modelled as state-space systems, the resulting class of control laws tends to focus quite heavily on the frequency domain, making the enforcement of time-domain constraints a rather onerous and unintuitive affair. In parallel with the previously mentioned line of enquiry, the distributed control problem has also been addressed using the Model Predictive Control (MPC) formalism \cite{RMD}. The latter is remarkable for its ability to leverage set-theoretical concepts \cite{STMC} to satisfy time-based restrictions for said class of systems \cite{BBM}. This particular avenue of research has, therefore, resulted in numerous design procedures \cite{R8A,R8B,R8D}.\vspace{-3mm}
		
		Distinct from the above-stated approaches, a new family of distributed control laws has been prototyped in \cite{plutonizare}, formalised in \cite{NRF}, and subsequently extended in \cite{aug_sparse}. This novel design framework, dubbed the \emph{Network Realisation Function} (NRF) approach, is chiefly based upon the celebrated \emph{Youla Parametrisation} (see, for example, Chapter 12 in \cite{zhou}) and it is notable for filling in the Youla-oriented theoretical gap highlighted in \cite{Luca3}, thus completing the SLS-IOP-NRF triad hinted at therein. Among the many benefits of NRF-based control (see \cite{NRF},\newpage and also \cite{aug_sparse}), we single out the desirable properties it ensures in formation-based applications, with the \emph{string stability} property featured in \cite{plutonizare} being noticeably more difficult to retrieve in the MPC framework \cite{plat_MPC_1,plat_MPC_2,plat_MPC_3,plat_MPC_4,plat_MPC_5}.\vspace{-4mm}
		
		\subsection{Motivation}\vspace{-4mm}
		
		Although efforts have been made to combine some of the techniques discussed in \cite{Luca3} with the MPC framework, several key factors detract from the obtained results. More specifically, the procedure proposed in \cite{DMPC1,DMPC2} has a distinctly monolithic character, in which both the control laws' sparsity constraints (which represent the core obstacle in the way of distributed control, and which are routinely tackled \emph{offline}) along with the time-based constraints placed upon the controlled network's variables are enforced \emph{simultaneously}, in an \emph{online setting}.\vspace{-3mm}
		
		The drawbacks mentioned above are notably mitigated in the so-called \emph{tube-based MPC} framework, adopted in \cite{R8A,R8B,R8D}. This advantage is enabled by a set of ancillary control laws that split the applied actuation between themselves and the MPC-generated command signals. Despite this promising perspective, said ancillary controllers are limited to being \emph{static} feedback matrices endowed with restrictive, \emph{block-diagonal} sparsity patterns.\vspace{-4mm}
		
		\subsection{Scope of Work}\label{subsec:scope}\vspace{-4mm}
		
		The results presented in the sequel notably employ the NRF-based design framework from \cite{NRF,aug_sparse} to address the challenges plaguing the previously mentioned works from distributed control literature. Inspired by the approach in \cite{R8A,R8B,R8D}, we propose a bipartite control architecture that divides the monolithic synthesis framework from \cite{DMPC1,DMPC2} into two steps. The first design step, covered in this paper, deals with the \emph{offline} satisfaction of sparsity constraints for the employed control laws. Simultaneously, this paper also addresses the problem of decoupling the target plant into dynamically independent areas, whose variables can subsequently be constrained in a distributed and computationally efficient manner. The latter goal forms the topic of our companion paper \cite{part2}.\vspace{-3mm}
		
		\begin{remark}
			Due to the fundamentally distinct natures of the components that comprise our control architecture, the proposed solution is split into two parts. By propagating this separation to the contents of our two publications (see also our companion paper \cite{part2}), we can coherently present the specificities of each component, in the context of their particular mathematical apparatus. As a result, both of our manuscripts are self-contained, while also addressing a larger problem with an overarching scope.
			\vspace{-4mm}
		\end{remark}
		
		Due to the manner in which we structure our proposed control architecture, the latter takes on a distinctive hierarchical interpretation (see \cite{arch_rev}), with the NRF-based first layer acting as a form of \emph{dynamical interface} between the regulated network and the control laws that form the high-level decision strategies. Indeed, while the NRF-based controllers act directly on the network's inputs, we point out that the second layer also acts upon the information being fed back to the first layer. In doing so, the second layer acts as a supervisory authority for the closed-loop interconnection between the network and the first layer, and the connections with the topic of \emph{reference governors} \cite{refgov} become immediately apparent. Moreover, the theoretical framework established in this paper is intended to ensure a high degree of flexibility in selecting the type of strategy to use for the second layer.\vspace{-4mm}
		
		\begin{remark}
			While the second-layer implementations from our companion paper \cite{part2} are firmly grounded in the MPC framework, we emphasise the fact that this paper's main results are intended to showcase the NRF framework's general applicability in the distributed setting. By leveraging its remarkable closed-loop model-matching properties (see \cite{NRF} and \cite{aug_sparse}), the results presented in the sequel can be used to enhance the feasibility and scalability of numerous control-oriented techniques which do not lend themselves gracefully to the distributed setting.\vspace{-4mm}
		\end{remark}
		
		\subsection{Contributions}\vspace{-4mm}
		
		The chief contributions of the work presented in this paper can be listed via the following three features:\vspace{-3mm}
		\begin{enumerate}
			\item[a)] the introduction of \emph{specialised implementations} for NRF-based control laws, whose sparsity patterns inherit the structure of their NRF pairs and enable communication-constrained hierarchical control;\vspace{0.5mm}
			
			\item[b)] the \emph{complete characterisation} of both exogenous-signal response and initial-condition evolution for an NRF-controlled closed-loop system, in terms of a rich and highly amenable parametrisation;\vspace{0.5mm}
			
			\item[c)] the formulation of a \emph{flexible design framework}, which employs tractable procedures to dynamically decouple a network into smaller areas, all while promoting the synthesis of distributed supervisors.
		\end{enumerate}\vspace{-3mm}
		A further point to emphasise is the fact that, by combining the implementations mentioned above with the MPC framework, one obtains a \emph{novel generalisation} of the tube MPC framework, in which the accompanying control laws are no longer static, but rather dynamical in nature. Indeed, the results presented in the sequel are the key components which enable the development of the aforementioned framework in our companion paper \cite{part2}.\vspace{-3mm}
		
		\subsection{Paper Structure}\vspace{-3mm}
		
		The rest of the paper is structured as follows. Section~\ref{sec:prelims} presents a set of preliminary notions, while Section~\ref{sec:arch} discusses our proposed two-layer architecture and its overarching objectives, with a focus on the first layer. Section~\ref{sec:NRF_theo} holds the main theoretical results concerning our specialised implementations for NRF-based control, and Section~\ref{sec:NRF_des} describes the overall synthesis framework, along with the means of interfacing with the employed upper-layer strategies. The theoretical results from the latter two sections are given practical interpretations through the numerical example in Section~\ref{sec:num_ex}, which revolves around the grid application from \cite{PT}. Finally, Section~\ref{sec:outro} offers a set of concluding remarks. In addition to this, we also include a pair of appendices: one focusing on auxiliary notions (Appendix~\ref{app:aux}), and one containing the proofs of this paper's main results (Appendix~\ref{app:proofs}).
		\vspace{-3mm}
		
	}
	
	\section{Preliminaries}\vspace{-3mm}
	\label{sec:prelims}
	
	\subsection{Notation and Definitions}\vspace{-3mm}
	\label{subsec:not}
	
	Let $\mathbb{N}$, $\mathbb{R}$ and $\mathbb{C}$ denote, respectively, the set of natural, real and complex numbers, while the the \emph{domain of stability} is henceforth denoted as $\mathbb{S}:=\{z\in\mathbb{C}\ \vert\ |z|<1\}$. Let $\mathcal{M}^{p\times m}$ be the set of all $p\times m$ matrices whose entries are scalar elements that are part of a set denoted by $\mathcal{M}$. Similarly, $\mathcal{M}^{p}$ denotes the set of vectors with dimension $p$ and entries in $\mathcal{M}$. For any matrix $M$, $\mathrm{row}_i(M)$ is its $i^\text{th}$ row, $\mathrm{col}_j(M)$ stands for its $j^\text{th}$ column and $\mathrm{elm}_{ij}(M):=\mathrm{row}_i(\mathrm{col}_j(M))$. Moreover, the \emph{transpose} of an arbitrary matrix $M$ is denoted by $M^\top$. For an $A\in\mathbb{C}^{p\times m}$ and an $E\in\mathbb{C}^{p\times m}$, the matrix polynomial $A-z E$ of indeterminate $z\in\mathbb{C}$ is called a \emph{(matrix) pencil}. The vector $e_i$ stands for the $i^\text{th}$ column of the identity matrix, with the latter's dimension being inferred from context. We denote by $\mathcal{R}(z)$ the set of real-rational functions of indeterminate $z\in\mathbb{C}$. A matrix with entries in $\mathcal{R}(z)$ is termed a Transfer Function Matrix (TFM), and all such TFMs will be denoted using boldface letters.
	\vspace{-3mm}

	\subsection{System Representations}
	\label{subsec:sys_nots}\vspace{-3mm}
	
	The class of systems considered in this paper are represented in the time domain by the set of equations\vspace{-3mm}
	\begin{subequations}
		\begin{align}
			x[k+1] =&\ Ax[k]+B_u\, u[k]+B_d\,d[k],\label{eq:ss_a}\\
			y[k]=&\ Cx[k]+D_u\, u[k]+D_d\,d[k]\label{eq:ss_b},
		\end{align}
	\end{subequations}\phantom{ }\vspace{-9mm}
	
	for any \textcolor{black}{$k\in\mathbb{Z}$ with $k\geq k_0\in\mathbb{Z}$}, where $k_0$ represents the initial time instant. For the representations of type \eqref{eq:ss_a}-\eqref{eq:ss_b}, which are referred to as \emph{state-space realisations}, $u$ is the realisation's \emph{controlled input vector}, $d$ its \emph{disturbance vector}, $y$ its \emph{output vector} and $x$ its \emph{state vector}.\vspace{-3mm}
	
	{\color{black}
		The type of network that our procedure aims to control is described by a more particular class of realisations than \eqref{eq:ss_a}-\eqref{eq:ss_b}. These systems also satisfy the fact that\vspace{-3mm}
		\begin{equation}\label{eq:ss_c}\tag{1c}
			C=I,\ D_u = O,\ D_d = O,\vspace{-3mm}
		\end{equation}
		implying that the states are available for measurement. Despite this fact, we \emph{do not also assume} that these readings are perfect (or otherwise exact), and, therefore, we consider that these measurements are affected by some additive measurement noise, henceforth denoted $\zeta[k]$, whose effect can be quantified or bounded in some sense.}\vspace{-3mm} 
	
	{\color{black}
		\begin{remark}\label{rem:dist_pres}
			The assumption that the network's states are available for measurement is made only in the context of the MPC-based framework from our companion paper \cite{part2}, for which this property is standard in literature. Note also that the previously introduced noise $\zeta$ could be assimilated into the disturbance vector $d$, yet doing so would render $D_d$ non-zero, which would encumber the formulation of our main results. The vector $\zeta$ will be kept independent of $d$, and its effect on the closed-loop system designed in the sequel will be compensated for separately.\vspace{-3mm}
		\end{remark}
	}

	For systems of type \eqref{eq:ss_a}-\eqref{eq:ss_c}, we denote by $x_{c}\in\mathbb{R}^{n_x}$ the initial condition of the state vector and we also consider $A\in\mathbb{R}^{n_x\times n_x}$, $B_u\in\mathbb{R}^{n_x\times n_u}$, $B_d\in\mathbb{R}^{n_x\times n_d}$, $C\in\mathbb{R}^{n_y\times n_x}$, while, more generically, we consider  $D_u\in\mathbb{R}^{n_y\times n_u}$ and $D_d\in\mathbb{R}^{n_y\times n_d}$. The positive integer $n_x$ will be referred to as the \emph{order} of the realisation, and we also denote any realisation of type \eqref{eq:ss_a}-\eqref{eq:ss_b} in compact form via a matrix quadruplet of type $\left(A,\begin{bmatrix}
		B_u & B_d
	\end{bmatrix}, C, \begin{bmatrix}
		D_u & D_d
	\end{bmatrix}\right)$. For every system which is represented via \eqref{eq:ss_a}-\eqref{eq:ss_b}, the pencil $A-z I_{n_x}$ is regular and is referred to as the realisation's \emph{pole pencil}. Additionally, the connection between a system described by \eqref{eq:ss_a}-\eqref{eq:ss_b} and its TFM is given by\vspace{-3mm}
	\begin{multline}\label{eq:TFM_def}
		\mathbf{G}(z)=\left[\scriptsize\begin{array}{c|cc}
			A-z I_{n_x} & B_u & B_d \\\hline C & D_u & D_d
		\end{array}\right]:=\\:=C(z I_{n_x}-A)^{-1}\begin{bmatrix}
			B_u & B_d
		\end{bmatrix}+\begin{bmatrix}
			D_u & D_d
		\end{bmatrix}.
	\end{multline}
	\phantom{ }\vspace{-9mm}
	
	Finally, given any $\mathbf{H}\in\mathcal{R}(z)^{n_y\times n_u}$ which is proper (see Appendix~\ref{app:aux}), the operation $y[k]=\mathbf{H}(z)\star u[k]$ denotes the time-response of $\mathbf{H}(z)$ to an input signal vector $u[k]$, which can be computed as follows\vspace{-3mm}
	\begin{equation}\label{eq:io_resp}
		\scriptsize\begin{array}{l}
			y[k]=\mathbf{H}(z)\star u[k] = D_Hu[k] + \sum_{\textcolor{black}{i\geq1}} C_H^{}A_H^{i-1}B_H^{}u[k-i],
		\end{array}\hspace{-1mm}\vspace{-3mm}\normalsize
	\end{equation}
	for any realisation $(A_H,B_H,C_H,D_H)$ of $\mathbf{H}(z)$. \textcolor{black}{Additionally, in accordance with the dynamics from \eqref{eq:ss_a}-\eqref{eq:ss_b}, we point out that all of the signal vectors considered in this paper are taken to be equal to zero for all $k<k_0$.}\vspace{-3mm}
	
	{\color{black}
		\begin{remark}\label{rem:db}
			Although we focus on discrete-time systems in this manuscript, we point out the fact that the NRF framework was developed in \cite{NRF} for both discrete- and continuous-time contexts. Thus, all of the results presented in the sequel can also be derived in the continuous-time context with only minor alterations. However, given that our companion paper \cite{part2} employs MPC (primarily formalised for discrete-time systems) as the design framework for our second layer, we have opted to omit the continuous-time analogues of our main results in the sequel. While the first layer can also be designed in the continuous-time setting (see \cite{NRF}) and the resulting closed-loop system can be discretised to fit into the scope of \cite{part2}, note that this option precludes the use of discrete-time-exclusive design techniques (such as \emph{deadbeat control}; see, for example, Section~10.8 in \cite{Kuo}) which are highly beneficial for the synthesis of the second layer.
			\vspace{-3mm}
		\end{remark}
	}

	\begin{figure*}[t]
		\centering
		\resizebox{.95\textwidth}{!}{
			\begin{tikzpicture}[scale=0.6]
				\draw [thick,rounded corners=1]  (-15,-0.5) rectangle +(45,1);
				\draw [thick]  (-2,0)   node {Interconnected};
				\draw [thick]  (5.5,0)   node {Dynamical};
				\draw [thick]  (13,0)   node {System};
				
				\draw [thick,rounded corners=1]  (-1.5-11,2.5) rectangle +(3,1.5);
				\draw [thick]  (0-11,3.25)   node {NRF $1$};
				\node at (4-11,1) {\textbf{Area} $\mathbf 1$};
				\draw [thick,rounded corners=1]  (-3.5-11,6) rectangle +(5,1.5);
				\draw [thick]  (-1-11,6.75)   node {MPC $1$};
				\draw[->,thick] (0.5-11,0.5)--(0.5-11,2.5);
				\node at (1-11,1.5) {$x_1$};
				\draw[->,thick] (-0.5-11,2.5)--(-0.5-11,0.5);
				\node at (-1-11,1.5) {$u_{f1}$};
				\draw[->,thick] (0.5-11,4)--(0.5-11,6);
				\node at (1.7-11,5) {$\{x_1,w_1\}$};
				\draw[->,thick] (-0.5-11,6)--(-0.5-11,4);
				\node at (-1.1-11,5) {$u_{s11}$};
				\draw[->,thick] (-2.5-11,6)--(-2.5-11,0.5);
				\node at (-3.2-11,3.25) {$u_{s21}$};
				\draw[->,thick] (1.5-11,2.8)--(16.5-11,2.8);
				\node at (3.7-11,2.3) {$u_{f1}$};
				\draw[->,thick] (16.5-11,3.6)--(1.5-11,3.6);
				\node at (8-11,4) {$\{x_{j\in\mathcal{N}_1,j\neq 1},u_{f(j\in\mathcal{N}_1,j\neq 1)},u_{s1(j\in\mathcal{N}_1,j\neq 1)}\}$};
				\draw[->,thick] (1.5-11,6.3)--(16.5-11,6.3);
				\node at (4.75-11,5.75) {$\{x_1,w_1,u_{s11}\}$};
				\draw[->,thick] (16.5-11,7.1)--(1.5-11,7.1);
				\node at (6-11,7.5) {$\{x_{j\in\mathcal{N}_1,j\neq 1},w_{j\in\mathcal{N}_1,j\neq 1}\}$};
				\draw[dash dot,thick] (-4.5-11,-1)--(-4.5-11,8);
				\draw[dash dot,thick] (14-11,-1)--(14-11,8);
				
				\node at (5.6,4.5) {$\cdots$};
				
				\draw [thick,rounded corners=1]  (-1.5+13,2.5) rectangle +(3,1.5);
				\draw [thick]  (0+13,3.25)   node {NRF $i$};
				\node at (4+13,1) {\textbf{Area} $\mathbf i$};
				\draw [thick,rounded corners=1]  (-3.5+13,6) rectangle +(5,1.5);
				\draw [thick]  (-1+13,6.75)   node {MPC $i$};
				\draw[->,thick] (0.5+13,0.5)--(0.5+13,2.5);
				\node at (1+13,1.5) {$x_i$};
				\draw[->,thick] (-0.5+13,2.5)--(-0.5+13,0.5);
				\node at (-1+13,1.5) {$u_{fi}$};
				\draw[->,thick] (0.5+13,4)--(0.5+13,6);
				\node at (1.6+13,5) {$\{x_i,w_i\}$};
				\draw[->,thick] (-0.5+13,6)--(-0.5+13,4);
				\node at (-1.1+13,5) {$u_{s1i}$};
				\draw[->,thick] (-2.5+13,6)--(-2.5+13,0.5);
				\node at (-3.2+13,3.25) {$u_{s2i}$};
				\draw[->,thick] (1.5+13,2.8)--(16.5+13,2.8);
				\node at (3.7+13,2.3) {$u_{fi}$};
				\draw[->,thick] (16.5+13,3.6)--(1.5+13,3.6);
				\node at (8+13,4) {$\{x_{j\in\mathcal{N}_i,j\neq i},u_{f(j\in\mathcal{N}_i,j\neq i)},u_{s1(j\in\mathcal{N}_i,j\neq i)}\}$};
				\draw[->,thick] (1.5+13,6.3)--(16.5+13,6.3);
				\node at (4.75+13,5.75) {$\{x_i,w_i,u_{s1i}\}$};
				\draw[->,thick] (16.5+13,7.1)--(1.5+13,7.1);
				\node at (6+13,7.5) {$\{x_{j\in\mathcal{N}_i,j\neq i},w_{j\in\mathcal{N}_i,j\neq i}\}$};
				\draw[dash dot,thick] (-4.5+13,-1)--(-4.5+13,8);
				\draw[dash dot,thick] (14+13,-1)--(14+13,8);
				
				\node at (29.5,4.5) {$\cdots$};
				
				\draw [thick,color=white,fill=white]  (29.5,-0.6) rectangle +(1,1.2);
				
				\node at (29.5,0) {$\cdots$};
		\end{tikzpicture}}\vspace{-3mm}
		\caption{High-level implementation scheme depicting the proposed distributed control strategy}\vspace{-1mm}
		\label{fig:scheme}
		\hrulefill\vspace{-2mm}
	\end{figure*}

	\subsection{Distributed Networks}
	\label{subsec:distrib_part}\vspace{-3mm}
	
	Although we consider that $x$ and $u$ are completely accessible to us, in terms of measurement and actuation, respectively, we do not assume that manipulating \emph{all} of these variables is possible from any single location in our physical network. Indeed, a particular feature of distributed networks is the fact that their controlled inputs and measurable state variables are often spatially distributed. This dispersal naturally separates the network into $N\in\mathbb{N}$ distinct areas, with $N>1$.\vspace{-3mm}
	
	To each of the network's $N$ areas, we now assign two sets of indices which denote the entries of $x$ and $u$ that are (uniquely) accessible for measurement or actuation in that area. Without loss of generality, we assume that these indices are assigned to the sets in ascending order, since the rows and columns of the matrices expressed in \eqref{eq:ss_a}-\eqref{eq:ss_b} may be permuted and then redenoted at will. Therefore, each area will be represented by the pair\vspace{-3mm}
	\begin{equation}\label{eq:trip}
		\mathcal{A}_i:=(\mathcal{A}_{xi},\mathcal{A}_{ui}),\,\forall\,i\in\{1:N\},\vspace{-3mm}
	\end{equation}
	which collects the two aforementioned index sets.\vspace{-3mm}
	
	To streamline the definition of the sets given in \eqref{eq:trip}, we will provide generic expressions via the placeholder subscript $\bullet$, which stands for any one of the subscripts $\{x,u\}$. Let each set contain a number of $n_{\bullet i}\in\mathbb{N}$ indices, where $n_{\bullet i}>0,\,\forall\,i\in\{1:N\}$, such that we define\vspace{-2mm}
	\begin{equation}\label{eq:net_part}
		\hspace{-1mm}\left\{\begin{aligned}
			\mathcal{A}_{\bullet i}:=\{\alpha_{\bullet i}+j\ \vert\ j\in1:n_{\bullet i}\},\,\forall\,i\in\{1:N\},\\
			\alpha_{\bullet 1}:=0,\, \alpha_{\bullet \ell}:=\alpha_{\bullet (\ell-1)}+n_{\bullet (\ell-1)},\,\forall\,\ell\in\{2:N\}.
		\end{aligned}\right.\vspace{-3mm}
	\end{equation}
	As an example\footnote{An additional partitioning will be showcased in Section~\ref{sec:num_ex}.}, a network with 12 states and 7 inputs could be partitioned as in \eqref{eq:trip}-\eqref{eq:net_part} via the following collection of disjoint sets $
	\{(\{1:2\},\{1\}),(\{3:5\},\{2:3\}),$ $(\{6:11\},\{4:5\}),(\{12\},\{6:7\})\}$. Additionally, to each partition in \eqref{eq:net_part}, we attach the following selection matrix
	\vspace{-3mm}
	\begin{equation}\label{eq:sel_mat}
		S_{\bullet i}:=\left[\scriptsize\begin{array}{ccc}
			e_{\alpha_{\bullet i}+1}&\dots&e_{\alpha_{\bullet i}+n_{\bullet i}}
		\end{array}\right]\in\mathbb{R}^{n_{\bullet}\times n_{\bullet i}},\,\forall\,i\in\{1:N\},\vspace{1mm}
	\end{equation}
	for which $x_i[k]:=S^\top_{{xi}}x[k]$, $x_{c{i}}:=S^\top_{{xi}}x_c$, $u_i[k]:=S^\top_{{ui}}u[k]$.\vspace{-3mm} 
	
	{\color{black}
		\begin{remark}\label{rem:part}
			The restriction mentioned above, regarding the disjoint nature of the area-based collections, is in no way restrictive from a practical standpoint. In most applications, this partitioning is based on spatial arguments: some of the system's states can only be measured at certain locations, whereas the actuators associated with the command signals are positioned at distinct locations throughout the network. Indeed, the notions of \emph{separation} and \emph{locality} are among the core topics addressed in distributed control literature \cite{matni1,DMPC1,DMPC2}. Crucially, however, we point out that the imposed separation has more to do with controller implementation than with system-theoretical concerns. To each localised set of actuators (corresponding to a set of inputs), we assign one of the subcontrollers designed in the sequel, whereas selecting an area's states merely designates them as the particular target of that area's second-layer subcontroller (as designed and implemented in our companion paper; see \cite{part2}).\vspace{-3mm}
		\end{remark}
	}	
	
	\section{The Control Architecture}\label{sec:arch}\vspace{-3mm}
	
	\subsection{Global Perspective and Objectives}\label{subsec:global_arch}\vspace{-3mm}
	
	The aim of this manuscript and of its companion paper \cite{part2} is to formalise the design framework of the two-layer architecture for the interconnected dynamical system illustrated in Figure~\ref{fig:scheme}, at the top of the next page. By identifying the subset of areas $\mathcal{N}_i\subseteq\{1:N\}$ which can transmit information to the network's $i^\text{th}$ area (and which satisfy $i\in\mathcal{N}_i$), for all $i\in\{1:N\}$, the communication graph of the network is now explicitly designated.\vspace{-3mm}
	
	{\color{black}
		\begin{remark}
			We point out that it is the neighbourhoods denoted as $\mathcal{N}_i$ which encode the system-theoretical concerns mentioned previously in Remark~\ref{rem:part}, rather than the areas in \eqref{eq:net_part}. These sets dictate the flow of information through the closed-loop system that will be designed in the sequel and, crucially, reinforce the claim (made in the aforementioned remark) that our partitioning is in no way restrictive. To this end, notice the fact that the states and the computed commands of the $i^\text{th}$ area can be freely shared with all areas whose index $j$ satisfies $i\in\mathcal{N}_j$. Thus, introducing this notion addresses the issue of using the aforementioned variables (for computational purposes) across multiple areas. On the other hand, note that the computed command signals may \emph{only} be applied to the actuators of their assigned area, and that the cross-coupling between these signals and the other areas' (state) variables is one of the key topics treated in the sequel.\vspace{-3mm}
		\end{remark}
	}

	In light of the above comments, the fundamental operating principle of our control scheme is the following:\vspace{-3mm}
	\begin{enumerate}
		\item[a)] the\, state-space-implemented\, NRF\, subcontrollers\newline receive network state information along with the command signals of other first-layer subcontrollers from their areas' neighbourhoods, and use these values to compute the signals denoted $u_{fi}$ in Figure~\ref{fig:scheme};\smallskip
		
		\item[b)] the MPC subcontrollers also receive network state information along with the state variables of the NRF subcontroller implementations, denoted $w_i$ in Figure~\ref{fig:scheme}, from their areas' neighbours to compute:\smallskip
		
		\begin{enumerate}
			\item[b1)] the command signals denoted by $u_{s1i}$ in Figure~\ref{fig:scheme}, which \emph{are broadcast} and combined additively with the state vectors $x_i$, before the later are fed to the NRF subcontrollers;\smallskip
			
			\item[b2)] the command signals denoted by $u_{s2i}$ in Figure~\ref{fig:scheme}, which \emph{remain local} and are combined additively with the NRF outputs $u_{fi}$, to obtain the control signals being sent to the actuators.\vspace{-3mm}
		\end{enumerate}
	\end{enumerate}
	
	\begin{remark}
		Regarding the control architecture's design, we point out that both the first and second layers employ feedback and feedforward components. However, their goals and their communication-based mechanisms are completely different. Whereas the first layer focuses on dynamical decoupling and disturbance rejection, the second layer concerns itself with constraint satisfaction and feasibility, in the sense of robust control invariance.\vspace{-3mm}
	\end{remark}
	
	Since the second layer is built on top of the closed-loop interconnection between the first layer and the network, it is natural to begin our discussion with the architecture and requirements of the first layer, along with the manner in which the latter interfaces with the network. Therefore, we defer all procedural discussions regarding the second layer (such as the standalone design algorithm) to our companion paper \cite{part2}, and we denote its global command signals as $u_{s1}[k]:=\tiny\begin{bmatrix}
		u_{s11}^\top[k] & \dots & u_{s1N}^\top[k]
	\end{bmatrix}^\top$ and $u_{s2}[k]:=\tiny\begin{bmatrix}
		u_{s21}^\top[k] & \dots & u_{s2N}^\top[k]
	\end{bmatrix}^\top$, formed by concatenating the area-based command signals shown in Figure~\ref{fig:scheme}.\vspace{-4mm}
	
	{\color{black}
		\subsection{Fundamentals of NRF-based Control}\label{subsec:NRF_basic}\vspace{-4mm}
		
		The NRF-based representation for distributed controller implementation was prototyped in \cite{plutonizare} and subsequently formalised in the most general setting in \cite{NRF}. This design framework aims to bypass the restrictions of well-established distributed control procedures (see \cite{NRF} and \cite{aug_sparse} for a comprehensive discussion on these limitations), by producing highly scalable control laws whose specialised implementations were only directly applicable (see \cite{SLS} along with \cite{JA}) to open-loop stable networks.\vspace{-4mm}
		
		The key concept around which this paradigm revolves is the so-called NRF pair. For a dynamical controller of dimension $n_u\times n_x$, whose TFM will henceforth be denoted as $\mathbf{K}(z)$, we introduce the factored representation\vspace{-4mm}
		\begin{equation*}
			\mathbf{K}(z) = (I_{n_u}-\mathbf{\Phi}(z))^{-1}\mathbf{\Gamma}(z),\vspace{-3mm}
		\end{equation*}
		for which $\mathrm{det}(I_{n_u}-\mathbf{\Phi}(z))\not\equiv 0$ and $\mathrm{elm}_{ii}({\Phi}(z))\equiv 0$, for all $i\in\{1:n_u\}$. The TFM pair $(\mathbf{\Phi}(z),\mathbf{\Gamma}(z))$, for which $\mathbf{\Phi}\in\mathcal{R}(z)^{n_u\times n_u}$ and $\mathbf{\Gamma}\in\mathcal{R}(z)^{n_u\times n_x}$ are proper and which forms an \emph{NRF representation} of $\mathbf{K}(z)$, allows for the computation of the signal $u[k]$ in \eqref{eq:ss_a}-\eqref{eq:ss_c} as follows\vspace{-3mm}
		\begin{equation}\label{eq:NRF_general}
			u[k] = \mathbf{\Phi}(z)\star u[k]+\mathbf{\Gamma}(z)\star x[k].\vspace{-3mm}
		\end{equation}
		Indeed, since $\mathrm{elm}_{ii}({\Phi}(z))\equiv 0$, for all $i\in\{1:n_u\}$, this enables a distinct feedback-feedforward expression of the resulting control laws, in which a particular command signal $\mathrm{row}_i(u[k])$ may always be computed via the use of a feedback component $x[k]$ alongside the feedforward of the command signals denoted $\mathrm{row}_j(u[k])$, which are assigned to other inputs of the network ($j\neq i$).\vspace{-3mm}
		
		\begin{remark}
			In the NRF framework, the problem of imposing sparsity on $\mathbf{K}(z)$ for a distributed implementation (a challenging objective, see \cite{NRF} and \cite{aug_sparse}) is relaxed into ensuring sparsity for its NRF pair $(\mathbf{\Phi}(z),\mathbf{\Gamma}(z))$. The latter goal is far more tractable, due to the way in which said pair may be computed (see, in particular, Appendix~\ref{app:aux}).\vspace{-3mm}
		\end{remark}
		One of the key properties of the control laws in \eqref{eq:NRF_general} is their ability to freely place within $\mathbb{C}$ the poles of all closed-loop maps. For some subset $\mathbb{C}_{g}\subseteq\mathbb{C}$ with desirable properties, we will refer to control laws of type \eqref{eq:NRF_general} which ensure that all closed-loop maps are described by proper TFMs and which place all of their poles (see Appendix~\ref{app:aux}, along with the main result of \cite{NRF}) in said subset as \emph{$\mathbb{C}_g$-allocating}.\vspace{-3mm}
		\begin{remark}\label{rem:NRF_stab}
			The choice of $\mathbb{C}_g$ underlines several key properties of the resulting NRF-based control laws. For example, choosing $\mathbb{C}_g\subseteq\mathbb{S}$ ensures that the effect of \emph{bounded} exogenous disturbance on the signals arising in closed-loop configuration \emph{is guaranteed to be bounded}.\vspace{-3mm}
		\end{remark}
	}
	
	\begin{figure}[t]
		\centering
		\begin{tikzpicture}[scale=0.225]
			\draw [thick] [->] (-2,0) -- (3.5,0);
			\draw [thick]  (4,0) circle(0.5);
			\draw [thick]  (4,0) node {$+$};
			\draw [thick] (3,1)   node {\bf{ }} (0,1) node {$\beta_x$};
			\draw [thick] [->] (4.5,0) -- (8,0);
			\draw [thick, rounded corners=1]  (8,-1.5) rectangle +(3,3);
			\draw [thick] (9.5,0)   node {{$\,\mathbf{K}_{\mathbf{D}}$}} ;
			\draw [thick] [->] (11.1,0) -- (15.5,0);
			\draw [thick]  (16,0) circle(0.5cm);
			\draw [thick]  (16,0) node {$+$};
			\draw [thick] (12.6,0.8)   node {$u_f$} ;
			\draw [thick] [->] (14,0) -- (14,4)--(10,4);
			\draw [thick] (18.6,0.8)   node {$u$} ;
			\draw [thick] (14,0)   node {$\bullet$} ;
			\draw [thick]  (14.8,0.7)   node {\bf{ }};
			\draw [thick] [->] (16,5) -- (16,0.5);
			\draw [thick]  (0,4)  node[anchor=south] {$\beta_f$};
			\draw [thick]  (9.5,4) circle(0.5);
			\draw [thick]  (9.5,4) node {$+$};
			\draw [thick] [->] (9.5,3.5) -- (9.5,1.5);
			\draw [thick] [->] (-2,4) -- (9,4);
			\draw [thick]  (16,5.1)  node[anchor=south] {$\beta_u$};
			\draw [thick] [->] (16.5,0) -- (21,0);
			\draw [thick,rounded corners=1]  (21,-1.5) rectangle +(3,3) ;
			\draw [thick]  (22.5,0)   node {{${\bf G}$}} ;
			\draw [thick]  (26,0.7)   node {$x$};
			\draw [thick] [->] (24,0) -- (32,0);
			\draw [thick] (28,0)   node {$\bullet$} ;
			\draw [thick] [->] (28,0) -- (28,-3) -- (4,-3)-- (4, -0.5);
			\draw [thick] [->] (22.5,5) -- (22.5,1.5);
			\draw [thick]  (22.5,5.1)  node[anchor=south] {$d$};
		\end{tikzpicture}\vspace{-2mm}
		\caption{Feedback loop of a network's model ${\bf G}(z)$ with the NRF-based implementation $\mathbf{K}_{\bf \mathbf{D}}(z)$\vspace{-2mm}}
		\label{fig:NRF_implem}
	\end{figure}
	
	{\color{black}
		
		These {$\mathbb{C}_g$-allocating} and NRF-based controllers are implemented as shown in Figure~\ref{fig:NRF_implem} above, in which the distributed controller is represented by the composite TFM\vspace{-3mm}
		\begin{equation}\label{eq:Kd_def}
			\mathbf{K}_{\mathbf{D}}(z):=\begin{bmatrix}
				\mathbf{\Phi}(z) & \mathbf{\Gamma}(z)
			\end{bmatrix}.\vspace{-3mm}
		\end{equation}
		Alongside those introduced in Section~\ref{subsec:sys_nots}, we consider a collection of additional exogenous signals, which we henceforth group into the following compound vectors:\vspace{-3mm}
		\begin{enumerate}
			\item[a)] the state-feedback disturbance vector, defined as $\beta_x[k]:=\zeta[k]+u_{s1}[k]+\beta_{s1}[k]$ and composed of the state measurement noise $\zeta[k]$, of the second-layer command vector $u_{s1}[k]$, and of the encoding errors associated with broadcasting the latter to the first-layer subcontrollers, denoted as $\beta_{s1}[k]$;\vspace{0.5mm}
			
			\item[b)] the plant-input disturbance vector, which is defined as $\beta_u[k]:=u_{s2}[k]+\beta_{s2}[k]$ and is composed of the second-layer command vector $u_{s2i}[k]$, and of the encoding errors associated with transmitting the latter to be summed up with $u_{fi}[k]$, denoted as $\beta_{s2}[k]$;\vspace{0.5mm}
			
			\item[c)] the NRF-feedforward disturbance vector denoted $\beta_f[k]$, which arises in the context of communication between neighbouring first-layer subcontrollers.\vspace{-3mm}
		\end{enumerate}
		We exclusively employ the compound vectors introduced above since, from the point of view of the first-layer control system, the dynamics by which the components themselves (such as the three which make up $\beta_x$) propagate in closed-loop configuration are the same. Additionally, we also concatenate all of these compound signals along with $d$ from \eqref{eq:ss_a}-\eqref{eq:ss_c} into the vector\vspace{-3mm}
		\begin{equation*}
			d_s[k]:=\scriptsize\begin{bmatrix}
				\beta_{x}^\top[k] & \beta_{u}^\top[k] & \beta_f^\top[k] & d^\top[k]
			\end{bmatrix}^\top.\vspace{-6mm}
		\end{equation*}

	}

	\subsection{Objectives of the First Layer}\label{subsec:NRF_obj}\vspace{-3mm}

	{\color{black} Although the ultimate goal of our endeavour is represented by a distributed constraint management policy, previous results from literature \cite{plat_MPC_4} have shown that employing ancillary control laws can ensure a dramatic increase in performance (and a reduction of conservativeness) over the case in which, for example, MPC-based strategies are applied \emph{directly and independently} to the controlled network \cite{plat_MPC_1,plat_MPC_2,plat_MPC_3}. Indeed, one of the core approaches for applying MPC in the distributed setting involves the careful design of said ancillary controllers \cite{R8A,R8B,R8D}. The main obstacle in the way of implementing this type of policy is the fact that, as shown in Figure~\ref{fig:scheme}, the second-layer subcontrollers do not exchange information regarding the values of $u_{s1i}[k]$ and $u_{s2i}[k]$.\vspace{-3mm} 
		
		\begin{remark}
			This lack of information is owed to the fact that all of these signals must be computed simultaneously by separate subcontrollers. Consensus-based strategies tackle this problem explicitly (see, for example, \cite{DMPCB}), by successively exchanging and negotiating the values of said signals, yet this approach is highly demanding from both a computational and communicational point of view.\vspace{-3mm}
		\end{remark}

		The resulting closed-loop system must therefore compensate for any cross-coupling induced by the control actions of other subcontrollers in the second layer, while exchanging information only via the network's available communication infrastructure. Thus, the primary role of our control architecture's first layer is to alleviate all of the design challenges stated above, by concurrently achieving the following objectives:\vspace{-3mm}}
	\begin{enumerate}
		\item[I)] it must be implementable in a distributed manner, using sparse state-space models that are computationally inexpensive to deploy (Section~\ref{subsec:ss_implems});\vspace{0.5mm}
		
		\item[II)] it must render the closed-loop maps of its interconnection as simple expressions, by means of a rich and highly tractable parametrisation (Section~\ref{subsec:theo_NRF});\vspace{0.5mm}
		
		\item[III)] it must enable the dynamical decoupling of the network areas discussed in Sections~\ref{subsec:distrib_part}~and~\ref{subsec:global_arch}, for the benefit of the second layer, through numerically sound procedures available in literature (Section~\ref{sec:NRF_des});\vspace{-3mm}
		
	\end{enumerate}
	
	\begin{figure*}
		\begin{equation}\label{eq:area_NRF}\tag{16}
			\mathbf{K}_{\mathbf{D}i}(z)=\left[\scriptsize\begin{array}{c|c}
				A_{wi}-z I_{n_{wi}} & B_{wi} \\ \hline C_{wi} & D_{wi}
			\end{array}\right]:= \left[\tiny\begin{array}{ccc|c}
				A_{r(\alpha_{ui}+1)}-z I_{n_{r(\alpha_{ui}+1)}} & & & B_{r(\alpha_{ui}+1)} \\
				& \ddots & & \vdots \\
				& & A_{r(\alpha_{ui}+n_{ui})}-z I_{n_{r(\alpha_{ui}+n_{ui})}} & B_{r(\alpha_{ui}+n_{ui})} \\ \hline 
				C_{r(\alpha_{ui}+1)} & & & D_{r(\alpha_{ui}+1)}\\
				& \ddots & & \vdots\\
				& & C_{r(\alpha_{ui}+n_{ui})} & D_{r(\alpha_{ui}+n_{ui})}\\
			\end{array}\right].\vspace{1mm}
		\end{equation}
		\hrulefill\vspace{-2mm}
	\end{figure*}
	
	The remainder of this paper presents solutions to all of the design challenges stated above.\vspace{-3mm}
	
	\section{Theoretical Results}\vspace{-3mm}
	\label{sec:NRF_theo}
	
	\subsection{State-space Implementations for NRF Control}\vspace{-3mm}
	\label{subsec:ss_implems}

	The main feature of the NRF framework is the fact that the command signal vector $u_f$ can be computed in a \emph{distributed} manner, by enforcing an appropriate sparsity pattern upon the controller's NRF pair (see \cite{NRF}). For the distributed setting described in Section~\ref{subsec:distrib_part}, it is possible to employ the TFM from \eqref{eq:Kd_def} to \emph{independently} compute\vspace{-3mm}
	\begin{equation}\label{eq:uf_implem}
		\hspace{-1mm}u_{fi}[k]:=S_{ui}^\top u_f[k] = \mathbf{K}_{\mathbf{D}i}(z)\star\scriptsize\begin{bmatrix}
			u_f[k]+\beta_f[k] \\ x[k]+\beta_{x}[k]
		\end{bmatrix},\normalsize\vspace{-3mm}
	\end{equation}
	for all $i\in\{1:N\}$, with the TFMs $\mathbf{K}_{\mathbf{D}i}(z):=S_{ui}^\top\mathbf{K}_{\mathbf{D}}(z)$ denoting each area's NRF-based subcontroller.

	Although the subcontrollers $\mathbf{K}_{\mathbf{D}i}(z)$ from \eqref{eq:uf_implem} can always be implemented \emph{separately} in the control scheme from Figure~\ref{fig:NRF_implem}, this does not immediately lead to tractable formulations for distributed constraint management. To be effective, these subcontrollers must be accompanied by state-space representations that mirror the sparsity patterns of their TFMs. The next constructive result, given below, addresses precisely this issue.\vspace{-3mm}

	\begin{proposition}\label{prop:ss_implem}
		Let the rows of the TFM from \eqref{eq:Kd_def} be described by the \emph{minimal} realisations\vspace{-3mm}
		\begin{equation}\label{eq:Kd_rows}
			\mathrm{row}_\ell\left(\mathbf{K}_{\mathbf{D}}(z)\right)=\scriptsize\left[ \begin{array}{c|c}
				\widehat A_{\ell}-z I_{n_{r\ell}} & \widehat B_{\ell} \\ \hline \widehat C_{\ell} & \widehat D_{\ell}
			\end{array}\right]\normalsize, \forall\, \ell\in\{1:n_u\},\vspace{-3mm}
		\end{equation}for which we define the polynomials\vspace{-3mm}
		\begin{equation}\label{eq:min_poly}
			\hspace{-1mm}\chi_\ell(z):=\det(zI_{n_{r\ell}}-\widehat A_{\ell})=z^{n_{r\ell}}+\textstyle\sum_{j=1}^{n_{r\ell}}a_{j\ell}z^{(n_{r\ell}-j)}.\hspace{-1mm}\vspace{-3mm}
		\end{equation}
		Then, by expressing the aforementioned row TFMs as\vspace{-3mm}
		\begin{equation}\label{eq:TFM_coefs}
			\mathrm{row}_\ell\left(\mathbf{K}_{\mathbf{D}}(z)\right)=\widehat D_{\ell}+\tfrac{1}{\chi_\ell(z)}\textstyle\sum_{j=1}^{n_{r\ell}}K_{j \ell} z^{(n_{r\ell}-j)},\vspace{-3mm}
		\end{equation}
		we have that:\vspace{-3mm}
		\begin{enumerate}
			\item[i)] The following identities hold\vspace{-3mm}
			\begin{equation}\label{eq:Kd_implem}
				\mathrm{row}_\ell\left(\mathbf{K}_{\mathbf{D}}(z)\right)=\left[\scriptsize\begin{array}{c|c}
					A_{r\ell}-z I_{n_{r\ell}} & B_{r\ell} \\ \hline C_{r\ell} & D_{r\ell}
				\end{array}\right],\vspace{-3mm}
			\end{equation}
			for all $\ell\in\{1:n_u\}$, where\vspace{-3mm}
			\begin{equation}\label{eq:mat_coef}
				\left\{\begin{aligned}
					\widetilde{A}_{r\ell}:=&\ \scriptsize\begin{bmatrix}
						-a_{1\ell} & \dots & -a_{(n_{r\ell}-1)\ell}
					\end{bmatrix}^\top,\\ A_{r\ell}:=&\ \scriptsize\begin{bmatrix}
						\widetilde{A}_{r\ell} & I_{n_{r\ell}-1}\\
						-a_{n_{r\ell}\ell} & O
					\end{bmatrix},\\
					\ B_{r\ell}:=&\ \scriptsize\begin{bmatrix}
						K_{1\ell}^\top &\dots& K_{n_{r\ell}\ell}^\top
					\end{bmatrix}^\top,\ C_{r\ell}:=e_1^\top,\\
					D_{r\ell}:=&\ \widehat D_{\ell}=\lim_{|z|\rightarrow\infty}\mathrm{row}_\ell\left(\mathbf{K}_{\mathbf{D}}(z)\right);
				\end{aligned}\right.\vspace{-3mm}
			\end{equation}
			
			\item[ii)] The realisations given in \eqref{eq:Kd_implem} are \emph{minimal};\medskip
			
			\item[iii)] The columns of $B_{r\ell}$ and $D_{r\ell}$ have \emph{the same sparsity pattern} as $\mathrm{row}_\ell\left(\mathbf{K}_{\mathbf{D}}(z)\right)$, which is to say that\vspace{-2mm}
			\begin{equation}
				\mathrm{elm}_{\ell j}\left(\mathbf{K}_{\mathbf{D}}(z)\right) \equiv 0 \Longrightarrow \left\{\begin{array}{l}
					\mathrm{col}_{j}(B_{r\ell}) = O,\\
					\mathrm{col}_{j}(D_{r\ell}) = O,
				\end{array}\right.\vspace{-2mm}
			\end{equation}
			for all $j\in\{1:n_x+n_u\}$;\medskip
			
			\item[iv)] The subcontrollers $\mathbf{K}_{\mathbf{D}i}(z)$ can be implemented, for all $i\in\{1:N\}$, via the structured realisations given in \eqref{eq:area_NRF}, which are located at the top of this page.\vspace{-6mm}
		\end{enumerate}
	\end{proposition}
	\begin{pf}
		See Appendix~\ref{app:proofs}.\vspace{-2mm}
	\end{pf}
	
	\begin{remark}\label{rem:sparse_real}
		The polynomials $\chi_\ell(z)$ can be computed in a numerically reliable manner by first effecting a change of coordinates in the realisation from \eqref{eq:Kd_implem} which brings $\widehat{A}_\ell$ to Real Schur Form. This representation can also be used to compute the constant real-valued row vectors $K_{j\ell}$.\vspace{-3mm}
	\end{remark}\stepcounter{equation}
	
	{\color{black} To highlight one of the key benefits of Proposition~\ref{prop:ss_implem}, we refer to Figure~\ref{fig:NRF_zoom} at the top of the next page, which illustrates\footnote{\color{black}Although the additive disturbance $\beta_{wi}$ has no \emph{direct} impact on the first layer's closed-loop system, it is nevertheless represented for the sake of completeness, as its effects on the second layer will be referred to in our companion paper \cite{part2}.} a more detailed and implementation-centric view of one of the NRF blocks that are depicted only schematically in Figure~\ref{fig:scheme}. Assuming that the following set of algebraic conditions has been enforced\vspace{-2mm}
		\begin{equation}\label{eq:first_sparse}
			\hspace{-3mm}\scriptsize\begin{array}{l}
				\mathbf{K}_{\mathbf{D}i}(z)\mathrm{diag}(S_{uj},S_{xj})\equiv O,\,\forall\,i\in\{1:N\},\,j\in\{1:N\}\setminus\,\mathcal{N}_i,
			\end{array}\normalsize\hspace{-3mm}\vspace{-2mm}
		\end{equation}
		the closed-loop system obtained by implementing \eqref{eq:area_NRF} as in Figure~\ref{fig:NRF_zoom} is \emph{equivalent} to the one depicted in Figure~\ref{fig:NRF_implem}, since all of the signals outside of $\mathcal{N}_i$ have no impact on the \emph{computation of $u_{fi}$} and, therefore, there is no need to transmit them. This fact forms the core feature of our distributed implementations, since it ensures that all of the benefits associated with the closed-loop system in Figure~\ref{fig:NRF_implem} (which will be amply discussed in the sequel) apply to the implementations depicted in Figure~\ref{fig:NRF_zoom}. Note that the satisfaction of \eqref{eq:first_sparse} will constitute one of the key points in this manuscript's central design procedure.
		
		\vspace{-3mm}
		
		\begin{figure*}[t]
			\centering
			\resizebox{.975\textwidth}{!}
			{
				\begin{tikzpicture}[scale=.5]
					\draw [thick,rounded corners=1]  (-2.5,-1) rectangle +(5,2);
					\node at (0,0) {\tiny\begin{tabular}{c}
							From the $i^\text{th}$ MPC\\ subcontroller
					\end{tabular}};
					
					\draw [thick,rounded corners=1]  (-2.5,-3.5) rectangle +(5,2);
					\node at (0,-2.5) {\tiny\begin{tabular}{c}
							From the $i^\text{th}$\\ area's sensors
					\end{tabular}};
					
					\draw [thick,rounded corners=1]  (-2.5,-8) rectangle +(5,4);
					\node at (0,-6) {\tiny\begin{tabular}{c}
							From the $j^\text{th}$ area\\ for all $j\in\mathcal{N}_i\setminus\{i\}$
					\end{tabular}};
					
					\draw [thick]  (5.5+0.5,-2.5) circle(0.25cm);
					\draw [thick]  (5.5+0.5,-2.5) node {$+$};
					
					\draw [thick,rounded corners=1]  (10,-8) rectangle +(5,7);
					\node at (12.5,-4.5) {\tiny\begin{tabular}{c}
							$\mathbf{K}_{\mathbf{D}i}(z)$\\ implemented\\ as in \eqref{eq:area_NRF}\\ while \eqref{eq:first_sparse} holds
					\end{tabular}};
					
					\draw [thick] [->] (2.5,-2.5) -- (5.25+0.5,-2.5);
					\node at (4.2,-2.2) {\scriptsize$x_i+\zeta_i$};
					\draw [thick] [->] (5.75+0.5,-2.5) -- (10,-2.5);
					
					\draw [thick] [->] (2.5,-0.5) -- (5.5+0.5,-0.5) -- (5.5+0.5,-2.25);
					\node at (4.2,-0.15) {\scriptsize$u_{s1i}+\beta_{s1i}$};
					\draw [thick] [->] (2.5,0.5) -- (19.25,0.5);
					\node at (4.2,0.85) {\scriptsize$u_{s2i}+\beta_{s2i}$};
					
					\draw [thick] [->] (2.5,-4.5) -- (5.25+0.5,-4.5);
					\node at (4.2,-4.2) {\scriptsize$x_j+\zeta_j$};
					\draw [thick]  (5.5+0.5,-4.5) circle(0.25cm);
					\draw [thick]  (5.5+0.5,-4.5) node {$+$};
					\draw [thick] [->] (5.75+0.5,-4.5) -- (10,-4.5);
					\draw [thick] [->] (2.5,-6) -- (5.5+0.5,-6) -- (5.5+0.5,-4.75);
					\node at (4.2,-6+0.35) {\scriptsize$u_{s1j}+\beta_{s1j}$};
					\draw [thick] [->] (2.5,-7.5) -- (10,-7.5);
					\node at (4.2,-7.5+0.35) {\scriptsize$u_{fj}+\beta_{fj}$};
					
					\draw [thick]  (19.5, 0.5) circle(0.25cm);
					\draw [thick]  (19.5, 0.5) node {$+$};
					\draw [thick] [->] (19.75,0.5) -- (28.5,0.5);
					
					\draw [thick,rounded corners=1]  (28.5,-0.5) rectangle +(5,2);
					\node at (31,0.5) {\tiny\begin{tabular}{c}
							To the $i^\text{th}$\\ area's actuators
					\end{tabular}};
					
					\draw [thick,rounded corners=1]  (28.5,-5.5) rectangle +(5,4.5);
					\node at (31,-3.25) {\tiny\begin{tabular}{c}
							To the $\ell^\text{th}$ area\\ for all $\ell\neq i$\\ such that $i\in\mathcal{N}_\ell$
					\end{tabular}};
					
					\draw [thick,rounded corners=1]  (28.5,-8) rectangle +(5,2);
					\node at (31,-7) {\tiny\begin{tabular}{c}
							To the $i^\text{th}$ MPC\\ subcontroller
					\end{tabular}};
					
					\draw [thick]  (17, -6.625) node {$w_{i}$};
					\draw [thick] [->] (15,-7) -- (19.25,-7);
					\draw [thick]  (19.5, -7) circle(0.25cm);
					\draw [thick]  (19.5, -7) node {$+$};
					\draw [thick] [->] (19.75,-7) -- (28.5,-7);
					\draw [thick] [->] (24,-7) -- (24,-4) -- (28.5,-4);
					\node at (24,-7.025) {$\bullet$};
					\draw [thick] [->] (19.5,-5.5) -- (19.5,-6.75);
					\draw [thick]  (19.5, -5) node {$\beta_{wi}$};
					
					\draw [thick]  (17, -2.125) node {$u_{fi}$};
					\draw [thick] [->] (15,-2.5) -- (23.75,-2.5);
					\node at (19.5,-2.525) {$\bullet$};
					\draw [thick] [->] (19.5,-2.5) -- (19.5,0.25);
					\draw [thick]  (24, -2.5) circle(0.25cm);
					\draw [thick]  (24, -2.5) node {$+$};
					\draw [thick] [->] (24.25,-2.5) -- (28.5,-2.5);
					\draw [thick] [->] (24,-1) -- (24,-2.25);
					\draw [thick]  (24, -0.5) node {$\beta_{fi}$};
					\draw [thick]  (24, 0.875) node {$u_{i}$};
					
				\end{tikzpicture}
			}\vspace{-1mm}
			\caption{Implementation scheme for the $i^\text{th}$ area's NRF-based subcontroller}\vspace{-1mm}
			\label{fig:NRF_zoom}
			\hrulefill\vspace{-1mm}
		\end{figure*}
		
		\begin{remark}\label{rem:com}
			We assume that the communication infrastructure of our control architecture is equipped with error-detection/correction mechanisms. Consequently, all communication errors affecting transmitted information can be modelled as disturbance signals which originate at the source of transmission, not at the receiving ends. Doing so will ensure that, when transmitting information from an arbitrarily chosen $i^\text{th}$ area, all the other areas whose index $j$ satisfies $i\in\mathcal{N}_j$ will receive \emph{the same information} from the $i^\text{th}$ area. It is this fact, alongside the sparsity conditions in \eqref{eq:first_sparse}, which ensures that the signal structures in Figures~\ref{fig:NRF_implem} and~\ref{fig:NRF_zoom} are the same, thus ensuring the equivalence between the two control schemes. 
			\vspace{-3mm}
		\end{remark}
		
		The realisations given in \eqref{eq:area_NRF} are the sought-after representations mentioned in objective $\mathrm I)$ from Section~\ref{subsec:NRF_obj}, since they can be deployed independently of each other, as per Figure~\ref{fig:NRF_zoom}, and since implementing discrete-time state-space systems reduces to a mere four matrix-vector multiplications and two vector-vector additions.\vspace{-3mm}
		
		\begin{remark}
			We hereby make a point of the fact that the only type of constraint handled during the design of the first layer is the communication constraint dictated by the area neighbourhoods $\mathcal{N}_i$, as formalised via \eqref{eq:first_sparse}. In addition to constraining the state variables appearing in \eqref{eq:ss_a}-\eqref{eq:ss_c}, the second layer will also enforce constraints on the partitions $u_i[k]$ of the command signal vector, as divided in \eqref{eq:trip}, by exploiting the particular structure of the matrices expressed in \eqref{eq:mat_coef}. Indeed, by leveraging Proposition~\ref{prop:ss_implem}, an MPC-based strategy (see, for example, the approach proposed in \cite{part2}) can be employed to compute time-invariant sets that constrain the individual entries of the state vectors belonging to the realisations in \eqref{eq:area_NRF}. By using $u_{s1}[k]$ to regulate the state dynamics of said realisations, the second layer deftly avoids the need to saturate the outputs of the first-layer subcontrollers in a blunt manner. This choice also frees up $u_{s2}[k]$ to handle plant-related constraint management in situations where the dynamical control action from \eqref{eq:uf_implem} may be too slow to act.\vspace{-3mm}
		\end{remark}
	}
	
	Having obtained the realisations targeted by objective I) in Section~\ref{subsec:NRF_obj}, it now becomes relevant to investigate the expressions of the closed-loop maps obtained by deploying the implementations from \eqref{eq:area_NRF}. Before doing so, however, we must introduce some additional notation. We denote the orders of the aforementioned realisations as $n_{wi}:=\textstyle\sum_{\ell=\alpha_{ui}+1}^{\alpha_{ui}+n_{ui}} n_{r\ell}$ and let $n_w:=\textstyle\sum_{i=1}^N n_{wi}$. Additionally, we denote the initial conditions of the subcontrollers given in \eqref{eq:area_NRF} as $w_{ci}\in\mathbb{R}^{n_{wi}},\,\forall\,i\in\{1:N\}$ and we define $w_c:=\begin{bmatrix}
		w_{c1}^\top&\dots&w_{cN}^\top
	\end{bmatrix}^\top$. With this notation at hand, we may now tackle objective $\mathrm{II})$ from Section~\ref{subsec:NRF_obj}.\vspace{-3mm}

	\subsection{Theoretical Guarantees of NRF-based Control}\label{subsec:theo_NRF}\vspace{-3mm}
	
	Yet another benefit of NRF-based control is the fact that, even in the distributed setting, \emph{all} closed-loop maps from exogenous disturbance \emph{and} initial conditions depend upon a freely tunable parameter (see Appendix~\ref{app:aux}). This flexibility is illustrated by the following result, which provides the theoretical framework for the \emph{complete} tuning of NRF-based closed-loop response. \vspace{-3mm}
	
	\begin{figure*}
		\begin{subequations}
			\begin{align}\label{eq:F_Q}\tag{19a}
				&\ \mathbf{F}_{\mathbf{Q}}(z):=\left[\begin{array}{c:c:c:c}
					\mathbf{N}(z)\widetilde{\mathbf{X}}_{\mathbf{Q}}(z) &
					\mathbf{N}(z)\widetilde{\mathbf{Y}}_{\mathbf{Q}}(z) &
					\mathbf{N}(z)\left(\widetilde{\mathbf{Y}}_{\mathbf{Q}}^{\mathrm{diag}}(z)-\widetilde{\mathbf{Y}}_{\mathbf{Q}}(z)\right) & 
					(\mathbf{N}(z)\widetilde{\mathbf{X}}_{\mathbf{Q}}(z)+I_{n_x})\mathbf{G}_d(z)\\\hdashline
					\mathbf{M}(z)\widetilde{\mathbf{X}}_{\mathbf{Q}}(z) &
					\mathbf{M}(z)\widetilde{\mathbf{Y}}_{\mathbf{Q}}(z)-I_{n_u} &
					\mathbf{M}(z)\left(\widetilde{\mathbf{Y}}_{\mathbf{Q}}^{\mathrm{diag}}(z)-\widetilde{\mathbf{Y}}_{\mathbf{Q}}(z)\right) & 
					\mathbf{M}(z)\widetilde{\mathbf{X}}_{\mathbf{Q}}(z)\mathbf{G}_d(z)
				\end{array}\right],\\
				\label{eq:I_Q}\tag{19b}
				&\begin{array}{ll}
					\ \mathbf{I}_{\mathbf{Q}}(z):=\left[\begin{array}{c:c}
						\mathbf{Y}(z)\mathbf{J}_1(z)+\mathbf{N}(z)\mathbf{Q}(z)\mathbf{J}_1(z)&\mathbf{N}(z)\mathbf{J}_2(z)\\\hdashline
						\mathbf{X}(z)\mathbf{J}_1(z)+\mathbf{M}(z)\mathbf{Q}(z)\mathbf{J}_1(z)&\mathbf{M}(z)\mathbf{J}_2(z)
					\end{array}\right],&\begin{array}{l}
						\mathbf{J}_1(z):=\widetilde{\mathbf{M}}(z)(zI_{n_x}-A)^{-1}z,\\
						\mathbf{J}_2(z):=\widetilde{\mathbf{Y}}_{\mathbf{Q}}^{\mathrm{diag}}(z)C_w(zI_{n_w}-A_w)^{-1}z.
					\end{array}
				\end{array}
			\end{align}
		\end{subequations}
		\begin{equation*}
			\vspace{-8mm}
		\end{equation*}
		\hrulefill\vspace{-2mm}
	\end{figure*}
	
	\begin{theorem}\label{thm:NRF_state}
		Let $\mathbb{C}_g\subseteq\mathbb{C}$ and let $\mathbb{C}_b:=\mathbb{C}\setminus\mathbb{C}_g$. Let a network described by \eqref{eq:ss_a}-\eqref{eq:ss_b} be split into $N$ areas, as in \eqref{eq:trip}-\eqref{eq:net_part}, and let it be connected, as in Figure~\ref{fig:NRF_implem}, to a set\newpage\noindent of NRF subcontrollers of type \eqref{eq:uf_implem}. Assume also that:\vspace{-3mm}
		\begin{enumerate}
			\item[A1)] The subrealisation $(A,B_u,I_{n_x},O)$ from \eqref{eq:ss_a}-\eqref{eq:ss_c} is $\mathbb{C}_b$-controllable;\smallskip
			
			\item[A2)] The collection \resizebox{.65\columnwidth}{!}{$(\mathbf{N}(z),\mathbf{M}(z),\mathbf{X}(z),\mathbf{Y}(z),\widetilde{\mathbf{N}}(z),\widetilde{\mathbf{M}}(z),$} $\widetilde{\mathbf{X}}(z),\widetilde{\mathbf{Y}}(z))$ is a doubly coprime factorisation over $\mathbb{C}_g$ of $\mathbf{G}_u(z):=(z I_{n_x}-A)^{-1}B_u$, as in Appendix~\ref{app:aux};\medskip
			
			\item[A3)] The pair $(\mathbf{\Phi}(z),\mathbf{\Gamma}(z))$ is a $\mathbb{C}_g$-allocating NRF pair, based on the aforementioned factorisation of $\mathbf{G}_u(z)$;\medskip
			
			\item[A4)] The control laws from \eqref{eq:uf_implem} are based on $(\mathbf{\Phi}(z),\mathbf{\Gamma}(z))$, and are implemented via the realisations from \eqref{eq:area_NRF};\vspace{-3mm}
		\end{enumerate}
		Then, by using \eqref{eq:area_NRF} to define $A_w:=\mathrm{diag}(A_{w1},\dots,A_{wN})$ along with $C_w:=\mathrm{diag}(C_{w1},\dots,C_{wN})$, the closed-loop dynamics of the network's states and of the NRF-based subcontrollers' outputs in Figure~\ref{fig:NRF_implem} are given\textcolor{black}{, $\forall\,k\geq k_0$,} by\vspace{-3mm}
		\begin{equation}\label{eq:cl_dyn}\tag{18}
			\scriptsize\begin{bmatrix}
				x[k]  \\  u_f[k]
			\end{bmatrix}=\mathbf{F}_{\mathbf{Q}}(z)\star d_s[k]+\mathcal{I}_{\mathbf{Q}}[k]
			{\scriptsize\begin{bmatrix}
					x_c \\  w_c
			\end{bmatrix}}
			\,,\vspace{-3mm}\normalsize
		\end{equation}
		where we have that:\vspace{-3mm}
		\begin{enumerate}
			\item[i)] The TFM denoted $\mathbf{F}_{\mathbf{Q}}(z)$ is expressed in \eqref{eq:F_Q} at the top of the next page, with $\widetilde{\mathbf{Y}}_{\mathbf{Q}}(z)$, $\widetilde{\mathbf{Y}}_{\mathbf{Q}}(z)$ and $\widetilde{\mathbf{Y}}_{\mathbf{Q}}^{\mathrm{diag}}(z)$ as defined in \eqref{eq:aux_NRF} from Appendix~\ref{app:aux}, while\vspace{-3mm}
			\begin{equation*}
				\mathbf{G}_d(z):=(z I_{n_x}-A)^{-1}B_d;\vspace{-3mm}
			\end{equation*}
			
			\item[ii)] \textcolor{black}{The time-varying matrix denoted $\mathcal{I}_{\mathbf{Q}}[k]$ is} the inverse $\mathcal{Z}$-transform of the TFM \textcolor{black}{$z^{-k_0}\mathbf{I}_{\mathbf{Q}}(z)$, where $\mathbf{I}_{\mathbf{Q}}(z)$ is defined} in \eqref{eq:I_Q}, at the top of the next page;\smallskip

			\item[iii)] The TFM $\mathbf{Q}(z)$ appearing in \eqref{eq:F_Q}-\eqref{eq:I_Q} is the $\mathbf{Q}$-parameter which designates the employed NRF pair;\smallskip
			
			\item[iv)] All the TFMs defined in \eqref{eq:F_Q}-\eqref{eq:I_Q} are proper and $\mathbb{C}_b$-bounded.\vspace{-7mm}
		\end{enumerate}
	\end{theorem}
	
	\begin{pf}
		See Appendix~\ref{app:proofs}.\vspace{-7mm}
	\end{pf}

	{\color{black} 
		\begin{remark}
			Notice that, as per the partitioned dynamics from \eqref{eq:cl_dyn}, the TFM labelled $\mathbf{F}_{\mathbf{Q}}(z)$ governs the forced response of the closed-loop system (with respect to exogenous signals), whereas $\mathcal{I}_{\mathbf{Q}}[k]$ models the influence of initial conditions on the said partitioned dynamics. We also point out that, while $x_c$ is almost always predetermined by the network's physical conditions, the vector $w_c$ can be freely assigned at the control scheme's initialisation. Indeed, while the zero vector is often the most convenient value, we hereby make no assumptions upon it: firstly, for the sake of mathematical generality, and secondly, due to the fact that the enforcement of constraints on the partitions of $w[k]$ (see the set-theoretical analysis in \cite{part2}) may not always be compatible with a zero-vector initialisation.\vspace{-2mm}
		\end{remark}
	}

	\begin{remark}
		We highlight the fact that assumptions A1) through A4) in Theorem~\ref{thm:NRF_state} are \emph{in no way restrictive}, from a practical standpoint. The closed-loop dynamics of the network cannot be tuned with respect to $\mathbb{C}_g$ unless A1) holds, while the factorisation and the NRF pair from assumptions A2) through A4) can be reliably obtained via classical system-theoretical results (see \cite{NRF} and \cite{aug_sparse}).\vspace{-2mm}
	\end{remark}
	
	{\color{black}
		We draw particular attention to the closed-loop maps being showcased in Theorem~\ref{thm:NRF_state}. In particular, we point out the fact that the $\mathbf{Q}$-parameter, which characterises all of these expressions, represents the \emph{main tuning parameter} of our control architecture's first layer, and the freedom in choosing it provides fertile grounds for the development of NRF-based design procedures. Combining this fact with the straightforward nature of the algebraic expressions obtained in \eqref{eq:F_Q}-\eqref{eq:I_Q}, we conclude that the distributed implementations proposed in Proposition~\ref{prop:ss_implem} fully address objective $\mathrm{II})$ from Section~\ref{subsec:NRF_obj}. Having done so, we now proceed to tackle the latter section's final objective, concerning the $\mathbf{Q}$-parameter's tuning.
		
		\vspace{-2mm}
	}
	
	\begin{remark}
		Since the aforementioned tuning aims to produce a $\mathbb{C}_g$-allocating NRF pair, of the kind used in Theorem~\ref{thm:NRF_state}, then the resulting $\mathbf{Q}$-parameter \emph{must} be a proper and $\mathbb{C}_b$-bounded TFM (see Appendix~\ref{app:aux} and \cite{NRF}). From a practical point of view, we also point out that certain choices of $\,\mathbb{C}_g$ are preferable. For example, by choosing $\,\mathbb{C}_g\subseteq\mathbb{S}$, it follows (see Appendix~\ref{app:aux}) that the state variables of the plant and those of the NRF subcontrollers can be kept bounded at all times, for any initial conditions of these systems and for any bounded signals in $d_s[k]$.\vspace{-3mm}
	\end{remark}

	\section{First Layer Design}\vspace{-3mm}
	\label{sec:NRF_des}
	
	\subsection{The Decoupling-based Approach}\vspace{-3mm}\label{subsec:decup}
	
	To formulate the first layer's design procedure, we need to account for the state variables of the implementations in \eqref{eq:area_NRF}. Therefore, aside from the $x$ and $u$ subscripts employed in \eqref{eq:net_part}-\eqref{eq:sel_mat}, we introduce the additional subscript $w$ to refer to the state vector of each area's NRF subcontroller, and we extend the index collection in \eqref{eq:trip} to\vspace{-3mm}
	\begin{equation*}
		\mathcal{A}_i:=(\mathcal{A}_{xi},\mathcal{A}_{ui},\mathcal{A}_{wi}),\,\forall\,i\in\{1:N\}.\vspace{-3mm}
	\end{equation*}
	This extension enables us to rewrite the closed-loop dynamics from Theorem~\ref{thm:NRF_state}, and then partition them into the areas discussed in Section~\ref{subsec:distrib_part}. As previously stated in Section~\ref{subsec:NRF_basic}, the second-layer command vectors $u_{s1i}[k]$ and $u_{s2i}[k]$ that appear in Figure~\ref{fig:scheme} are folded into $\beta_x[k]$ and $\beta_u[k]$, respectively. Thus, we will henceforth refer to the quadruplet given by $\{S_{xi}^\top\beta_x^{}[k],S_{ui}^\top\beta_u^{}[k],x_{ci},w_{ci}\}$ as the \emph{$i^\text{th}$ area's variables}, from the point of view of that area's local second-layer subcontroller.\vspace{-2mm}
	
	All of these new terms enable us to express the dynamics of the $i^\text{th}$ area as a combination between the effect of exogenous disturbance, given by a TFM's input-output response as in \eqref{eq:io_resp}, and that of area cross-coupling, with said combination being expressed as follows\stepcounter{equation}\vspace{-2mm}
	\begin{equation}\label{eq:dyn_i}
		\resizebox{.975\columnwidth}{!}{$\begin{bmatrix}
				x_i[k] \\  u_{fi}[k]
			\end{bmatrix}=\left(Z_i^\top\mathbf{F}_{\mathbf{Q}}(z)\scriptsize\begin{bmatrix}
				O \\  I_{(n_u+n_d)}
			\end{bmatrix}\right)\star\begin{bmatrix}
				\beta_f[k] \\ d[k]
			\end{bmatrix} + \displaystyle\sum_{j=1}^N\mathcal{E}_{ij}[k],$}
	\end{equation}
	where $Z_i:=\mathrm{diag}(S_{xi},S_{ui})$. The terms $\mathcal{E}_{ij}[k]$ may be represented as a sum between the input-output response of another TFM and of a free response term, which depends upon the initial conditions of the coupled area, to obtain\vspace{-2mm}
	\begin{equation}\label{eq:Ei_def}
		\resizebox{.975\columnwidth}{!}{$\mathcal{E}_{ij}[k]:=\left(Z_i^\top\mathbf{F}_{\mathbf{Q}}(z)\begin{bmatrix}
				Z_j\\O
			\end{bmatrix}\right)\star\begin{bmatrix}
				S_{xj}^\top\beta_x[k]\\S_{uj}^\top\beta_u[k]
			\end{bmatrix}+ Z_i^\top \mathcal{I}_{\mathbf{Q}}[k]{Z}_{cj}\begin{bmatrix}
				x_{cj}\\w_{cj}
			\end{bmatrix}\hspace{-1mm},$}
	\end{equation}
	which is the \emph{effect} of the $j^\text{th}$ area's variables on the $i^\text{th}$ area, and in which the time-domain expression from the right-hand term depends upon ${Z}_{cj}:=\mathrm{diag}(S_{xj},S_{wj})$.\vspace{-2mm}

	\begin{remark}
		The specific formulations obtained in this section facilitate the construction of the prediction models which are employed in \cite{part2} for the distributed synthesis of our top-layer constraint management policies.\vspace{-2mm}
	\end{remark}
	
	In light of \eqref{eq:dyn_i}-\eqref{eq:Ei_def}, the final objective of the first layer, meant to facilitate the second layer's design and implementation, now becomes apparent. With respect to the communication neighbourhoods introduced in Section~\ref{subsec:global_arch}, the $i^\text{th}$ area's NRF subcontroller must:\vspace{-2mm}
	\begin{enumerate}
		\item[a)] only employ information originating in $\mathcal{N}_i$;\medskip
		\item[b)] handle the effect of $\beta_f[k]$ and $d[k]$ according to some pre-specified design objective.\medskip
		\item[c)] minimise the effect of $S_{xj}^\top\beta_x^{}[k]$ and of $S_{uj}^\top\beta_u^{}[k],$ for all indices $j\in\{1:N\}\setminus\{i\}$;\medskip
		\item[d)] minimise the effect of $x_{cj}$ and of $w_{cj},$ for all indices $j\in\{1:N\}\setminus\mathcal{N}_i$.\vspace{-2mm}
	\end{enumerate}
	
	We now proceed to formalise all of these requirements.\vspace{-3mm}
	
	\begin{figure}[H]
		\centering
		\resizebox{\columnwidth}{!}{\begin{tikzpicture}[scale=.5]
				\draw [thick,rounded corners=1]  (-2,-1) rectangle +(2,2);
				\node at (-1,0) {$\mathbf{T}_{di}$};
				\draw [thick] [->] (-8,0.5) -- (-2,0.5);
				\draw [thick] [->] (-8,-0.5) -- (-2,-0.5);
				\node at (-5,1) {$\beta_f$};
				\node at (-5,-1) {$d$};
				
				\draw [thick,rounded corners=1]  (-2,-4.5) rectangle +(2,2);
				\node at (-1,-3.5) {$\mathbf{T}_{uii}$};
				\draw [thick] [->] (-8,-3) -- (-2,-3);
				\draw [thick] [->] (-8,-4) -- (-2,-4);
				\node at (-5,-2.5) {$u_{s1i}+\beta_{s1i}+\zeta_i$};
				\node at (-5,-4.5) {$u_{s2i}+\beta_{s2i}$};
				
				\draw [thick,rounded corners=1]  (-2,-8) rectangle +(2,2);
				\node at (-1,-7) {$\mathbf{T}_{cij}$};
				\draw [thick] [->] (-8,-6.5) -- (-2,-6.5);
				\draw [thick] [->] (-8,-7.5) -- (-2,-7.5);
				\node at (-5,-6) {$x_{cj},\,j\in\mathcal{N}_i\setminus\{i\}$};
				\node at (-5,-8) {$w_{cj},\,j\in\mathcal{N}_i\setminus\{i\}$};
				
				\draw [thick, dashed] (-8.5,-1.5) -- (13.5,-1.5);
				\draw [thick, dashed] (-8.5,-5.25) -- (13.5,-5.25);

				\draw [thick]  (2,-3.5) circle(0.25cm);
				\draw [thick]  (2,-3.5) node {$+$};
				
				\draw [thick] [->] (0,-3.5) -- (1.75,-3.5);
				\draw [thick] [->] (2.25,-3.5) -- (13,-3.5);
				\draw [thick] [->] (0,0) -- (2,0) -- (2,-3.25);

				\node at (9.25,-2.5) {$\scriptsize\begin{bmatrix}
						x_i^\top & u_{fi}^\top
					\end{bmatrix}^\top$};
				
				\draw [thick]  (2,-7) circle(0.25cm);
				\draw [thick]  (2,-7) node {$+$};
				
				\draw [thick] [->] (0,-7) -- (1.75,-7);
				\draw [thick] [->] (2,-6.75) -- (2,-3.75);
				
				\draw [thick,rounded corners=1]  (4,-8) rectangle +(2,2);
				\node at (5,-7) {$\mathbf{T}_{cii}$};
				
				\draw [thick,rounded corners=1]  (7.5,-7) rectangle +(3,1);
				\node at (9,-6.5) {$\tiny\begin{bmatrix}
						I_{n_{xi}} & O
					\end{bmatrix}$};
				
				\draw [thick] [->] (4,-7) -- (2.25,-7);
				\draw [thick] [->] (7.5,-6.5) -- (6,-6.5);
				\draw [thick] [->] (13,-7.5) -- (6,-7.5);
				\node at (12,-3.55) {$\bullet$};
				\draw [thick] [->] (12,-3.5) -- (12,-6.5) -- (10.5,-6.5);
				
				\node at (9,-8) {$w_{ci}$};
		\end{tikzpicture}}
		\caption{The $i^\text{th}$ area's desired closed-loop response, as designated by the choice of TFMs in \eqref{eq:des_d_a}-\eqref{eq:des_c_d}\vspace{-3mm}}
		\label{fig:des_resp}
	\end{figure}
	
	\subsection{Formulating the Model-Matching Problem}\label{subsec:MM_form}\vspace{-3mm}
	
	The most straightforward of the four stated requirements is the one from point a), which places restrictions upon the sparsity pattern of the pair $(\mathbf{\Phi}(z),\mathbf{\Gamma}(z))$. This communication constraint, first introduced in \eqref{eq:first_sparse}, can be written in a more explicit form via the pair conditions\vspace{-2mm}
	\begin{subequations}
		\begin{align}\label{eq:con_NRF_a}
			\hspace{-1mm}S_{ui}^\top\mathbf{\Phi}(z)S_{{uj}}\equiv&\ O,\forall\,i\in\{1:N\},\, j\in\{1:N\}\hspace{-0.5mm}\setminus\hspace{-0.5mm}\mathcal{N}_i,\\
			S_{ui}^\top\mathbf{\Gamma}(z)S_{{xj}}\equiv&\ O,\forall\,i\in\{1:N\},\, j\in\{1:N\}\hspace{-0.5mm}\setminus\hspace{-0.5mm}\mathcal{N}_i,\label{eq:con_NRF_b}
		\end{align}
	\end{subequations}
	\phantom{ }\vspace{-11mm}
	
	which form an \emph{exact model-matching problem} (see \cite{aug_sparse}). \vspace{-2mm}

	{\color{black}
		\begin{remark}
			We point out that the above-mentioned \emph{exact model-matching problem} refers to the fact that all of the block-entries appearing in \eqref{eq:con_NRF_a}-\eqref{eq:con_NRF_b} \emph{must be identically zero matrices}, at the conclusion of the synthesis procedure. One computationally appealing way of enforcing this stark restriction is to parametrise a class of $\mathbb{C}_g$-allocating NRF pairs which have the desired sparsity patterns, via a closed-form expression that depends upon a \emph{freely tunable parameter}. This tractable approach, previously employed in \cite{aug_sparse}, has been included in Appendix~\ref{app:aux} as a beneficial technique that can be integrated into the design procedure described in the current section.\vspace{-3mm}
		\end{remark}
	}
	
	The requirements stated in points b)-d) from Section~\ref{subsec:decup} can be tackled by first selecting, for each network area, a triplet of TFMs which describes its closed-loop response. These triplets separate distinctly into: those which concern the disturbance vectors $\beta_f[k]$ and $d[k]$, namely\vspace{-2mm}
	\begin{subequations}
		\begin{align}\label{eq:des_d_a}
			&\mathbf{T}_{di}\in\mathcal{R}(z)^{(n_{xi}+n_{ui})\times(n_u+n_d)},\,\forall\,i\in\{1:N\},\\
			&\mathbf{T}_{di}(z)\text{ proper and $\mathbb{C}_b$-bounded},\,\forall\,i\in\{1:N\},\label{eq:des_d_b}
		\end{align}
	\end{subequations}
	\phantom{ }\vspace{-10mm}

	those concerning the disturbance channels associated with the command signals computed by the second layer\vspace{-2mm}
	\begin{subequations}
		\begin{align}\label{eq:des_u_a}
			&\mathbf{T}_{ui}\in\mathcal{R}(z)^{(n_{xi}+n_{ui})\times (n_x+n_u)},\,\forall\,i\in\{1:N\},\\\label{eq:des_u_b}
			&\mathbf{T}_{ui}(z)\text{ proper and $\mathbb{C}_b$-bounded},\,\forall\,i\in\{1:N\},\\\label{eq:des_u_c}
			&\mathbf{T}_{u{ij}}(z):=\mathbf{T}_{ui}(z)Z_j,\,\forall\,i,j\in\{1:N\},\\
			&\mathbf{T}_{u{ij}}(z)\equiv O,\forall\,i\in\{1:N\},\forall\, j\in\{1:N\}\setminus\{i\},\label{eq:des_u_d}
		\end{align}
	\end{subequations}
	\phantom{ }\vspace{-10mm}
	
	and those concerning the initial conditions of the areas\vspace{-2mm}
	\begin{subequations}
		\begin{align}\label{eq:des_c_a}
			&\mathbf{T}_{c{i}}\in\mathcal{R}(z)^{(n_{xi}+n_{ui})\times (n_x+n_w)},\,\forall\,i\in\{1:N\},\\\label{eq:des_c_b}
			&\mathbf{T}_{c{i}}(z)\text{ proper and $\mathbb{C}_b$-bounded},\,\forall\,i\in\{1:N\},\\\label{eq:des_c_c}
			&\mathbf{T}_{c{ij}}(z):=\mathbf{T}_{c{i}}(z){Z}_{cj},\,\forall\,i,j\in\{1:N\},\\
			&\mathbf{T}_{c{ij}}(z)\equiv O,\forall\,i\in\{1:N\},\forall\, j\in\{1:N\}\setminus\,\mathcal{N}_i.\label{eq:des_c_d}
		\end{align}
	\end{subequations}
	\phantom{ }\vspace{-10mm}

	{\color{black}
		
		Figure~\ref{fig:des_resp} shows a schematic representation of the TFMs introduced in \eqref{eq:des_d_a}-\eqref{eq:des_c_d}, and how they interact to produce the desired closed-loop response of the $i^\text{th}$ area.\vspace{-4mm}

		\begin{remark}
			Note that, as per the identities given in \eqref{eq:des_u_d} and in \eqref{eq:des_c_d}, only the TFMs which are not identically zero appear in Figure~\ref{fig:des_resp}, thus further emphasising the decoupling-oriented nature of our approach. Notice, moreover, that the desired closed-loop response separates naturally into the three components that are depicted in Figure~\ref{fig:des_resp}, separated by a series of dashed lines. Firstly, we have the middle component, which shows the feedforward-like action of the local second-layer subcontroller, and the lower component, which highlights the feedback-like behaviour of the initial conditions of the $i^\text{th}$ and $j^\text{th}$ areas, for all $j\in\mathcal{N}_i\setminus\{i\}$. Lastly, we have the upper component, which handles disturbance propagation and whose effects should be suppressed as thoroughly as possible.\vspace{-4mm}
		\end{remark}
		
		Once the closed-loop response TFMs are selected, the design problem reduces to minimising the difference between the TFMs appearing in \eqref{eq:dyn_i}-\eqref{eq:Ei_def} (along with the TFMs $Z_i^\top\mathbf{I}_{\mathbf{Q}}(z){Z}_{cj}$, which represent the $\mathcal{Z}$-transforms belonging to $Z_i^\top \mathcal{I}_{\mathbf{Q}}[k+k_0]{Z}_{cj}$) and the ones from \eqref{eq:des_d_a}-\eqref{eq:des_c_d}, in terms of a suitable system norm. To provide a highly flexible design framework, we formulate our model-matching expressions in terms of a generic system norm. However, we point out that the overall complexity of the resulting optimisation problem depends largely on the chosen norm. Notably, two popular choices in system-theoretical literature are the $\mathcal{H}_2$ and the $\mathcal{H}_\infty$ norms, which entail the following considerations: \vspace{-4mm}
		\begin{enumerate}
			\item[a)] For the $\mathcal{H}_2$ norm, we point out that the reasoning used in Theorem~29 of \cite{Rotko} can also be employed in the discrete-time case. Additionally, when combined with the vectorisation-based parametrisation described in Appendix~\ref{app:aux}, said norm yields a \emph{convex formulation} that can be tackled \emph{analytically} using classical theory (see Sections~17.6 and~21.5 in \cite{zhou}).
			
			\item[b)] For the $\mathcal{H}_\infty$ norm, the related problem is more intricate than the one from point a), but it nevertheless remains numerically tractable. One solution is to employ the vectorisation-based parametrisation from Appendix~\ref{app:aux} in conjunction with the \emph{Bounded Real Lemma} to adapt the norm optimisation technique proposed in \cite{aug_sparse}, thus resulting in a \emph{convex} and iterative procedure with \emph{guaranteed convergence}. In this case, the Semidefinite-Programming-based formulations which result from employing the Bounded Real Lemma have been shown to exhibit rapid convergence (see section V.B in \cite{aug_sparse}).\vspace{-4mm}
		\end{enumerate}

	}
	
	\begin{figure*}
		\begin{equation}\label{eq:MM_prob}\tag{28}
			\begin{array}{l}
				\min\limits_{\mathbf{Q}(z),\gamma_{di},\gamma_{u{ij}},\gamma_{c{ij}}} \textstyle\sum_{i=1}^N\bigg(\tau_{di}\gamma_{di} + \textstyle\sum_{j=1}^N(\tau_{u{ij}}\gamma_{u{ij}}+\tau_{c{ij}}\gamma_{c{ij}})\bigg),\\\vspace{-5mm}\\
				\text{subject to }
				\left\{\begin{aligned}
					&\text{\eqref{eq:con_NRF_a}-\eqref{eq:con_NRF_b}, \eqref{eq:con_pert}-\eqref{eq:con_init} and \eqref{eq:con_gamma},}\\
					&\mathbf{Q}\in\mathcal{R}(z)^{n_u\times n_x}\text{ strictly proper and }\mathbb{C}_b\text{-bounded}.
				\end{aligned}\right.
			\end{array}\normalsize
		\end{equation}
		\hrulefill\vspace{-3mm}
	\end{figure*}

	Regardless of the chosen norm, we introduce scalar upper bounds ${\gamma}_{d{i\phantom{j}}}$, ${\gamma}_{u{ij}}$ and ${\gamma}_{c{ij}}$, where $i,j\in\{1:N\}$, for each separate model-matching subproblem such that\vspace{-3mm}
	\begin{subequations}
		\begin{align}\label{eq:con_pert}
			&\hspace{-3mm}\left\|Z_i^\top\mathbf{F}_{\mathbf{Q}}(z)\tiny\hspace{-1mm}\begin{bmatrix}
				O \\  I_{(n_u+n_d)}
			\end{bmatrix}\hspace{-1mm}-\mathbf{T}_{di}(z)\right\|\leq \gamma_{di},\\
			\label{eq:con_cmd}
			&\hspace{-3mm}\left\|Z_i^\top\mathbf{F}_{\mathbf{Q}}(z)\tiny\hspace{-1mm}\begin{bmatrix}
				Z_j\\O
			\end{bmatrix}\hspace{-1mm}-\mathbf{T}_{u{ij}}(z)\right\|\leq \gamma_{u{ij}},\forall\,j\in\{1:N\},\hspace{-2mm}\\
			\label{eq:con_init}
			&\hspace{-3mm}\left\|Z_i^\top\mathbf{I}_{\mathbf{Q}}(z){Z}_{cj}-\mathbf{T}_{c{ij}}(z)\right\|\leq \gamma_{c{ij}},\forall\,j\in\{1:N\},
		\end{align}
	\end{subequations}
	\phantom{ }\vspace{-10mm}
	
	for each area index $i\in\{1:N\}$.\newpage
	As shown in the sequel, the first-layer design problem reduces to a generic optimisation problem, whose aim is to simultaneously minimise all of the upper bounds which appear in the right-hand terms of \eqref{eq:con_pert}-\eqref{eq:con_init}. Before proceeding to the computational details of our synthesis procedure, we point out that our approach shares several notable features with the SLS-based technique from \cite{DMPC1}-\cite{DMPC2}. In both of these instances, the focus rests upon designing the network's closed-loop response in accordance with several structural restrictions. In addition to this, the \emph{d-locality constraints} from \cite{DMPC1}-\cite{DMPC2} closely match our own area-based sparsity patterns from \eqref{eq:des_d_a}-\eqref{eq:des_c_d}, while the feasibilities of these two approaches are also closely intertwined, since both the SLS \cite{SLS} and the Youla Parametrisation (upon which our proposed technique is based, see \cite{NRF}) yield the set of all achievable $\mathbb{C}_b$-bounded closed-loop maps (see also \cite{Luca3}). On the other hand, our procedure leverages the fact that the NRF-based first layer can \emph{always be designed and implemented offline}, thus relegating all online computation to our significantly less resource-demanding second layer. This approach stands in stark contrast to the one from \cite{DMPC1}-\cite{DMPC2}, in which sparsity constraints must be factored into the \emph{online computation}, alongside any time-based constraints.\vspace{-3mm}
	
	\subsection{Solving the Model-Matching Problem}\vspace{-3mm}

	By choosing a set of constant weights $\tau_{di},\tau_{u{ij}},\tau_{c{ij}}\geq0$, for all $i,j\in\{1:N\}$, and by imposing a set of admissible upper bounds on all of the closed-loop norm values\vspace{-3mm}
	\begin{equation}\label{eq:con_gamma}
		\left\{
		\begin{aligned}
			&0\leq\gamma_{d{i\phantom{j}}}\leq\overline{\gamma}_{d{i}},\\
			&0\leq\gamma_{u{ij}}\leq\overline{\gamma}_{u{ij}},\forall\,j\in\{1:N\},\\
			&0\leq\gamma_{c{ij}}\leq\overline{\gamma}_{c{ij}},\forall\,j\in\{1:N\},
		\end{aligned}
		\right.\vspace{-3mm}
	\end{equation}
	for each $i\in\{1:N\}$, the full design of our architecture's first layer can be expressed via the model-matching problem in \eqref{eq:MM_prob}, which is located at the top of this page.\vspace{-3mm}
	
	{\color{black}	
		\begin{remark}\label{rem:feas2}
			A straightforward means of obtaining the $\overline{\gamma}_{\bullet}$ values from \eqref{eq:con_gamma}, which are feasible for the problem in \eqref{eq:MM_prob}, is to exploit the parametrisation of structured NRF-based control laws proposed in Appendix~\ref{app:aux}. Since the $\mathbf{Q}$-parameter can be expressed as $\mathbf{Q}(z)=\mathbf{Q}_0(z)+\widehat{\mathbf{Q}}(z)$, for which $\widehat{\mathbf{Q}}(z)\equiv O$ marks the central object of said class (see Appendix~\ref{app:aux}), taking $\mathbf{Q}(z)=\mathbf{Q}_0(z)$, choosing $\gamma_{\bullet}$ values which satisfy the resulting inequalities in \eqref{eq:con_pert}-\eqref{eq:con_init}, and then setting $\overline{\gamma}_\bullet=\gamma_{\bullet}$ in each component of \eqref{eq:con_gamma} will provide a feasible solution from which to start. Similarly, the $\tau_\bullet$ weights can all be set to $0$ except for those that correspond to subproblems associated with the TFMs from \eqref{eq:des_u_d} and \eqref{eq:des_c_d}, for which a value of $1$ can be attributed. In this manner, the control laws which result from the initial tuning proposed above will focus exclusively on area decoupling, for the (exclusive) benefit of the second layer.
			
			\vspace{-3mm}
	\end{remark}}

	We summarise the discussion on our first-layer synthesis procedure by providing the high-level design algorithm located directly above this paragraph, which concisely describes the technique presented in this section.\vspace{-9mm}
	
	\begin{algorithm}[t]
		
		\KwData{A network described by \eqref{eq:ss_a}-\eqref{eq:ss_b}}\smallskip
		
		\KwResult{NRF-based control laws of type \eqref{eq:Kd_def}-\eqref{eq:uf_implem}}\smallskip
		
		\textbf{Initialisation:} Partition the network into areas as described in \eqref{eq:trip}-\eqref{eq:net_part};\smallskip
		
		\textbf{Step 1:} Impose NRF communication constraints from \eqref{eq:con_NRF_a}-\eqref{eq:con_NRF_b} in accordance with the area partition;\smallskip
		
		\textbf{Step 2:} Check the feasibility of \eqref{eq:con_NRF_a}-\eqref{eq:con_NRF_b};\smallskip
		
		\eIf{the desired sparsity pattern is feasible}{
			go to \textbf{Step 3};\smallskip
		}{
			go to \textbf{Initialisation} and choose a more compact area distribution, by grouping up independent areas;\smallskip
		}
		
		\textbf{Step 3:} Choose closed-loop response TFMs \eqref{eq:des_d_a}-\eqref{eq:des_c_d};\smallskip
		
		\textbf{Step 4:} Formulate the norm-based expressions of the model-matching problem as in \eqref{eq:con_pert}-\eqref{eq:con_init};\smallskip
		
		\textbf{Step 5:} Compute the values of the $\overline{\gamma}_\bullet$ bounds as in Remark~\ref{rem:feas2} and solve the problem described in \eqref{eq:MM_prob};\smallskip
		
		\eIf{a suitable solution is found}{
			use $\mathbf{Q}(z)$ to form the NRF pair as in Appendix~\ref{app:aux};\smallskip
		}{
			go to \textbf{Step 3} and adjust the closed-loop TFMs;\smallskip
		}
		\caption{First-layer design procedure}
		\label{alg:NRF_des}
	\end{algorithm}
	\phantom{ }\vspace{-2mm}

	{\color{black}
		\begin{remark}\label{rem:else_branch}
			We point out that, for some practical applications, the \textbf{\emph{else}} branch that follows \textbf{\emph{Step 2}} in Algorithm~\ref{alg:NRF_des} may reach a point at which grouping up additional individual areas into larger ones may not be possible, due to the physical limitations of the network. Thus, when choosing the initial partition from \eqref{eq:trip}-\eqref{eq:net_part}, the most relaxed area separation which can be supported by the available implementation resources \emph{should also be designated}. This relaxed partition can be employed as a \emph{feasibility check} for the entire design of the first layer, and also as a best-case scenario for the degree of closed-loop performance that can be achieved for the application, given the fundamental restrictions of the distributed setting.
			
			\vspace{-3mm}
		\end{remark}
		
		\begin{figure*}
			{\color{black}
				\begin{equation}\label{eq:area_prob}\tag{32}
					\begin{aligned}
						&\min\limits_{u_{s1i}[k],\dots,u_{s1i}[k+T_i-1],u_{s2i}[k],\dots,u_{s2i}[k+T_i+\overline T_i-1]}\textstyle\sum_{t=1}^{T_i+\overline T_i}g_{t}(\xi_i[k+t],u_{s1i}[k+t-1],u_{s2i}[k+t-1]),\\
						&\text{subject to}\left\{
						\begin{aligned}
							&\xi_i[k+t] = A_{si} \xi_i[k+t-1] + B_{s1i} u_{s1i}[k+t-1] + B_{s2i} u_{s2i}[k+t-1],\,\forall\,t\in\{1:T_i+\overline T_i\},\\
							&\xi_i[k+t]\in\Xi_{it},\,\forall\,t\in\{1:T_i\},\ \xi_i[k]=O,\\
							&u_{s1i}[k+t-1]\in{\mathcal{U}}_{s1i},\, u_{s2i}[k+t-1]\in{\mathcal{U}}_{s2i},\,\forall\,t\in\{1:T_i\}.
						\end{aligned}\right.
					\end{aligned}
				\end{equation}
			}\hrulefill\vspace{-2mm}
		\end{figure*}
		
		Before giving constructive details on how the first layer's closed-loop system can be interfaced with our architecture's second layer, we touch upon some of the concepts discussed in Remark~\ref{rem:else_branch}. In doing so, we first address a key trade-off that must be carefully managed when employing our design procedure, and which is fundamentally inherent to the broader distributed control setting.
		
		More specifically, we refer to the fact that, as the degree of fragmentation increases in the partitioning from \eqref{eq:trip}-\eqref{eq:net_part}, more and more control authority has to be sacrificed to ensure that all of these areas are decoupled from each other, in the considered closed-loop configuration. Moreover, as the granularity of the partitioning increases, the constraints in \eqref{eq:con_NRF_a}-\eqref{eq:con_NRF_b} generally become more stringent, as some of the smaller, newly-designated areas may be unable to exchange information with one another. \vspace{-2mm}
		
		Despite all of these complexities, a positive aspect of this fragmentation is the fact that the number of variables which have to be constrained by each second-layer subcontroller decreases significantly, thereby lowering the computational costs of implementing them. Should the second layer be based upon an MPC-like technique, as is the case in our companion paper, this design choice can have a dramatic impact on the overall computational cost of our proposed two-layer control architecture, since implementing MPC-based policies is several orders of magnitude more demanding than deploying their NRF-based counterparts (see the numerical example in \cite{part2} for a concrete breakdown of said computational cost). This particular aspect further motivates the decoupling-based approach proposed in the current body of work.\vspace{-2mm}
		
		\subsection{Interfacing with the Second Layer}\vspace{-2mm}
		
		As showcased in Figure~\ref{fig:scheme}, the NRF-based control laws proposed in this paper are meant to be accompanied by a set of distributed constraint management policies which enforce restrictions of the following type\vspace{-2mm}\stepcounter{equation}
		\begin{equation}\label{eq:explic_constr}
			x_i[k]\in\mathcal{X}_i,\,u_i[k]\in\mathcal{U}_i,\,\forall\,k\geq k_0,\vspace{-2mm}
		\end{equation}
		where $\mathcal{X}_i$ and $\mathcal{U}_i$ designate a pair of time-invariant and application-specific sets (see, in particular, \cite{STMC}) given by the $i^\text{th}$ area's design requirements, and $k_0$ is some initial moment from which the constraints must be enforced. Recalling Figure~\ref{fig:NRF_implem} and expressing the identities\vspace{-2mm}
		\begin{equation*}
			u_i[k]=u_{fi}[k]+u_{s2i}[k]+S_{ui}^\top\beta_{s2}[k],\,\forall\,i\in\{1:N\},\vspace{-2mm}
		\end{equation*}
		it follows that the constraints from \eqref{eq:explic_constr} may be enforced by exploiting an area-based partition of the closed-loop evolution obtained in \eqref{eq:cl_dyn} to constrain $x_i[k]$ and $u_{fi}[k]$.\vspace{-2mm}
		
		Given the flexibility afforded by our framework, several distinct techniques can be employed to implement the second layer of our control architecture (recall the discussion in Section~\ref{subsec:scope}). Due to its particular popularity in literature, our companion paper \cite{part2} employs MPC-based policies for the upper layer, and, before concluding this section, we proceed to discuss the manner in which these two control systems interact. The interplay between the two layers revolves around an area's predic-\newpage\noindent
		tion model, which can be obtained by expressing
		\vspace{-2mm}
		\begin{equation}\label{eq:disc_real}
			\hspace{-1mm}\left[\tiny\begin{array}{cc}
				S^\top_{{xi}} & O \\\hdashline
				O & S^\top_{{ui}}
			\end{array}\right]\hspace{-1mm}\mathbf{F}_{\mathbf{Q}}(z)\hspace{-1mm}\left[\tiny\begin{array}{c:c}
				S_{{xi}} & O\\O & S_{{ui}}\\ O & O
			\end{array}\right]\hspace{-1mm}=\hspace{-1mm}\left[\tiny\begin{array}{c|c:c}
				A_{si}-zI_{n_{si}} & B_{s1i} & B_{s2i}\\\hline C_{xi} & O & O\\\hdashline C_{ui} & O & O
			\end{array}\right]\hspace{-0.5mm},\hspace{-0.5mm}\vspace{-2mm}
		\end{equation}
		for all $i\in\{1:N\}$, and then	left-multiplying \eqref{eq:cl_dyn} with the matrix $Z_i^\top$, in order to obtain the following dynamics\vspace{-2mm}
		\begin{subequations}
			\begin{align}\label{eq:disc_cl_dyn_a}
				\xi_i[k+1] =A_{si} \xi_i[k] + &\ B_{s1i} u_{s1i}[k]+B_{s2i} u_{s2i}[k],\\
				\label{eq:disc_cl_dyn_b}
				x_i[k] =C_{xi} \xi_i[k] + &\ (\psi_{x i}[k] + \theta_{x i}[k] + \delta_{x i}[k]),\\
				u_{fi}[k] =C_{ui} \xi_i[k] + &\ (\psi_{ui}[k] + \theta_{ui}[k] + \delta_{ui}[k]),\label{eq:disc_cl_dyn_c}\\
				\label{eq:disc_cl_dyn_d}
				&\hspace{-28mm}{\color{black}k\in\mathbb{Z}},\ k\geq k_0\in\mathbb{Z},\ \xi_i[k_0]=O,\,\forall\,i\in\{1:N\},
			\end{align}
		\end{subequations}
		\phantom{}
		
		\vspace{-10mm}
		\begin{remark}
			Notice that a realisation of the type given in \eqref{eq:disc_real} can \emph{always} be obtained for the TFM on the left-hand side of said identity due to the choice in \eqref{eq:DCF_gain_inf} and to $\mathbf{Q}(z)$, obtained by solving \eqref{eq:MM_prob}, also being strictly proper.\vspace{-3mm}
		\end{remark}

		The quantities obtained in \eqref{eq:disc_cl_dyn_a}-\eqref{eq:disc_cl_dyn_d} are as follows:\vspace{-3mm}
		\begin{enumerate}
			\item[a)] $\xi_i[k]$ is the state vector of the realisation obtained in \eqref{eq:disc_real} and which \emph{always} has zero initial conditions at initial time $k_0$, recalling the identity from \eqref{eq:io_resp};\smallskip
			
			\item[b)] the contribution of exogenous disturbance to the area dynamics is given by \vspace{-3mm}
			\begin{equation*}
				\left\{
				\begin{aligned}
					\psi_i[k]:=&\ Z_i^\top\mathbf{F}_{\mathbf{Q}}(z)\star d_s[k],\\
					\psi_{x i}[k]:=&\ \tiny\begin{bmatrix}
						I_{n_{xi}} & O
					\end{bmatrix}\psi_i[k],\ 
					\psi_{ui}[k]:=\ \tiny\begin{bmatrix}
						O & I_{n_{ui}}
					\end{bmatrix}\psi_i[k],
				\end{aligned}
				\right.\vspace{-3mm}
			\end{equation*}
			\item[c)] the contribution of the \emph{neighbouring} areas' initial conditions to the area dynamics is given by\vspace{-3mm}
			\begin{equation*}
				\left\{\begin{aligned}
					&\theta_i[k]:=\textstyle\sum_{j\in\mathcal{N}_i}Z_i^\top \mathcal{I}_{\mathbf{Q}}[k]Z_{cj}\tiny\begin{bmatrix}
						x_{cj}^\top& w_{cj}^\top
					\end{bmatrix}^\top,\normalsize\\
					&\theta_{x i}[k]:=\tiny\begin{bmatrix}
						I_{n_{xi}} &O
					\end{bmatrix}\theta_i[k],\ \theta_{ui}[k]:=\tiny\begin{bmatrix}
						O & I_{n_{ui}}
					\end{bmatrix}\theta_i[k],
				\end{aligned}
				\right.\vspace{-3mm}
			\end{equation*}
			\item[d)] the \emph{residual} contributions of initial conditions from outside of the $i^\text{th}$ area's neighbourhood and of cross-coupling with other areas' MPC subcontrollers is\vspace{-3mm}
			\begin{equation*}
				\left\{\begin{aligned}
					&\delta_i[k]:=\hspace{-2mm}\sum_{j\in\{1:N\}\setminus\{i\}}\hspace{-2mm}Z_i^\top\mathbf{F}_{\mathbf{Q}}(z)\tiny\begin{bmatrix}
						Z_j\\O
					\end{bmatrix}\star\begin{bmatrix}
						u_{s1j}[k] \\ u_{s2j}[k]
					\end{bmatrix}+\\&\qquad\qquad\quad\,+\sum_{j\in\{1:N\}\setminus\,\mathcal{N}_i}Z_i^\top \mathcal{I}_{\mathbf{Q}}[k]Z_{cj}\tiny\begin{bmatrix}
						x_{cj}\\w_{cj}
					\end{bmatrix},\\
					&\delta_{x i}[k]:=\tiny\begin{bmatrix}
						I_{n_{xi}} & O
					\end{bmatrix}\delta_i[k],\ \delta_{ui}[k]:=\tiny\begin{bmatrix}
						O & I_{n_{ui}}
					\end{bmatrix}\delta_i[k],
				\end{aligned}\right.\vspace{-3mm}
				\normalsize
			\end{equation*}
		\end{enumerate}
		With the aid of the quantities defined above and of the prediction model from \eqref{eq:disc_cl_dyn_a}-\eqref{eq:disc_cl_dyn_d}, it becomes possible to formulate the optimisation problem from \eqref{eq:area_prob}, based on the \emph{output MPC} formalism (see Chapter 5 in \cite{RMD}) and located at the top of this page. In this newly introduced framework, the two local command signals $u_{s1i}[k]$ and $u_{s2i}[k]$ are themselves constrained to a pair of sets, $\mathcal{U}_{s1i}$ and $\mathcal{U}_{s2i}$, respectively, which become the main tuning parameters of the second-layer subcontrollers (see \cite{part2}).\vspace{-3mm}
		
		\begin{remark}
			The construction of the sets $\Xi_{it}$ in \eqref{eq:area_prob} is one of the main problems tackled in our companion paper \cite{part2}, whose main results revolve around ensuring the (recursive, see Chapter~5 in \cite{RMD}) feasibility of the problem in \eqref{eq:area_prob} via the previously mentioned sets. Additionally, we point out that \emph{no terminal-set-based mechanisms} are required to ensure said feasibility, which is yet another significant benefit of our proposed two-layer approach. Moreover, note that the prediction horizon in \eqref{eq:area_prob} is split in two, for an additional degree of flexibility: $T_i$ denotes the constrained horizon, whereas $\overline{T}_i$ is the unconstrained one (see \cite{part2} for a discussion on the benefits of this split).
			\vspace{-3mm}
		\end{remark}
		
		The focus in this manuscript on area-based decoupling becomes all the more justified when considering the identities in \eqref{eq:disc_cl_dyn_b} and \eqref{eq:disc_cl_dyn_c}. The unmeasurable quantities $\delta_{\bullet i}$ and $\psi_{\bullet i}$ must be made manageable by the first layer's closed-loop response, so that by taking into account $\theta_{\bullet i}$ and computing the two command signals in \eqref{eq:disc_cl_dyn_a}, each second-layer subcontroller must be able to ensure the satisfaction of \eqref{eq:explic_constr}. Although not explicitly required, possessing information about the pairs of sets in \eqref{eq:explic_constr} can have a beneficial impact when tuning the components of the optimisation problem from \eqref{eq:MM_prob}, as detailed in Section~\ref{subsec:MM_form}. Indeed, knowing that $x_i[k]$ and $u_i[k]$ must be contained in their respective sets, one may implicitly derive a set of constraints or suitable bounds for $\delta_{\bullet i}$ and $\psi_{\bullet i}$, such that their additive effect on the local state and command vectors still renders the constraints in \eqref{eq:explic_constr} feasible. Thus, the sets from \eqref{eq:explic_constr} can be used to implicitly inform the choices made in \eqref{eq:des_d_a}-\eqref{eq:des_c_d} and \eqref{eq:con_gamma} (see also Section~4.5 in \cite{zhou}, for a series of norm-based formulations which may prove helpful in making said choices).\vspace{-4mm}
		
		Given the highly theoretical nature of Sections~\ref{sec:NRF_theo} and~\ref{sec:NRF_des}, we now proceed to illustrate the main features of the results presented therein via a practical example.\vspace{-3mm}
		
		\section{Numerical Example}\label{sec:num_ex}\vspace{-3mm}
		
		The two-layer approach discussed in this paper (and expanded upon in \cite{part2}) is a suitable control technique for the power grid application in \cite{PT}, which we employ as a benchmark in order to showcase the efficacy of our proposed method. In this section, we elaborate upon the design procedure used to obtain distributed control laws for the benchmark in \cite{PT} (by methodically applying Algorithm~\ref{alg:NRF_des}), we transition from the recurrence-equation-based implementations in \cite{PT} to the specialised state-space realisations from Proposition~\ref{prop:ss_implem} and, finally, we illustrate all of the beneficial properties of the resulting closed-loop system, as guaranteed by Theorem~\ref{thm:NRF_state}.\vspace{-3mm}
		
		\begin{remark}
			In addition to the application presented in the sequel, the NRF-based control laws discussed in this paper are also suitable for vehicle-based platooning applications, whose dynamics and network topology (see \cite{plutonizare}) are radically different to those of the power transmission system from \cite{PT}. Therefore, the numerical results obtained here and in the platooning-based example from \cite{part2} show that our proposed framework is more than capable of tackling generic problems in the distributed setting.\vspace{-3mm}
		\end{remark}
		
		\subsection{The Controlled Network}\label{subsec:model}\vspace{-3mm}
		
		\begin{figure}[t]
			\centering
			\resizebox{.925\columnwidth}{!}{
				\begin{tikzpicture}[scale=0.15]
					\draw [thick,rounded corners=1]  (0,0) rectangle +(8,4);
					\draw [thick]  (4,2)   node {\footnotesize Node 1};
					\draw [thick,rounded corners=1]  (0,-30) rectangle +(8,4);
					\draw [thick]  (4,-28)   node {\footnotesize Node 2};
					\draw [thick,rounded corners=1]  (40,0) rectangle +(8,4);
					\draw [thick]  (44,2)   node {\footnotesize Node 3};
					\draw [thick,rounded corners=1]  (40,-30) rectangle +(8,4);
					\draw [thick]  (44,-28)   node {\footnotesize Node 4};
					\draw [thick,rounded corners=1]  (20,-15) rectangle +(8,4);
					\draw [thick]  (24,-13)   node {\footnotesize Node 5};
					\draw [thick] [->] (8,3)--(40,3);
					\draw [thick]  (24,4.25)   node {\scriptsize $\ell_{31}\cdot \delta_1$};
					\draw [thick] [->] (40,1)--(8,1);
					\draw [thick]  (24,-0.25)   node {\scriptsize $\ell_{13}\cdot \delta_3$};
					\draw [thick] [->] (8,-27)--(40,-27);
					\draw [thick]  (24,-19.75-6)   node {\scriptsize $\ell_{42}\cdot \delta_2$};
					\draw [thick] [->] (40,-29)--(8,-29);
					\draw [thick]  (24,-24.25-6)   node {\scriptsize $\ell_{24}\cdot \delta_4$};
					\draw [thick] [->] (3,0)--(3,-26);
					\draw [thick]  (-0.5,-13)   node {\scriptsize $\ell_{21}\cdot \delta_1$};
					\draw [thick] [->] (5,-26)--(5,0);
					\draw [thick]  (8.5,-13)   node {\scriptsize $\ell_{12}\cdot \delta_2$};
					\draw [thick] [->] (43,0)--(43,-26);
					\draw [thick]  (39.5,-13)   node {\scriptsize $\ell_{42}\cdot \delta_2$};
					\draw [thick] [->] (45,-26)--(45,0);
					\draw [thick]  (48.5,-13)   node {\scriptsize $\ell_{24}\cdot \delta_4$};
					\draw [thick] [->] (-4,2)--(0,2);
					\draw [thick]  (-2,3.25)   node {\footnotesize $u_1$};
					\draw [thick] [->] (-4,-28)--(0,-28);
					\draw [thick]  (-2,-26.75)   node {\footnotesize $u_2$};
					\draw [thick] [->] (52,2)--(48,2);
					\draw [thick]  (50.25,3.25)   node {\footnotesize $u_3$};
					\draw [thick] [->] (52,-28)--(48,-28);
					\draw [thick]  (50.25,-26.75)   node {\footnotesize $u_4$};
					\draw [thick] [->] (24,-19)--(24,-15);
					\draw [thick]  (24,-20)   node {\footnotesize $u_5$};
					\draw [thick] [->] (4,8)--(4,4);
					\draw [thick]  (6,6.5)   node {\footnotesize $d_1$};
					\draw [thick] [->] (4,-34)--(4,-30);
					\draw [thick]  (6,-32.5)   node {\footnotesize $d_2$};
					\draw [thick] [->] (44,8)--(44,4);
					\draw [thick]  (42,6.5)   node {\footnotesize $d_3$};
					\draw [thick] [->] (44,-34)--(44,-30);
					\draw [thick]  (42,-32.5)   node {\footnotesize $d_4$};
					\draw [thick] [->] (24,-7)--(24,-11);
					\draw [thick]  (24,-6)   node {\footnotesize $d_5$};
					\draw [thick] [->] (6,0)--(20,-12);
					\draw [thick]  (10,-7)   node {\scriptsize $\ell_{51}\cdot \delta_1$};
					\draw [thick] [->] (21,-11)--(8,0);
					\draw [thick]  (17,-4)   node {\scriptsize $\ell_{15}\cdot \delta_5$};
					\draw [thick] [->] (40,0)--(27,-11);
					\draw [thick]  (31,-4)   node {\scriptsize $\ell_{53}\cdot \delta_3$};
					\draw [thick] [->] (28,-12)--(42,0);
					\draw [thick]  (38,-7)   node {\scriptsize $\ell_{35}\cdot \delta_5$};
					\draw [thick] [->] (42,-26)--(28,-14);
					\draw [thick]  (38,-19)   node {\scriptsize $\ell_{54}\cdot \delta_4$};
					\draw [thick] [->] (27,-15)--(40,-26);
					\draw [thick]  (31,-22)   node {\scriptsize $\ell_{45}\cdot \delta_5$};
					\draw [thick] [->] (8,-26)--(21,-15);
					\draw [thick]  (17,-22)   node {\scriptsize $\ell_{52}\cdot \delta_2$};
					\draw [thick] [->] (20,-14)--(6,-26);
					\draw [thick]  (10,-19)   node {\scriptsize $\ell_{25}\cdot \delta_5$};
				\end{tikzpicture}
			}\vspace{-2mm}
			\caption{Interconnection topology of the grid's dynamics}\vspace{-1mm}
			\label{fig:topo}
		\end{figure}
		
		The power grid takes the form of a mesh, whose $N=5$ nodes are governed by the swing dynamics akin to those used in the numerical examples from \cite{DMPC1,DMPC2}, and given by\vspace{-3mm}\stepcounter{equation}
		\begin{equation}\label{eq:node_dyn}
			x_i[k+1]=\textstyle\sum_{j=1}^N A_{ij}x_j[k] + B_i (u_i[k] + d_i[k]),\vspace{-3mm}
		\end{equation}
		where we have that \textcolor{black}{$k_0=0$ and that}:\vspace{-4mm}
		\begin{enumerate}
			\item[a)] the $i^\text{th}$ node's state vector $x_i[k]=\scriptsize\begin{bmatrix}
				\delta_i[k] & \omega_i[k]
			\end{bmatrix}^\top$ is composed of the node's rated electrical frequency $\omega_i[k]$, and of the corresponding electrical angle $\delta_i[k]$;\smallskip
			
			\item[b)] the controllable power injection of the $i^\text{th}$ node, which enables actuation, is denoted by $u_i[k]$;\smallskip
			
			\item[c)] the $i^\text{th}$ node's input disturbance injection is $d_i[k]$;\smallskip
			
			\item[d)] the sampling time of the dynamics is $T_s=0.2$ s;\smallskip
			
			\item[e)] the coefficient matrices are expressed, $\forall\,i\in\{1:N\}$, as $B_i :=\scriptsize\begin{bmatrix} 0 & 1 \end{bmatrix}^\top$ and as\vspace{-3mm}
			\begin{equation*}
				A_{ij} :=\left\{\hspace{-1mm}
				\begin{array}{l}
					\scriptsize\begin{bmatrix}
						1 & T_s \\ -{h_i}{T_s}\sum_{q=1}^N \ell_{iq} & 1-{h_i}d_i{T_s}
					\end{bmatrix},\ i=j,\\\vspace{-3mm} \\
					\scriptsize\begin{bmatrix}
						0 & 0 \\ {h_i}\ell_{ij}{T_s} & 0
					\end{bmatrix},\ i\neq j.
				\end{array}
				\right.\vspace{-3mm}
			\end{equation*}
		\end{enumerate}
		We employ the same numerical coefficients as those given in Table~1 from \cite{PT}, as well as the same grid topology, shown in Figure~\ref{fig:topo}, while concatenating the node-based vectors $x_i$, $u_i$ and $d_i$ into their global counterparts from \eqref{eq:ss_a}-\eqref{eq:ss_c} and forming the grid's state-space realisation, by block-concatenating the matrices from point e) above.\vspace{-3mm}
		
		Naturally, we assign a separate area to each of the individual nodes in the same order as indexed in \eqref{eq:node_dyn}, with the resulting partition yielding the following index sets $\mathcal{A}_i=(\{2i-1,2i\},{i}),\,\forall\,i\in\{1:N\}$, as per \eqref{eq:trip}. Crucially, the network topology shown in Figure~\ref{fig:topo} informs the area-based communication networks, which are\vspace{-3mm}
		\begin{equation}\label{eq:comms}
			\begin{array}{c}
				\mathcal{N}_1=\{1,2,3,5\},\ \mathcal{N}_2=\{1,2,4,5\},\ \mathcal{N}_3=\{1,3,4,5\},\\
				\mathcal{N}_4=\{2,3,4,5\},\ \mathcal{N}_5=\{1,2,3,4,5\}.\\\vspace{-10.5mm}
			\end{array}\normalsize\vspace{1mm}
		\end{equation}
		Having clarified the application's communication infrastructure, we proceed to the design phase of the first layer.\vspace{-3mm}
		
		\begin{figure*}\color{black}
			\begin{equation}\label{eq:Q_param}
				\left\{
				\begin{aligned}
					\mathbf{Q}_0(z) =& \tiny\left[\begin{array}{rrrrrrrrrr}
						0.0000  &       0.0000 &  -0.0605   &      0.0000  & -0.0232      &   0.0000     &    0.0000     &    0.0000 &  -0.0125    &     0.0000\\
						-0.0972    &     0.0000    &     0.0000    &     0.0000    &     0.0000    &     0.0000  & -0.3305     &    0.0000  & -0.1081    &     0.0000\\
						-0.0323     &    0.0000    &     0.0000    &     0.0000    &     0.0000      &   0.0000  & -0.1362     &    0.0000  & -0.0320     &    0.0000\\
						0.0000     &    0.0000 &  -0.2366     &    0.0000  & -0.1127   &      0.0000     &    0.0000    &     0.0000  & -0.0569     &    0.0000\\
						0.0000     &    0.0000    &     0.0000    &     0.0000     &    0.0000     &    0.0000     &    0.0000     &    0.0000      &   0.0000      &   0.0000
					\end{array}\right],\ \mathbf{H}_{\mathcal{N}}(z)=\scriptsize\left[\begin{array}{c|c}
						-zI_4&B_{\mathcal{N}}\\\hline
						I_4&D_{\mathcal{N}}\\
						O&O
					\end{array}\right],\\
					B_{\mathcal{N}} =& \left[\tiny\begin{array}{rrrrrrrrrrr}
						0.0000     &    0.0000 &  -0.9167 &  -0.1833  & -0.3517  & -0.0703     &    0.0000    &     0.0000  & -0.1897  & -0.0379\\
						-0.2692 &  -0.0538    &     0.0000     &    0.0000    &     0.0000    &     0.0000 &  -0.9154 &  -0.1831  & -0.2994  & -0.0599\\
						-0.2247 &  -0.0449     &    0.0000     &    0.0000 &        0.0000     &    0.0000 &  -0.9486 &  -0.1897  & -0.2227  & -0.0445\\
						0.0000    &     0.0000 &  -0.8822 &  -0.1764  & -0.4202  & -0.0840   &      0.0000    &     0.0000  & -0.2123 &  -0.0425
					\end{array}\right],\\
					D_{\mathcal{N}} =& \left[\tiny\begin{array}{rrrrrrrrrr}
						0.0000    &     0.0000  &  0.9167    &     0.0000 &   0.3517    &     0.0000    &     0.0000      &   0.0000  &  0.1897 &        0.0000\\
						0.2692  &  0.0000    &     0.0000    &     0.0000  &       0.0000     &    0.0000  &  0.9154    &     0.0000  &  0.2994 &  0.0000\\
						0.2247 &   0.0000    &     0.0000    &     0.0000  &       0.0000   &      0.0000 &   0.9486     &    0.0000  &  0.2227   &      0.0000\\
						0.0000    &     0.0000  &  0.8822    &     0.0000  &  0.4202    &     0.0000     &    0.0000   &      0.0000  &  0.2123 &        0.0000\\
					\end{array}\right].
				\end{aligned}
				\right.\vspace{-3mm}
			\end{equation}\hrulefill\vspace{-2mm}
		\end{figure*}
		
		\subsection{Designing the First Layer}\vspace{-3mm}
		
		To obtain our control laws, we refer to Appendix~\ref{app:aux} in order to compute a factorisation of the type described in Section~\ref{subsec:DCF}, which is then used to express the sparsity-promoting parametrisation stated in Section~\ref{subsec:CONPRAS}. Said factorisation is obtained by setting $\mathbb{C}_g=\mathbb{S}$ along with $\mathbb{C}_b=\mathbb{C}\setminus\mathbb{S}$ and by first defining the two feedback matrices\vspace{-3mm}
		\begin{equation*}
			F:=(B^\top B)^{-1}B^\top(\mathrm{diag}(A_{11},A_{22},A_{33},A_{44},A_{55})-A),\vspace{-3mm}
		\end{equation*}
		along with\vspace{-3mm}
		\begin{equation*}
			L:=\mathrm{diag}(A_{\mathrm{db}},A_{\mathrm{db}},A_{\mathrm{db}},A_{\mathrm{db}},A_{\mathrm{db}})-A,\vspace{-3mm}
		\end{equation*}
		where $A_{\mathrm{db}}:=\tiny\begin{bmatrix}
			1 & T_s \\ \frac{-1}{T_s}& -1
		\end{bmatrix}$. Then, it suffices to use point \textbf{(a)} from Theorem~II.1 in \cite{aug_sparse} to produce said factorisation.\vspace{-3mm}
		\begin{remark}\label{rem:db_ex}
			The reason behind choosing the feedbacks $F$ and $L$ as stated above is made plain by recalling the decoupling-based properties we wish to obtain in closed-loop configuration, thereby explaining the block-diagonal structures of the state matrices $A+BF$ and $A+LC$, which broadly govern the closed-loop dynamics. Notice, moreover, that all the eigenvalues of $A+LC$ will be placed in $0$, which allows us to leverage the benefits discussed in Remark~\ref{rem:db} and to obtain NRF-based subcontrollers having all their poles located in $z=0$. Doing so also ensures that the operations described in Proposition~\ref{prop:ss_implem} may be performed with a high degree of numerical accuracy.\vspace{-3mm}
		\end{remark}
		By leveraging the results in Section~\ref{subsec:CONPRAS}, the NRF-based control laws that only exchange information according to the communication neighbourhoods from \eqref{eq:comms} may be parametrised via the following family of $\mathbf{Q}$-parameters\vspace{-3mm}
		\begin{equation*}
			\resizebox{.975\columnwidth}{!}{$\mathbf{Q}(z) = \mathbf{Q}_0(z) + \mathrm{diag}(\mathbf{x}_1(z),\mathbf{x}_2(z),\mathbf{x}_3(z),\mathbf{x}_4(z),1)\mathbf{H}_{\mathcal{N}}(z),$}\vspace{-3mm}
		\end{equation*}
		where $\mathbf{Q}_0(z)$ and
		$\mathbf{H}_{\mathcal{N}}(z)$ are given in \eqref{eq:Q_param}, located at the top of this page, and $\mathbf{x}_i(z)\in\mathcal{R}(z)$, with $i\in\{1:4\}$, are \emph{any} four proper and $\mathbb{C}_b$-bounded transfer functions.\vspace{-3mm}
		
		In the interest of obtaining low-complexity implementations, we will constrain these four transfer functions to be \emph{constant scalars} when configuring the constraints from \eqref{eq:MM_prob}. Moreover, we choose the $\mathcal{H}_\infty$ norm as the systemic norm from \eqref{eq:con_pert}-\eqref{eq:con_init} and we eschew the implementa-\newpage\noindent tion of the constraints from \eqref{eq:con_init} and \eqref{eq:con_gamma}, while setting:\vspace{-3mm}
		\begin{enumerate}
			\item[a)] $\mathbf{T}_{di}(z)\equiv\ O$,\smallskip
			\item[b)] $\mathbf{T}_{uij}(z)\equiv \tfrac{1}{z}I_3$, for $i=j$,\smallskip
			\item[c)] $\mathbf{T}_{uij}(z)\equiv O$, for $i\neq j$,\smallskip
			\item[d)] $\tau_{di}=\tau_{u{ij}}=1,\,\tau_{c{ij}}=0$,\vspace{-3mm}
		\end{enumerate}
		for all $i,j\in\{1:N\}$.\vspace{-3mm}
		
		Solving \eqref{eq:MM_prob} as indicated\footnote{\color{black}The numerical routines from \cite{aug_sparse} are available at the following link: \texttt{https://github.com/AndreiSperila/CONPRAS}} in point b) from the ordered list presented in Section~\ref{subsec:MM_form} yields the optimal solution (for the constant $\mathbf{x}$-parameters) given by\vspace{-3mm}
		\begin{equation}\label{eq:opt_sol}
			\left\{\scriptsize\begin{array}{ll}
				\mathbf{x}_1(z)=0.2682,&\quad\mathbf{x}_2(z)=0.1436,\\\mathbf{x}_3(z)=0.3610,&\quad\mathbf{x}_4(z)=0.0660,
			\end{array}\right.\vspace{-3mm}
		\end{equation}
		which successfully reproduces the control laws from \cite{PT}.\vspace{-3mm}
		\begin{remark}\label{rem:PT}
			The time-based simulations presented in \cite{PT} showcase the high degree of closed-loop performance ensured by the above-mentioned control laws, especially when it comes to containing the effects of the disturbance signals $d_i[k]$ to the states of their corresponding areas. Additionally, the computational benefits afforded by this solution, in tandem with the inexpensive, \emph{linear-programming}-based second layer proposed in \cite{PT}, are highly desirable when compared to the implementation costs of other similar techniques, such as the one in \cite{DMPC1,DMPC2}.\vspace{-3mm}
		\end{remark}
		In light of facts stated in Remark~\ref{rem:PT}, we proceed to focus in the sequel only on showcasing the theoretical aspects of the resulting closed-loop system, and we refer to the numerical example from \cite{PT} for practical considerations.

		\vspace{-3mm}
		\subsection{Specialised Implementations and Guarantees}
		\vspace{-3mm}
		
		With the $\mathbf{Q}$-parameter now computed, and the factorisation from Section~\ref{subsec:DCF} at hand, it is possible to compute the TFM from \eqref{eq:Kd_def} via the identities given in Section~\ref{subsec:DCF2NRF} from Appendix~\ref{app:aux}. Once this is done and row-realisations as in \eqref{eq:Kd_rows} are computed, we apply the indications from Remark~\ref{rem:sparse_real} to first bring the state matrices of these realisations to their Real Schur Forms, and then to $\widehat{A}_\ell=\tiny\begin{bmatrix}
			0 & 1\\ 0 & 0
		\end{bmatrix},\,\forall\,\ell\in\{1:5\}$. As anticipated in Remark~\ref{rem:db_ex}, a simple computation yields $\chi_\ell(z)=z^2,\,\forall\,\ell\in\{1:5\}$.\vspace{-3mm}
		\begin{figure*}\color{black}
			\begin{equation}\label{eq:B_rl}
				\resizebox{.925\textwidth}{!}{$\left\{\begin{aligned}
						B_{r1}=&\tiny\left[\begin{array}{rrrrrrrrrrrrrrrr}
							\phantom{-}\textbf{0.0000}  &  \phantom{-}0.0000  &  \phantom{-}0.0000  &  \phantom{-}\textbf{0.0000}  &  \phantom{-}0.0000  &  \phantom{-}0.0000  &  \phantom{-}0.0000  & -0.0605  & -0.0121  & -0.0232  & -0.0046  &  \phantom{-}\textbf{0.0000}  & \phantom{-}\textbf{0.0000}  & -0.0125  & -0.0025\\
							\phantom{-}\textbf{0.0000}  &  \phantom{-}0.0037  & -0.0056  &  \phantom{-}\textbf{0.0000}  & -0.0072  & -0.0002  &  \phantom{-}0.0000  &  \phantom{-}0.0155  &  \phantom{-}0.0064  & -0.0268  & -0.0107  &  \phantom{-}\textbf{0.0000}  &  \phantom{-}\textbf{0.0000}  & -0.0346  & -0.0142
						\end{array}\right]\normalsize,\\
						B_{r2}=&\tiny\left[\begin{array}{rrrrrrrrrrrrrrrr}
							0.0000  &  \phantom{-}\textbf{0.0000}  &  \phantom{-}\textbf{0.0000}  &  \phantom{-}0.0000  &  \phantom{-}0.0000  & -0.0972  & -0.0194  &  \phantom{-}0.0000  &  \phantom{-}0.0000  &  \phantom{-}\textbf{0.0000}  &  \phantom{-}\textbf{0.0000}  & -0.3305  & -0.0661  & -0.1081  & -0.0216\\
							-0.0219  &  \phantom{-}\textbf{0.0000}  &  \phantom{-}\textbf{0.0000}  &  \phantom{-}0.0125  & -0.0185  & -0.1075  & -0.0431  &  \phantom{-}0.0005  &  \phantom{-}0.0000  &  \phantom{-}\textbf{0.0000}  &  \phantom{-}\textbf{0.0000}  &  \phantom{-}0.0549  &  \phantom{-}0.0237  & -0.0877  & -0.0362
						\end{array}\right]\normalsize,\\
						B_{r3}=&\tiny\left[\begin{array}{rrrrrrrrrrrrrrrr}
							0.0000  &  \phantom{-}\textbf{0.0000}  &  \phantom{-}\textbf{0.0000}  &  \phantom{-}0.0000  &  \phantom{-}0.0000  & -0.0323  & -0.0065  &  \phantom{-}\textbf{0.0000}  &  \phantom{-}\textbf{0.0000}  &  \phantom{-}0.0000  &  \phantom{-}0.0000  & -0.1362  & -0.0272  & -0.0320  & -0.0064\\
							-0.0133  &  \phantom{-}\textbf{0.0000}  &  \phantom{-}\textbf{0.0000}  &  \phantom{-}0.0067  & -0.0151  & -0.0651  & -0.0260  &  \phantom{-}\textbf{0.0000}  &  \phantom{-}\textbf{0.0000}  & -0.0004  &  \phantom{-}0.0000  &  \phantom{-}0.0290  &  \phantom{-}0.0126  & -0.0717  & -0.0295
						\end{array}\right]\normalsize,\\
						B_{r4}=&\tiny\left[\begin{array}{rrrrrrrrrrrrrrrr}
							\phantom{-}\textbf{0.0000}  &  \phantom{-}0.0000  &  \phantom{-}0.0000  &  \phantom{-}\textbf{0.0000}  &  \phantom{-}0.0000  &  \phantom{-}\textbf{0.0000}  &  \phantom{-}\textbf{0.0000}  & -0.2366  & -0.0473  & -0.1127  & -0.0225  &  \phantom{-}0.0000  &  \phantom{-}0.0000  & -0.0569  & -0.0114\\
							\phantom{-}\textbf{0.0000}  &  \phantom{-}0.0117  & -0.0124  &  \phantom{-}\textbf{0.0000}  & -0.0240  &  \phantom{-}\textbf{0.0000}  &  \phantom{-}\textbf{0.0000}  &  \phantom{-}0.0495  &  \phantom{-}0.0204  & -0.0594  & -0.0236  &  \phantom{-}0.0005  &  \phantom{-}0.0000  & -0.1140  & -0.0469
						\end{array}\right]\normalsize,\\
						B_{r5}=&\tiny\left[\begin{array}{rrrrrrrrrrrrrrrr}
							0.0000  &  \phantom{-}0.0000  &  \phantom{-}0.0000  &  \phantom{-}0.0000  &  \phantom{-}0.0000  &  \phantom{-}0.0000  &  \phantom{-}0.0000  &  \phantom{-}0.0000  &  \phantom{-}0.0000  &  \phantom{-}0.0000  &  \phantom{-}0.0000  &  \phantom{-}0.0000  &  \phantom{-}0.0000  &  \phantom{-}0.0000  &  \phantom{-}0.0000\\
							-0.0100  & -0.0084  & -0.0114  & -0.0111  &  \phantom{-}0.0000  & -0.0512  & -0.0195  & -0.0386  & -0.0147  & -0.0560  & -0.0218  & -0.0530  & -0.0210  & -0.0054  &  \phantom{-}0.0000
						\end{array}\right]\normalsize.
					\end{aligned}\right.$}\vspace{-3mm}
			\end{equation}\hrulefill\vspace{-2mm}
			\begin{equation}\label{eq:TFM_ex}
				\resizebox{.925\textwidth}{!}{
					$Z_1^\top\mathbf{F}_{\mathbf{Q}}(z)\scriptsize\begin{bmatrix}
						Z_4\\O
					\end{bmatrix}\normalsize=\left[\tiny\begin{array}{ccccccccc|ccc}
						\phantom{-}0.0722-z  & -0.0217  &  \phantom{-}0.1805  & -0.0536  &  \phantom{-}0.0509  & -0.1049  & -0.0283  &  \phantom{-}0.0462  &  \phantom{-}0.0000  & -0.0188  & -0.0020  & -0.0018\\
						-0.1633  &  \phantom{-}0.9360-z  & -0.2555  & -0.0070  &  \phantom{-}0.0066  & -0.0139  & -0.0037  &  \phantom{-}0.0062  &  \phantom{-}0.0000  & -0.0018  &  \phantom{-}0.0112  & -0.0108\\
						0.1349  &  \phantom{-}0.4076  &  \phantom{-}0.7141-z  & -0.2157  & -0.0157  &  \phantom{-}0.0264  &  \phantom{-}0.0087  & -0.0113  &  \phantom{-}0.0000  & -0.0899  &  \phantom{-}0.0130  & -0.0283\\
						-0.0588  & -0.1965  & -0.1778  &  \phantom{-}0.1978-z  & -0.3499  & -0.0085  & -0.0032  &  \phantom{-}0.0036  &  \phantom{-}0.0000  &  \phantom{-}0.3442  &  \phantom{-}0.1566  & -0.0777\\
						0.0937  & -0.1123  & -0.1077  & -0.2457  &  \phantom{-}0.8120-z  & -0.2149  &  \phantom{-}0.0022  & -0.0037  &  \phantom{-}0.0000  &  \phantom{-}0.2046  &  \phantom{-}0.0731  & -0.0283\\
						-0.0996  & -0.0650  & -0.0170  & -0.2051  &  \phantom{-}0.2285  &  \phantom{-}0.9396-z  &  \phantom{-}0.0330  &  \phantom{-}0.0060  &  \phantom{-}0.0000  & -0.0801  &  \phantom{-}0.0004  & -0.0146\\
						-0.0545  &  \phantom{-}0.0018  &  \phantom{-}0.0247  &  \phantom{-}0.0606  & -0.0193  & -0.0159  &  \phantom{-}0.9650-z  & -0.1843  &  \phantom{-}0.0000  &  \phantom{-}0.0034  & -0.0094  &  \phantom{-}0.0098\\
						0.0486  &  \phantom{-}0.0009  & -0.0220  & -0.0340  & -0.0156  &  \phantom{-}0.0167  &  \phantom{-}0.2293  &  \phantom{-}0.9783-z  &  \phantom{-}0.0000  &  \phantom{-}0.0422  &  \phantom{-}0.0246  & -0.0162\\
						0.0561  &  \phantom{-}0.5271  &  \phantom{-}0.5568  & -0.3219  & -0.1533  &  \phantom{-}0.4775  & -0.0724  &  \phantom{-}0.0366  &  -z  &  \phantom{-}0.2874  &  \phantom{-}0.0574  & -0.0009\\\hline
						-0.0482  & -0.0058  &  \phantom{-}0.0103  & -0.0028  &  \phantom{-}0.0033  & -0.0048  & -0.0019  &  \phantom{-}0.0021  & -0.0006  &  \phantom{-}0.0000  &  \phantom{-}0.0000  &  \phantom{-}0.0000\\
						0.2408  &  \phantom{-}0.0292  & -0.0517  &  \phantom{-}0.0141  & -0.0165  &  \phantom{-}0.0241  &  \phantom{-}0.0097  & -0.0107  &  \phantom{-}0.0028  &  \phantom{-}0.0000 &   \phantom{-}0.0000  &  \phantom{-}0.0000\\
						-0.2386  & -0.0077  &  \phantom{-}0.0536  & -0.0442  &  \phantom{-}0.0743  &  \phantom{-}0.0099  & -0.0320  & -0.0084  &  \phantom{-}0.0055  &  \phantom{-}0.0000  &  \phantom{-}0.0000  &  \phantom{-}0.0000
					\end{array}\right].$
				}\vspace{-3mm}
			\end{equation}
			\hrulefill\vspace{-2mm}
		\end{figure*}
		
		Effecting the rest of the numerical computations yields the fact that $A_{r\ell}=\widehat{A}_\ell$ and $D_{r\ell}=O$, for each $\ell\in\{1:5\}$, with the $B_{r\ell}$ matrices being given in \eqref{eq:B_rl}, located at the top of this page. A cursory inspection of the latter confirms the fact that point $iii)$ of Proposition~\ref{prop:ss_implem} holds, with the obtained state-space implementations adhering to the communication neighbourhoods in \eqref{eq:comms} and with the inherited zero-entries being highlighted in \eqref{eq:B_rl} via bold font. Moreover, by inspecting the minimality conditions given in Section~\ref{subsec:struc} from Appendix~\ref{app:aux} (said tests need only be performed at $z=0$), it directly follows that point $ii)$ of Proposition~\ref{prop:ss_implem} holds, as well.\vspace{-3mm}
		
		By deploying these specialised implementations for the power grid discussed in Section~\ref{subsec:model}, one readily obtains the closed-loop response described in the statement of Theorem~\ref{thm:NRF_state}. Given the large number of TFMs that determine said closed-loop response, we choose a single illustrative example from among them, and we focus in the sequel on showcasing all of its beneficial properties.\vspace{-3mm}
		
		Consider the TFM $Z_1^\top\mathbf{F}_{\mathbf{Q}}(z)\begin{bmatrix}
			Z_4^\top & O
		\end{bmatrix}^\top$, which governs the coupling between the $4^\text{th}$ area's second-layer inputs and the $1^\text{st}$ area's variables, \emph{i.e.}, its two states and its first-layer command signal. Said TFM is indeed proper, given that it is possible to obtain the state-space realisation from \eqref{eq:TFM_ex}, located at the top of this page. Moreover, the TFM is also $\mathbb{C}_b$-bounded, since the spectral radius of its state matrix is equal to $0.9983$ (recall that $\mathbb{C}_b=\mathbb{C}\setminus\mathbb{S}$).\vspace{-3mm}
		
		\begin{remark}\label{rem:no_comm}
			Note that the communication neighbourhoods from \eqref{eq:comms} do not allow any information exchange between areas 1 and 4. Therefore, given that no \emph{direct} feedback- or feedforward-based strategies can be employed to readily compensate for the interplay between the two areas' dynamics, the desire to attenuate the contribution of the TFM from \eqref{eq:TFM_ex} becomes all the more crucial.\vspace{-3mm}
		\end{remark}
		
		Finally, we inspect the norm-based model-matching condition from \eqref{eq:con_cmd} for $i=1$ and for $j=4$, to get that the optimal solution given by \eqref{eq:opt_sol} ensures a value of $\gamma_{u14}=0.8201$. We conclude by pointing out that, since\vspace{-3mm}
		\begin{equation*}
			\left\|\begin{array}{l}
				\begin{bmatrix}
					I_2 & O
				\end{bmatrix}Z_1^\top\mathbf{F}_{\mathbf{Q}}(z)\begin{bmatrix}
					Z_4^\top & O
				\end{bmatrix}^\top
			\end{array}\right\|_\infty=3\cdot10^{-12},\vspace{-3mm}
		\end{equation*}
		our control laws almost perfectly desensitise the two states of the $1^\text{st}$ area to the second-layer commands of the $4^\text{th}$ one, with the bulk of the residual coupling impacting $u_{f1}$. Not only does this align with the comments from Remark~\ref{rem:no_comm}, but it is also beneficial, since the latter signal can be acted upon directly and adjusted via $u_{s21}$.
		
	}

	\vspace{-3mm}
	\section{Conclusions}\vspace{-3mm}
	\label{sec:outro}
	
	Through judicious choices for the desensitised closed-loop target TFMs and for the associated system norms in the model-matching design procedure, this paper proposes a set of NRF-based control laws which facilitate the development and implementation of distributed supervisory policies, and which are characterised by strong theoretical guarantees. Specific to our chosen approach, the configuration and the implementation of these NRF-based control laws can be performed entirely in an offline setting, while also being independent of the design procedure for the top layer of our proposed control architecture. Thus, the second layer may benefit from all the system-theoretical properties provided by the first layer, in order to efficiently handle the remaining constraints.
	
	\vspace{-3mm}
	
	\begin{ack}\vspace{-3mm}
		The authors would like to thank both \c Serban Sab\u au and Cristian Oar\u a for the many fruitful discussions which ultimately led to the elaboration of this manuscript.\vspace{-3mm}
	\end{ack}

	\bibliographystyle{plain}
	\bibliography{manuscript}

@article{PT,
	title = {{Grid Phase Synchronization via Distributed Control}},
	journal = {To appear in Proc. of the $16^{th}$ IEEE PowerTech Conference},
	pages = {1--6},
	year = {2025},
	author = {Speril{\u a}, Andrei and Iovine, Alessio and Olaru, Sorin and Panciatici, Patrick},
	note = {Available Online: https://hal.science/hal-05056697}
}

@book{Kuo,
	title =     {{Digital Control Systems}},
	author =    {Benjamin Kuo},
	publisher = {Harcourt College Publishers},
	year =      {1980},
}

@article{VNULL,
	title = {{On Computing Nullspace Bases - a Fault Detection Perspective}},
	journal = {In Proc. of the $17^{th}$ IFAC World Congress},
	pages = {6295--6300},
	year = {2008},
	author = {A. Varga},
	
}

@article{VSOLVE,
	title={Computation of least order solutions of linear rational equations},
	author={A. Varga},
	journal={{In Proc. of the International Symposium on Mathematical Theory of Networks and Systems}},
	year={2004}
}

@article{VCOVER,
	title={Reliable algorithms for computing minimal dynamic covers for descriptor systems},
	author={A. Varga},
	journal={{In Proc. of the International Symposium on Mathematical Theory of Networks and Systems}},
	year={2004}
}

@article{JA,  author={Tseng, Shih-Hao and Anderson, James},  journal={In Proc. of the 2020 American Control Conf.},   title={{Deployment Architectures for Cyber-Physical Control Systems}},   year={2020},  pages={5287--5294},  doi={10.23919/ACC45564.2020.9147953}}

@article{arch_rev,
title = {{Architectures for distributed and hierarchical Model Predictive Control – A review}},
journal = {Journal of Process Control},
volume = {19},
number = {5},
pages = {723--731},
year = {2009},
doi = {https://doi.org/10.1016/j.jprocont.2009.02.003},
author = {Riccardo Scattolini},
}

@INPROCEEDINGS{R8D,

  author={Hernandez, Bernardo and Trodden, Paul},

  booktitle={2016 UKACC $11^\text{th}$ International Conference on Control}, 

  title={Distributed model predictive control using a chain of tubes}, 

  year={2016},

  pages={1--6},

  doi={10.1109/CONTROL.2016.7737610}}

@article{R8A,
title = {{Distributed predictive control: A non-cooperative algorithm with neighbor-to-neighbor communication for linear systems}},
journal = {Automatica},
volume = {48},
number = {6},
pages = {1088--1096},
year = {2012},
doi = {https://doi.org/10.1016/j.automatica.2012.03.020},
author = {Marcello Farina and Riccardo Scattolini},
}

@article{plat_MPC_4,
author = {Feng, Yangyang and Yu, Shuyou and Chen, Hao and Li, Yongfu and Shi, Shuming and Yu, Jianhua and Chen, Hong},
year = {2023},
month = {06},
pages = {1--21},
title = {Distributed MPC of vehicle platoons with guaranteed consensus and string stability},
volume = {13},
journal = {Scientific Reports},
doi = {10.1038/s41598-023-36898-4}
}

@ARTICLE{plat_MPC_5,

  author={van Nunen, Ellen and Reinders, Joey and Semsar-Kazerooni, Elham and van de Wouw, Nathan},

  journal={IEEE Transactions on Intelligent Vehicles}, 

  title={{String Stable Model Predictive Cooperative Adaptive Cruise Control for Heterogeneous Platoons}}, 

  year={2019},

  volume={4},

  number={2},

  pages={186--196},

  doi={10.1109/TIV.2019.2904418}}

@ARTICLE{plat_MPC_3,

  author={Dunbar, William B. and Caveney, Derek S.},

  journal={IEEE Transactions on Automatic Control}, 

  title={{Distributed Receding Horizon Control of Vehicle Platoons: Stability and String Stability}}, 

  year={2012},

  volume={57},

  number={3},

  pages={620--633},

  doi={10.1109/TAC.2011.2159651}}

@ARTICLE{plat_MPC_2,

  author={Wang, Jiange and Li, Xiaolei and Park, Ju H. and Guo, Ge},

  journal={IEEE Transactions on Intelligent Transportation Systems}, 

  title={{Distributed MPC-Based String Stable Platoon Control of Networked Vehicle Systems}}, 

  year={2023},

  volume={24},

  number={3},

  pages={3078--3090},

  doi={10.1109/TITS.2022.3221382}}

@ARTICLE{plat_MPC_1,

  author={Zheng, Yang and Li, Shengbo Eben and Li, Keqiang and Borrelli, Francesco and Hedrick, J. Karl},

  journal={IEEE Transactions on Control Systems Technology}, 

  title={{Distributed Model Predictive Control for Heterogeneous Vehicle Platoons Under Unidirectional Topologies}}, 

  year={2017},

  volume={25},

  number={3},

  pages={899--910},

  doi={10.1109/TCST.2016.2594588}}

@article{R8B,
title = {Distributed predictive control with minimization of mutual disturbances},
journal = {Automatica},
volume = {77},
pages = {31--43},
year = {2017},
doi = {https://doi.org/10.1016/j.automatica.2016.11.023},
author = {Paul A. Trodden and J.M. Maestre},
}

@article{refgov,
	title = {{Reference and command governors for systems with constraints: A survey on theory and applications}},
	journal = {Automatica},
	volume = {75},
	pages = {306--328},
	year = {2017},
	issn = {0005-1098},
	doi = {https://doi.org/10.1016/j.automatica.2016.08.013},
	author = {{Emanuele Garone, Stefano Di Cairano and Ilya Kolmanovsky}},
}

@book{STMC,
	title =     {{Set-Theoretic Methods in Control, Second Edition}},
	author =    {Franco Blanchini and Stefano Miani},
	publisher = {Birkhäuser},
	year =      {2015},
}

@book{DMPCB,
	title =     {{Distributed Model Predictive Control Made Easy}},
	author =    {Jos\'{e} Maestre and Rudy Negenborn},
	publisher = {Springer},
	year =      {2014},
	series =    {Intelligent Systems, Control and Automation: Science and Engineering},
}

@misc{part2,
	title={{Network-Realised Model Predictive Control -- Part~II: Distributed Constraint Management}}, 
	author={Andrei Speril\u{a} and Alessio Iovine and Sorin Olaru and Patrick Panciatici},
	year={2025},
	note = {Available Online: https://arxiv.org/abs/2502.13073}}

@book{BBM,
	title={{Predictive Control for Linear and Hybrid Systems}},
	author={Borrelli, Francesco and Bemporad, Alberto and Morari, Manfred},
	year={2017},
	publisher={Cambridge University Press}
}

@book{RMD,
	title={{Model Predictive Control: Theory, Computation, and Design, Second Edition}},
	author={Rawlings, James and Mayne, David and Diehl, Moritz},
	year={2020},
	publisher={Nob Hill Publishing}
}

@ARTICLE{DMPC2,
	
	author={Alonso, Carmen Amo and Li, Jing Shuang and Matni, Nikolai and Anderson, James},
	
	journal={IEEE Transactions on Control of Network Systems}, 
	
	title={{Distributed and Localized Model Predictive Control—Part II: Theoretical Guarantees}},
	
	year={2023},
	
	volume={10},
	
	number={3},
	
	pages={1113--1123},
	
	doi={10.1109/TCNS.2023.3262650}}

@ARTICLE{DMPC1,
	
	author={Alonso, Carmen Amo and Li, Jing Shuang and Anderson, James and Matni, Nikolai},
	
	journal={IEEE Transactions on Control of Network Systems}, 
	
	title={{Distributed and Localized Model-Predictive Control–Part I: Synthesis and Implementation}}, 
	
	year={2023},
	
	volume={10},
	
	number={2},
	
	pages={1058--1068},
	
	doi={10.1109/TCNS.2022.3219770}}

@article{SI,  author={Furieri, Luca and Zheng, Yang and Papachristodoulou, Antonis and Kamgarpour, Maryam},  journal={IEEE Transactions on Control of Network Systems},   title={{Sparsity Invariance for Convex Design of Distributed Controllers}},   year={2020},  volume={7},  number={4},  pages={1836--1847},  doi={10.1109/TCNS.2020.3002429}}

@article{matni1,
	Author = {Yuh-Shyang Wang  and Nikolai Matni  and John C. Doyle },
	Journal = { IEEE Transactions on Automatic Control},
	Number = {12},
	Pages = {4234--4249},
	Title = {{Separable and Localized System-Level Synthesis for Large-Scale Systems}},
	Volume = {63},
	Year = {2018}}

@article{Luca1,  
	author={Furieri, Luca and Zheng, Yang and Papachristodoulou, Antonis and Kamgarpour, Maryam},  
	journal={IEEE Control Systems Letters},   
	title={{An Input–Output Parametrization of Stabilizing Controllers: Amidst Youla and System Level Synthesis}},   
	year={2019},  
	volume={3},  
	number={4},  
	pages={1014--1019},
	doi={10.1109/LCSYS.2019.2920205}}

@article{Luca2,
	title = {{System-level, input–output and new parameterizations of stabilizing controllers, and their numerical computation}},
	journal = {Automatica},
	volume = {140},
	pages = {110211},
	year = {2022},
	author = {Yang Zheng and Luca Furieri and Maryam Kamgarpour and Na Li},
	doi = {https://doi.org/10.1016/j.automatica.2022.110211}
}

@article{Luca3,
	author={Zheng, Yang and Furieri, Luca and Papachristodoulou, Antonis and Li, Na and Kamgarpour, Maryam},
	journal={IEEE Transactions on Automatic Control}, 
	title={{On the Equivalence of Youla, System-Level, and Input–Output Parameterizations}}, 
	year={2021},
	volume={66},
	number={1},
	pages={413--420},
	doi={10.1109/TAC.2020.2979785}}

@article{aug_sparse,
	author={Speril\u{a}, Andrei and Oar\u{a}, Cristian and Ciubotaru, Bogdan and Sab\u{a}u, {\c{S}}erban},
	journal={{IEEE Transactions on Automatic Control}}, 
	title={{Distributed Control of Descriptor Networks: A Convex Procedure for Augmented Sparsity}}, 
	year={2023},
	volume={68},
	number={12},
	pages={8067--8074},
	doi={10.1109/TAC.2023.3301949}
}

@book{zhou,
	title={{Robust and Optimal Control}},
	author={Zhou, Kemin and Doyle, John and Glover, Keith},
	year={1996},
	publisher={Prentice-Hall}
}

@book{Kai,
	title={{Linear Systems}},
	author={Kailath, Thomas},
	publisher={Prentice-Hall},
	year={1980}
}

@book{gantmacher,
	title={{The Theory of Matrices}},
	author={Gantmacher, Feliks},
	year={1959},
	publisher={American Math. Society}
}

@article{Rotko,  
	author={Rotkowitz, Michael and Lall, Sanjay},  
	journal={IEEE Transactions on Automatic Control},   
	title={{A Characterization of Convex Problems in Decentralized Control}},   year={2006},  
	volume={51},  
	number={2},  
	pages={274--286},  
	doi={10.1109/TAC.2005.860365}
}

@article{plutonizare,  
	author={Sab\u{a}u, {\c{S}}erban and Oar\u{a}, Cristian and Warnick, Sean and Jadbabaie, Ali},  
	journal={IEEE Transactions on Automatic Control},   
	title={{Optimal Distributed Control for Platooning via Sparse Coprime Factorizations}},   
	year={2017},  
	volume={62},  
	number={1},  
	pages={305--320},  
	doi={10.1109/TAC.2016.2572002}
}

@article{NRF,
	author={Sab\u{a}u, {\c{S}}erban and Speril\u{a}, Andrei and Oar\u{a}, Cristian and Jadbabaie, Ali},
	journal={{IEEE Transactions on Automatic Control}}, 
	title={{Network Realization Functions for Optimal Distributed Control}}, 
	year={2023},
	volume={68},
	number={12},
	pages={8059--8066},
	doi={10.1109/TAC.2023.3298549}
}

@article{SLS,  
	author={Wang, Yuh-Shyang and Matni, Nikolai and Doyle, John},  journal={IEEE Transactions on Automatic Control},   
	title={{A System-Level Approach to Controller Synthesis}},   
	year={2019},  
	volume={64},  
	number={10},  
	pages={4079--4093},  
	doi={10.1109/TAC.2018.2890753}
}
	
	\vspace{-3mm}
	\phantom{ }\vspace{-9mm}
	\appendix
	\section{System-Theoretical Notions}\label{app:aux}\vspace{-3mm}
	
	\subsection{Matrix Pencils}\vspace{-3mm}
	
	A matrix pencil $A-z E$ which is square and which satisfies $\det(A-z E)\not\equiv 0$ is called \emph{regular}. Whenever $E$ equals the identity matrix, said pencil is guaranteed to be regular. With respect to a set $\mathbb{C}_g\subseteq\mathbb{C}$, we say that $ A-zE$ is \emph{$\mathbb{C}_g$-admissible} if $E$ is invertible and if all of the pencil's generalised eigenvalues (see \cite{gantmacher}) are located in $\mathbb{C}_g$.\vspace{-3mm}
	
	\subsection{Stability of State-Space Systems}\vspace{-3mm}
	
	We briefly discuss here the stability properties of networks which are described by \eqref{eq:ss_a}-\eqref{eq:ss_b}. A crucial property is the fact that if the pole pencils of such systems are $\mathbb{S}$-admissible, then said systems are \emph{asymptotically stable}, \emph{i.e.}, their state variables tend to $0$ from all finite initial conditions while remaining bounded at all times, when $u\equiv O$ and $d\equiv O$. For such systems, another remarkable property is the fact that, given any finite initial condition and any bounded $u$ and $d$, both $x$ and $y$ in \eqref{eq:ss_a}-\eqref{eq:ss_b} remain bounded at all times.\vspace{-3mm}
	
	\subsection{Structural Properties of State-Space Systems}\label{subsec:struc}\vspace{-3mm}
	
	We now refer to the structural properties of a generic (sub)realisation $(A,B_u,C,D_u)$ of type \eqref{eq:ss_a}-\eqref{eq:ss_b} and we do so from a purely algebraic standpoint, as our treatment of the subject is meant as a discrete-time counterpart of the continuous-time case presented in Section~3.2 of \cite{zhou} (see also Chapter 21 of \cite{zhou}). A system of the this type is called \emph{controllable at $z\in \mathbb{C}$} if it satisfies $\mathrm{rank}\begin{bmatrix}
		A-z I_{n_x} & B_u
	\end{bmatrix} = n_x$. With respect to a set $\mathbb{C}_b\subseteq\mathbb{C}$, a system \eqref{eq:ss_a}-\eqref{eq:ss_b} is called \emph{$\mathbb{C}_b$-controllable} if it is controllable at all $z\in\mathbb{C}_b$. Such a system is called \emph{observable at $z$} if $\mathrm{rank}\begin{bmatrix}
		A^\top-z I_{n_x} & C^\top
	\end{bmatrix}^\top = n_x$, and it is called \emph{$\mathbb{C}_b$-observable} if it is observable at all $z\in\mathbb{C}_b$. A system that is both $\mathbb{C}_b$-controllable and $\mathbb{C}_b$-observable will be called \emph{$\mathbb{C}_b$-irreducible}, while one that is both controllable and observable at all $z\in\mathbb{C}$ is called \emph{minimal} (see \cite{Kai}).\vspace{-3mm}
	
	\subsection{Matrices of Rational Functions}\vspace{-3mm}
	
	A TFM $\mathbf{G}\in\mathcal{R}(z)^{n_y\times n_u}$ for which $\lim_{|z|\rightarrow\infty}\mathbf{G}(z)$ contains only finite entries is called \emph{proper}. In addition to this, when $\lim_{|z|\rightarrow\infty}\mathbf{G}(z)=O$, the TFM si called \emph{strictly proper}. We now introduce a similar property, with respect to a set $\mathbb{C}_b\subseteq\mathbb{C}$. A TFM $\mathbf{G}\in\mathcal{R}(z)^{n_y\times n_u}$ that is \emph{$\mathbb{C}_b$-bounded} is one for which $\lim_{z\rightarrow\mu}\mathbf{G}(z)$ contains only finite entries, $\forall\,\mu\in\mathbb{C}_b$. Moreover, any proper TFM can be represented via a state space realisation, as in \eqref{eq:TFM_def}. The set of all proper TFMs which are bounded on $\mathbb{C}\setminus\mathbb{S}$ is denoted by $\mathcal{RH}_\infty(z)$. For a TFM $\mathbf{G}\in\mathcal{RH}_\infty(z)$, $\mathbf{G}(z)$ is called \emph{stable} and its $\mathcal{H}_\infty$ norm can be computed as $\|\mathbf{G}\|_\infty=\sup_{z\in\partial\mathbb{S}}\overline{\sigma}(\mathbf{G}(z))$, {\color{black} where $\overline{\sigma}(\cdot)$ is the largest singular value and where we define $\partial\mathbb{S}:=\{z\in\mathbb{C}\ \vert\ |z|=1\}$.}\vspace{-3mm}
	
	\subsection{Doubly Coprime Factorisations}\label{subsec:DCF}\vspace{-3mm}
	
	To begin with, we select a partition of the complex plane via $\mathbb{C}_g\subseteq\mathbb{C}$ along with its complement $\mathbb{C}_b:=\mathbb{C}\setminus\mathbb{C}_g$, and we also consider a network described by a realisation of type \eqref{eq:ss_a}-\eqref{eq:ss_c}. We denote the TFM of this system as $\mathbf{G}(z):=\begin{bmatrix}
		\mathbf{G}_u(z) & \mathbf{G}_d(z)
	\end{bmatrix}$, where the partition into $\mathbf{G}_u\in\mathcal{R}(z)^{n_x\times n_u}$ and $\mathbf{G}_d\in\mathcal{R}(z)^{n_x\times n_d}$ is conformal with that of the pair $(B_u,B_d)$ appearing in \eqref{eq:ss_a}-\eqref{eq:ss_c}. We say that a collection of eight TFMs $(\mathbf{N}(z),\mathbf{M}(z),\mathbf{X}(z),$ $\mathbf{Y}(z),\widetilde{\mathbf{N}}(z),\widetilde{\mathbf{M}}(z),\widetilde{\mathbf{X}}(z),\widetilde{\mathbf{Y}}(z))$ is a \emph{Doubly Coprime Factorisation} (DCF) over $\mathbb{C}_g$ of $\mathbf{G}_u(z)$ if:\vspace{-4mm}
	\begin{enumerate}
		\item[a)] $\mathbf{M}(z)$ along with $\widetilde{\mathbf{M}}(z)$ are invertible and they satisfy $\mathbf{G}_u(z)=\mathbf{N}(z)\mathbf{M}^{-1}(z)=\widetilde{\mathbf{M}}^{-1}(z)\widetilde{\mathbf{N}}(z)$;\smallskip
		
		\item[b)] all eight TFMs are proper and $\mathbb{C}_b$-bounded;\smallskip
		
		\item[c)] the eight TFMs satisfy the (B\' ezout-like) identity\vspace{-3mm}
		\begin{equation}\label{eq:Bezout}
			\hspace{-6mm}\scriptsize\begin{bmatrix}
				\phantom{-}\widetilde{\mathbf{Y}}(z) & -\widetilde{\mathbf{X}}(z)\\
				-\widetilde{\mathbf{N}}(z) & \phantom{-}\widetilde{\mathbf{M}}(z)
			\end{bmatrix}
			\begin{bmatrix}
				\mathbf{M}(z) & \mathbf{X}(z) \\\mathbf{N}(z) & \mathbf{Y}(z)
			\end{bmatrix}=\begin{bmatrix}
				I_{n_u} & O \\ O & I_{n_x}
			\end{bmatrix}.\normalsize\vspace{-3mm}
		\end{equation}
	\end{enumerate}

	\subsection{DCF-based NRF Pairs}\label{subsec:DCF2NRF}\vspace{-3mm}
	
	We state here some of the most important results from \cite{NRF} and \cite{aug_sparse}. For a network described by \eqref{eq:ss_a}-\eqref{eq:ss_c} and partitioned as in \eqref{eq:trip}-\eqref{eq:net_part}, the NRF pairs to which we refer in this paper are produced by first obtaining a DCF over $\mathbb{C}_g$ of $\mathbf{G}_u(z)$, as previously described, and then forming\vspace{-3mm}
	\begin{equation}\label{eq:aux_NRF}
		\hspace{-3mm}\left\{\begin{aligned}
			\widetilde{\mathbf{Y}}_{\mathbf{Q}}(z):=&\ \widetilde{\mathbf{Y}}(z)+{\mathbf{Q}}(z)\widetilde{\mathbf{N}}(z),\\
			\widetilde{\mathbf{X}}_{\mathbf{Q}}(z):=&\ \widetilde{\mathbf{X}}(z)+{\mathbf{Q}}(z)\widetilde{\mathbf{M}}(z),\\
			\widetilde{\mathbf{Y}}_{\mathbf{Q}}^{\mathrm{diag}}(z):=&\ \tiny\begin{array}{l}
				\mathrm{diag}\big(\mathrm{elm}_{11}\big(\widetilde{\mathbf{Y}}_{\mathbf{Q}}(z)\big),\dots,\mathrm{elm}_{n_un_u}\big(\widetilde{\mathbf{Y}}_{\mathbf{Q}}(z)\big)\big)
			\end{array},
		\end{aligned}\right.\vspace{-3mm}
	\end{equation}
	for a $\mathbf{Q}\in\mathcal{R}(z)^{n_u\times n_x}$ that is proper, $\mathbb{C}_b$-bounded and which ensures that $\widetilde{\mathbf{Y}}_{\mathbf{Q}}(z)$ and $\widetilde{\mathbf{Y}}_{\mathbf{Q}}^{\mathrm{diag}}(z)$ have proper inverses. For such a $\mathbf{Q}(z)$, an NRF pair is given by\vspace{-3mm}
	\begin{equation}\label{eq:NRF_def}
		\left\{\begin{aligned}
			\mathbf{\Phi}(z):=&\ I_{n_u}-\big(\widetilde{\mathbf{Y}}_{\mathbf{Q}}^{\mathrm{diag}}(z)\big)^{-1}\widetilde{\mathbf{Y}}_{\mathbf{Q}}(z),\\
			\mathbf{\Gamma}(z):=&\ \big(\widetilde{\mathbf{Y}}_{\mathbf{Q}}^{\mathrm{diag}}(z)\big)^{-1}\widetilde{\mathbf{X}}_{\mathbf{Q}}(z).
		\end{aligned}
		\right.\vspace{-3mm}
	\end{equation}
	For all systems of type \eqref{eq:ss_a}-\eqref{eq:ss_c}, it follows that both $\mathbf{N}(z)$ and $\widetilde{\mathbf{N}}(z)$ are strictly proper. When considering a partitioning as in \eqref{eq:trip}-\eqref{eq:net_part}, it is \emph{always} possible to obtain a DCF over $\mathbb{C}_g$ for a system of type \eqref{eq:ss_a}-\eqref{eq:ss_c} in which\vspace{-3mm}
	\begin{equation}\label{eq:DCF_gain_inf}
		\lim\limits_{|z|\rightarrow\infty}\widetilde{\mathbf{Y}}(z)=I_{n_u}\text{ and }\lim\limits_{|z|\rightarrow\infty}\widetilde{\mathbf{X}}(z)=O.\vspace{-3mm}
	\end{equation}
	By employing such a DCF, it is straightforward to check that any strictly proper $\mathbf{Q}(z)$ ensures the fact that both $\widetilde{\mathbf{Y}}_{\mathbf{Q}}^{}(z)$ and $\widetilde{\mathbf{Y}}_{\mathbf{Q}}^{\mathrm{diag}}(z)$ have proper inverses, and also that the resulting NRF pair will be strictly proper.\vspace{-3mm}
	
	In this paper, we consider only DCFs over $\mathbb{C}_g$ which satisfy \eqref{eq:DCF_gain_inf}. We point out that there is no loss of generality in doing so, since our method is based upon the Youla Parametrisation, whose theoretical guarantees hold when using \emph{any} DCF over $\mathbb{C}_g$ of our network (see \cite{NRF}, and also \cite{aug_sparse} for state-space formulas). It is for this reason that we attribute in Section~\ref{subsec:NRF_basic} the name {$\mathbb{C}_g$-allocating} to all of the NRF pairs obtained in the manner presented above. Finally, to conclude this discussion, we will additionally impose that $\lim_{|z|\rightarrow\infty}\mathbf{Q}(z)=O$, since having a strictly proper NRF pair greatly aids in implementing the first layer of our control architecture.\vspace{-3mm}
	
	{\color{black}Moving on, we now give a brief overview of the computational details from Sections~III and~IV of \cite{aug_sparse} on how to form an entire class of ${\mathbf{Q}}$-parameters which ensure that the TFMs from \eqref{eq:NRF_def} have a desired sparsity pattern.}\vspace{-3mm}
	
	{\color{black}
		\subsection{Parametrisation of Sparse NRF Pairs}\label{subsec:CONPRAS}\vspace{-3mm}
		
		One of the key features of an NRF pair $(\mathbf{\Phi}(z),\mathbf{\Gamma}(z))$ computed as in \eqref{eq:NRF_def} is the fact that it \emph{inherits the sparsity pattern} of the TFM pair $\big(\widetilde{\mathbf{Y}}_{\mathbf{Q}}(z),\widetilde{\mathbf{X}}_{\mathbf{Q}}(z)\big)$ introduced in \eqref{eq:aux_NRF}, due to the diagonal structure of $I_{n_u}$ and $\widetilde{\mathbf{Y}}_{\mathbf{Q}}^{\mathrm{diag}}(z)$. Since the first two TFMs defined in \eqref{eq:aux_NRF} are affine expressions of $\mathbf{Q}(z)$, it becomes possible to employ the techniques discussed in \cite{VSOLVE,VCOVER,VNULL} and summarised in \cite{aug_sparse} (see Section~III.B therein) to form an entire class of $\mathbb{C}_g$-allocating NRF pairs having a desired sparsity pattern.\vspace{-3mm} 
		
		In order to explain this sparsity-enforcing procedure, we now introduce the vectorisation operator which, for any $\mathbf{H}\in\mathcal{R}(z)^{n_y\times n_u}$, is given by $\mathrm{vec}(\mathbf{H}(z)):=\mathbf{h}\in\mathcal{R}(z)^{n_yn_u}$ that satisfies the identity\vspace{-3mm}
		\begin{equation*}
			\scriptsize\begin{array}{l}
				\mathrm{elm}_{ij}(\mathbf{H}(z))=\mathrm{row}_{i+(j-1)n_y}(\mathbf{h}(z)),\,\forall\,i\in\{1:n_y\},\,j\in\{1:n_u\}.
			\end{array}\vspace{-3mm}
		\end{equation*}
		Moving on, consider some matrix $\mathcal{S}\in\mathbb{R}^{n_u\times(n_u+n_x)}$ which is binary, \emph{i.e.}, $\mathrm{elm}_{ij}(\mathcal{S})\in\{0,1\}$ for all $i\in\{1:n_u\}$ and all $j\in\{1:(n_u+n_x)\}$, and which also satisfies $elm_{ii}(\mathcal{S})=0$ for all $i\in\{1:n_u\}$. Recalling the TFM defined in \eqref{eq:Kd_def}, Proposition~III.1 in \cite{aug_sparse} shows that it is possible to enforce the sparsity-promoting implication\vspace{-3mm}
		\begin{multline*}
			\mathrm{elm}_{ij}(\mathcal{S})=0\Longrightarrow\mathrm{elm}_{ij}(\mathbf{K}_{\mathbf{D}}(z))\equiv0,\\\,\forall\,i\in\{1:n_u\},\,j\in\{1:n_u+n_y\},
		\end{multline*}\phantom{}\vspace{-9mm}
		
		by forming the NRF pair as stated in \eqref{eq:aux_NRF}-\eqref{eq:NRF_def} via a Youla parameter given by $\mathbf{Q}(z)=\mathbf{Q}_0(z)+\widehat{\mathbf{Q}}(z)$, where:\vspace{-3mm}
		\begin{enumerate}
			\item[a)] The TFM $\mathbf{Q}_0(z)$ is a proper and $\mathbb{C}_b$-bounded solution of the linear system of equations\vspace{-3mm}
		\end{enumerate}
		\begin{equation}\label{eq:ctr_sol}
			\hspace{-1mm}\scriptsize\begin{array}{l}
				\mathbf{F}_1\big(\mathcal{S}, \widetilde{\mathbf{N}}(z), \widetilde{\mathbf{M}}(z)\big)\mathrm{vec}(\mathbf{Q}_0(z))+\mathbf{F}_2\big(\mathcal{S}, \widetilde{\mathbf{X}}(z), \widetilde{\mathbf{Y}}(z)\big)\equiv O;
			\end{array}\normalsize\hspace{-2mm}\vspace{-3mm}
		\end{equation}
		\begin{enumerate}
			\item[b)] The TFM $\mathbf{F}_1(z)$ appearing in \eqref{eq:ctr_sol} is constructed from the entries of $\mathcal{S}$, $\widetilde{\mathbf{N}}(z)$ and $\widetilde{\mathbf{M}}(z)$, while $\mathbf{F}_2(z)$ is constructed from those of $\mathcal{S}$, $\widetilde{\mathbf{X}}(z)$ and $\widetilde{\mathbf{Y}}(z)$;\smallskip
			
			\item[c)] The TFM $\widehat{\mathbf{Q}}(z)$ encodes within it the parametrisation's degree of freedom, being expressed as\vspace{-3mm}
			\begin{equation}\label{eq:basis_span}
				\mathrm{vec}\left(\widehat{\mathbf{Q}}(z)\right)=\mathbf{B}_{\mathrm{Ker}}(z)\mathbf{x}(z),\vspace{-3mm}
			\end{equation}
			where $\mathbf x(z)$ can be freely chosen as \emph{any} column-TFM that is both proper and $\mathbb{C}_b$-bounded;\smallskip
			
			\item[d)] The TFM $\mathbf{B}_{\mathrm{Ker}}(z)$ appearing in \eqref{eq:basis_span} is proper, $\mathbb{C}_b$-bounded, and its columns form a basis which spans the right null space of $\mathbf{F}_1(z)$ from \eqref{eq:ctr_sol}.\vspace{-3mm}
		\end{enumerate}
		
		\begin{remark}
			For the sake of clarity, we point out that the exact expressions of $\mathbf{F}_1(z)$ and of $\mathbf{F}_2(z)$ are rather involved (from an algebraic point of view) and, in order to promote the overall readability of this appendix, we refer to Proposition~III.1 in \cite{aug_sparse} for the precise formulations of these TFMs. The key takeaway here is the fact that, once the DCF over $\mathbb{C}_g$ in \eqref{eq:Bezout} is settled upon, proper and $\mathbb{C}_b$-bounded solutions for the equations in \eqref{eq:ctr_sol} are straightforward to compute, along with the basis-TFM from \eqref{eq:basis_span}, via the numerical procedures mentioned in Remark~III.2 of \cite{aug_sparse}. Notice, moreover, that although the aforementioned procedures are discussed within the latter remark in the context of continuous-time systems, TFM-based model-matching in the discrete-time setting amounts to a mere remapping of the complex plane, in which only the definitions of $\mathbb{C}_g$ and $\mathbb{C}_b$ are different.\vspace{-3mm}
		\end{remark}
		
		Due to the splitting of $\mathbf{Q}(z)$ into $\mathbf{Q}_0(z)$ and $\widehat{\mathbf{Q}}(z)$, which satisfy \eqref{eq:ctr_sol} and \eqref{eq:basis_span}, respectively, notice the fact that any $\mathbf{Q}$-parametrised NRF pair with sparsity constraints can be alternatively expressed via the newly introduced $\mathbf{x}$-dependent formulation, which \emph{implicitly ensures} the enforcement of a desired sparsity pattern for the resulting control laws. Thus freed from sparsity constraints, a norm-based optimisation procedure inspired by the one proposed in Section~IV of \cite{aug_sparse} can be employed to tackle the central problem of Section~\ref{sec:NRF_des}, as formulated in \eqref{eq:MM_prob}.

		\vspace{-3mm}
	}
	
	\section{Proofs and Auxiliary Results}\label{app:proofs}\vspace{-3mm}
	
	We begin with the result given in Section~\ref{subsec:ss_implems}, which is based upon several reinterpretations of classical state-space theory, as stated in \cite{zhou}, for the distributed setting. The proof of the result concentrates on the case in which no row of $\mathbf{K}_{\mathbf{D}}(z)$ is constant ($\mathrm{row}_\ell\left(\mathbf{K}_{\mathbf{D}}(z)\right)\equiv D_{r\ell}\,,$ for some $\ell\in\{1:n_u\}$) since, for those constant rows, the result's theoretical guarantees become trivial.\vspace{-3mm}
	
	\textbf{Proof of Proposition~\ref{prop:ss_implem}}\vspace{-3mm}
	
	Point i) follows by direct application of the canonical realisations discussed in Section~6.1 of \cite{Kai}. More specifically, the rows of the TFM $\mathbf{K}_{\mathbf{D}}(z)$ can always be brought to the forms given in \eqref{eq:TFM_coefs} through numerically reliable procedures (for more details, see Remark~\ref{rem:sparse_real}). By using the row vectors $K_{j\ell}\in\mathbb{R}^{1\times(n_x+n_u)}$ from \eqref{eq:TFM_coefs} along with the scalars $a_{j\ell}$ from \eqref{eq:min_poly}, we construct the state-space realisations given in \eqref{eq:Kd_implem}-\eqref{eq:mat_coef}, which represent the so-called \emph{observable canonical forms} of each $\mathrm{row}_\ell\left(\mathbf{K}_{\mathbf{D}}(z)\right)$.\vspace{-3mm}
	
	Point ii) follows by simply noticing that the realisations obtained in \eqref{eq:Kd_implem} have the same orders as the ones from \eqref{eq:Kd_rows}, which are minimal by their very construction.\vspace{-3mm}
	
	To prove point iii), we begin by pointing out the fact that, as per \eqref{eq:Kd_implem}, the following identity holds\vspace{-3mm}
	\begin{equation}\label{eq:Kd_elm}
		\mathrm{elm}_{\ell j}(\mathbf{K}_{\mathbf{D}}(z)) = \left[\scriptsize\begin{array}{c|c}
			A_{r\ell}-z I_{n_{r\ell}} & \mathrm{col}_j(B_{r\ell}) \\ \hline C_{r\ell} & \mathrm{col}_j(D_{r\ell})
		\end{array}\right].\vspace{-3mm}
	\end{equation}
	If $\mathrm{elm}_{\ell j}(\mathbf{K}_{\mathbf{D}}(z))\equiv 0$, then we directly obtain the fact that $\mathrm{col}_j(D_{r\ell})=\lim_{|z|\rightarrow\infty}\mathrm{elm}_{\ell j}(\mathbf{K}_{\mathbf{D}}(z))=0$. Additionally, if $\mathrm{elm}_{\ell j}(\mathbf{K}_{\mathbf{D}}(z))\equiv 0$, then we can also represent this transfer function as follows\vspace{-3mm}
	\begin{equation}\label{eq:Kd_elm_zero}
		\mathrm{elm}_{\ell j}(\mathbf{K}_{\mathbf{D}}(z)) = \left[\scriptsize\begin{array}{c|c}
			A_{r\ell}-z I_{n_{r\ell}} & O \\ \hline C_{r\ell} & 0
		\end{array}\right].\vspace{-3mm}
	\end{equation}
	Given that the realisations from \eqref{eq:Kd_elm} and from \eqref{eq:Kd_elm_zero} are state-space representations of the same TFM, it follows that $C_{r\ell}^{}A_{r\ell}^{q}\mathrm{col}_j(B_{r\ell}^{})=0$ for all $q\in\mathbb{N}$ (see, for example, the proof of Theorem~3.16 in \cite{zhou}).\vspace{-3mm}
	
	\begin{figure*}
		\begin{equation}\label{eq:ser_G_hat}\tag{B.6}
			\widehat{\mathbf{G}}_u(z) =\ T(zI_{n_x}-A)^{-1}T^{-1}TB_u\mathbf{M}_u(z)
			= \left[\tiny\begin{array}{ccc|c}
				A_{11}-zI_{n_1} & A_{12} & A_{13} & B_{u1}\\
				O      & A_{22}-zI_{n_2} & A_{23} & B_{u2}\\
				O      & O      & A_{33}-zI_{n_3} & O\\\hline
				I_{n_1}& O      & O      & O\\
				O      & I_{n_2}& O      & O\\
				O      & O      & I_{n_3} & O\\
			\end{array}\right]\left[\tiny\begin{array}{c|c}
				A_M-zI_{n_M} & B_M \\\hline C_M & D_M
			\end{array}\right].
		\end{equation}
		\hrulefill\vspace{-3mm}
		\begin{equation}\label{eq:comb}\tag{B.9}
			C(zI_{n_x}-A)^{-1}B_u\mathbf{M}_u(z)=\left[\tiny\begin{array}{cc|c}
				A_{22}-zI_{n_2} & B_{u2}C_M & B_{u2}D_M\\
				O      & A_M-zI_{n_M}       & B_M\\\hline
				C_2    & O         & O
			\end{array}\right] = \left[\tiny\begin{array}{cc|c}
				\widehat{A}_{11}-zI_{n_{h1}} & \widehat{A}_{12} & \widehat{B}_{1}\\
				O                & \widehat{A}_{22}-zI_{n_{h2}} & O\\\hline
				{C}_{t1}         & {C}_{t2}         & O
			\end{array}\right],\vspace{-3mm}
		\end{equation}
		\hrulefill\vspace{-2mm}
	\end{figure*}
	
	Denoting $\mathcal O_\ell:=\scriptsize\begin{bmatrix}
		C_{r\ell}^\top & (C_{r\ell}A_{r\ell})^\top & \dots & (C_{r\ell}A_{r\ell}^{n_{r\ell}-1})^\top
	\end{bmatrix}^\top\normalsize$, we have obtained that $\mathcal O_\ell\,\mathrm{col}_j(B_{r\ell}^{})=O$. To show that $\mathrm{col}_j(B_{r\ell}^{})=O$ is the only solution to this system of equations, recall from point ii) of this result the minimality of the realisations from \eqref{eq:Kd_implem}. Thus, we must have that all of these realisations are also observable or, equivalently (see Theorem~3.3 in \cite{zhou}), that $\mathcal O_\ell$ has full column rank and, therefore, $\mathcal O_\ell\,\mathrm{col}_j(B_{r\ell}^{})=O\Longleftrightarrow\mathrm{col}_j(B_{r\ell}^{})=O$.\vspace{-3mm}
	
	Finally, point iv) follows by noticing that\vspace{-3mm}
	\begin{equation}\label{eq:row_concat}
		\hspace{-1mm}\mathbf{K}_{\mathbf{D}i}(z) = \tiny\begin{bmatrix}
			\mathrm{row}_{\alpha_{ui}+1}(\mathbf{K}_{\mathbf{D}}(z)) \\ \vdots \\ \mathrm{row}_{\alpha_{ui}+n_{ui}}(\mathbf{K}_{\mathbf{D}}(z))
		\end{bmatrix},\,\forall\,i\in\{1:N\},\normalsize\vspace{-3mm}
	\end{equation}
	and that, by applying \eqref{eq:TFM_def} to the realisations located on the right-hand side of \eqref{eq:area_NRF} and by recalling the identities from \eqref{eq:Kd_implem}, we retrieve precisely the TFMs from the right-hand side of \eqref{eq:row_concat}, with the realisations from \eqref{eq:area_NRF} inheriting the sparsity patterns showcased in point iii).\vspace{-3mm}

	\begin{figure*}
		\begin{equation}\label{eq:Z_not}\tag{B.12}
			\left\{\begin{array}{cccc}
				\mathbf{X}_s(z):=\mathcal{Z}\{x[n]\},&
				\mathbf{U}(z):=\mathcal{Z}\{u[n]\},&
				\mathbf{D}(z):=\mathcal{Z}\{d[n]\},&
				\mathbf{B}_u(z):=\mathcal{Z}\{\beta_u[n]\},\\
				\mathbf{W}(z):=\mathcal{Z}\{w[n]\}&
				\mathbf{U}_f(z):=\mathcal{Z}\{u_f[n]\},&
				\mathbf{B}_f(z):=\mathcal{Z}\{\beta_f[n]\},&
				\mathbf{B}_x(z):=\mathcal{Z}\{\beta_x[n]\}.
			\end{array}\right.
		\end{equation}
		\hrulefill\vspace{-3mm}
		\begin{subequations}
			\begin{align}\tag{B.13a}\label{eq:cl_dyn_freq_a}
				\mathbf{X}_s(z)=&\ (zI_{n_x}-A)^{-1}B_u(\mathbf{U}_f(z)+\mathbf{B}_u(z))+(zI_{n_x}-A)^{-1}B_d\ \mathbf{D}(z)+z(zI_{n_x}-A)^{-1}\textcolor{black}{\left(z^{-k_0}x_{c}\right)}\,,\\\nonumber
				\mathbf{W}(z)=&\ (zI_{n_w}-A_w)^{-1}B_w\hspace{-1mm} \tiny\begin{bmatrix}
					I_{n_u} \\ O
				\end{bmatrix}\hspace{-2mm}\scriptsize\begin{array}{l}
					(\mathbf{U}_f(z)+\mathbf{B}_f(z))
				\end{array}\hspace{-1mm}+(zI_{n_w}-A_w)^{-1}B_w \hspace{-1mm}\tiny\begin{bmatrix}
					O \\ I_{n_x}
				\end{bmatrix}\hspace{-2mm}\scriptsize\begin{array}{l}
					(\mathbf{X}_{s}(z)+\mathbf{B}_x(z))
				\end{array}\hspace{-1mm}+z(zI_{n_w}-A_w)^{-1}\textcolor{black}{\left(z^{-k_0}w_{c}\right)}\,,\tag{B.13b}\label{eq:cl_dyn_freq_b}\\
				\mathbf{U}_{f}(z) =&\ C_w \mathbf{W}(z) + D_w \hspace{-1mm}\tiny\begin{bmatrix}
					I_{n_u} \\ O
				\end{bmatrix}\hspace{-1mm} (\mathbf{U}_{f}(z)+\mathbf{B}_{f}(z)) + D_w \hspace{-1mm}\tiny\begin{bmatrix}
					O \\ I_{n_x}
				\end{bmatrix}\hspace{-1mm} (\mathbf{X}_{s}(z)+\mathbf{B}_x(z)),\tag{B.13c}\label{eq:cl_dyn_freq_c}
			\end{align}
		\end{subequations}
		\begin{equation*}
			\vspace{-8mm}
		\end{equation*}
		\hrulefill\vspace{-3mm}
		\begin{multline}\label{eq:U_f_cl}
			(I_{n_u}-\mathbf{\Phi}(z)-\mathbf{\Gamma}(z)\mathbf{G}_u(z))\mathbf{U}_f(z) = \mathbf{\Phi}(z)\mathbf{B}_f(z) + \mathbf{\Gamma}(z)\mathbf{B}_{x}(z) + \mathbf{\Gamma}(z)\mathbf{G}_u(z)\mathbf{B}_{u}(z) +\\+ \mathbf{\Gamma}(z)\mathbf{G}_d(z)\mathbf{D}(z) + z\mathbf{\Gamma}(z)(zI_{n_x}-A)^{-1}\textcolor{black}{\left(z^{-k_0}x_{c}\right)} + zC_w(zI_{n_w}-A_w)^{-1}\textcolor{black}{\left(z^{-k_0}w_{c}\right)}.\tag{B.14}
		\end{multline}
		\begin{equation*}
			\vspace{-10mm}
		\end{equation*}
		\hrulefill\vspace{-2mm}
		\end{figure*}
	
	Before moving on to the proof of Theorem~\ref{thm:NRF_state}, we state the following auxiliary result.\vspace{-3mm}
	
	\begin{lemma}
		\label{lem:pole_pencil} Let $\mathbb{C}_g\subseteq\mathbb{C}$ along with $\mathbb{C}_b:=\mathbb{C}\setminus\mathbb{C}_g$, and consider a system given by \eqref{eq:ss_a}-\eqref{eq:ss_b}. For the latter realisation, let $\mathbf{G}_u(z):=C(z I_{n_x}-A)^{-1}B_u+D_u$ and let $\mathbf{N}_u(z)\in\mathcal{R}(z)^{n_y\times n_u}$ along with $\mathbf{M}_u(z)\in\mathcal{R}(z)^{n_u\times n_u}$ be two TFMs. Assume that the following statements hold:\vspace{-3mm}
		\begin{enumerate}
			\item[A1)] The subrealisation $(A,B_u,C,D_u)$ from \eqref{eq:ss_a}-\eqref{eq:ss_b} is $\mathbb{C}_b$-irreducible;
			
			\item[A2)] Both $\mathbf{N}_u(z)$ and $\mathbf{M}_u(z)$ are $\mathbb{C}_b$-bounded and proper, $\mathbf{M}_u(z)$ is invertible and also the following identity $\mathbf{G}_u(z)=\mathbf{N}_u(z)\mathbf{M}_u^{-1}(z)$ holds.\vspace{-3mm}
		\end{enumerate}
		Then, the TFM defined as\vspace{-3mm}
		\begin{equation}\label{eq:G_tilde}\tag{B.4}
			\widetilde{\mathbf{G}}_u(z):=(z I_{n_x}-A)^{-1}B_u\,\mathbf{M}_u(z)\vspace{-3mm}
		\end{equation}
		is $\mathbb{C}_b$-bounded.\vspace{-6mm}
	\end{lemma}
	\begin{pf}
		We begin by pointing out that if $A-zI_{n_x}$ happens to be a $\mathbb{C}_g$-admissible pencil, then the proof is trivial. Similarly, if $\mathbf{M}_u(z)$ from Assumption~A2) is a constant matrix, then it follows that $\mathbf{G}_u(z)$ is a $\mathbb{C}_b$-bounded TFM. In combination with Assumption~A1), this implies (see, for example, the proof of Theorem~III.6 in \cite{NRF}) that $A-zI_{n_x}$ is $\mathbb{C}_g$-admissible, from which the conclusion of our result's statement follows directly.\vspace{-3mm}
		
		We now consider the case in which $A-zI_{n_x}$ is not $\mathbb{C}_g$-admissible and $\mathbf{M}_u(z)$ is not a constant matrix, and for which the proof reduces to the application of coordinate transformations to the state-space system which describes $\mathbf{G}_u(z)$. We do so in order to show that the TFM from \eqref{eq:G_tilde} can be described by a realisation with a $\mathbb{C}_g$-admissible pole pencil, thus making the latter TFM $\mathbb{C}_b$-bounded. We begin by applying a transformation (as described in Section~3.3 of \cite{zhou}) given by the nonsingular matrix $T\in\mathbb{R}^{n_x\times n_x}$, in order to bring the realisation mentioned in Assumption~A1) to the form\vspace{-3mm}
		\begin{equation}\label{eq:G_trans}
			\mathbf{G}_u(z)=\left[\tiny\begin{array}{ccc|c}
				A_{11}-zI_{n_1} & A_{12} & A_{13} & B_{u1}\\
				O      & A_{22}-zI_{n_2} & A_{23} & B_{u2}\\
				O      & O      & A_{33}-zI_{n_3} & O\\\hline
				O      & C_2    & C_3    & D_u
			\end{array}\right],\vspace{-3mm}
		\end{equation}
		where the system $\left(\tiny\begin{bmatrix}
			A_{22} & A_{23} \\ O & A_{33}
		\end{bmatrix},\begin{bmatrix}
			B_{u2}^\top & O
		\end{bmatrix}^\top, \begin{bmatrix}
			C_2 & C_3
		\end{bmatrix}, O\right)$ is observable and $(A_{22}, B_{u2},	C_2, O)$ is minimal, and where the matrix blocks $A_{11}$ and $A_{33}$ may vanish, depending on the structural properties of the initial realisation describing $\mathbf{G}_u(z)$. Moving forward, let $(A_M,B_M,C_M,D_M)$ be a minimal realisation, of order $n_M\geq 1$, for $\mathbf{M}_u(z)$.\vspace{-4mm} 
		
		We now define $\widehat{\mathbf{G}}_u(z):=T\widetilde{\mathbf{G}}_u(z)$, which can be expressed via the series connection between the two state-space systems showcased in \eqref{eq:ser_G_hat} and located at the top of this page. By first eliminating the $n_3$ uncontrollable modes in the left-hand realisation from the rightmost equality in \eqref{eq:ser_G_hat}, followed by the computation of the resulting series interconnection, we get the fact that\vspace{-3mm}\stepcounter{equation}
		\begin{equation}\label{eq:G_hat_before}
			\hspace{-3mm}\widehat{\mathbf{G}}_u(z) =\hspace{-0.5mm} \left[\tiny\begin{array}{ccc|c}
				A_{11}-zI_{n_1} & A_{12} & B_{u1}C_M & B_{u1}D_M\\
				O      & A_{22}-zI_{n_2} & B_{u2}C_M & B_{u2}D_M\\
				O      & O      & A_{M}-zI_{n_M} & B_M\\\hline
				I_{n_1}& O      & O      & O\\
				O      & I_{n_2}& O      & O\\
				O      & O      & O      & O\\
			\end{array}\right]\hspace{-0.5mm}.\hspace{-2mm}\vspace{-3mm}
		\end{equation}
		Due to the invertibility of $T$, it follows by standard state-space theory that $\widetilde{\mathbf{G}}_u(z)=T^{-1}\widehat{\mathbf{G}}_u(z)$ is $\mathbb{C}_b$-bounded if and only if so is $\widehat{\mathbf{G}}_u(z)$. To show that the  TFM from \eqref{eq:G_hat_before} is indeed $\mathbb{C}_b$-bounded, note that the identity from Assumption~A2) can be multiplied to the right by $\mathbf{M}_u(z)$, in order to rewrite it as follows\vspace{-3mm}
		\begin{equation}\label{eq:A2_trans}
			C(zI_{n_x}-A)^{-1}B_u\mathbf{M}_u(z)=\mathbf{N}_u(z)-D_u\mathbf{M}_u(z),\vspace{-3mm}
		\end{equation}
		with the term located on the right-hand side of the equality from \eqref{eq:A2_trans} being itself a $\mathbb{C}_b$-bounded TFM. By employing $C(zI_{n_x}-A)^{-1}B_u=C_2(zI_{n_x}-A_{22})^{-1}B_{u2}$ along with the minimal realisation of $\mathbf{M}_u(z)$, and by repeating the same procedure as the one applied for $\widehat{\mathbf{G}}_u(z)$, we obtain the state-space representation given in the middle term of \eqref{eq:comb}, which is located at the top of this page. For the aforementioned state-space system, we now compute a coordinate transformation given by an invertible matrix, which we denote $T_c\in\mathbb{R}^{(n_2+n_M)\times(n_2+n_M)}$ and which brings the respective realisation to the form showcased in the right-hand term of \eqref{eq:comb}, where the system given by the realisation $(\widehat{A}_{11},\widehat{B}_{1},C_{t1},O)$ is controllable.\vspace{-3mm}\stepcounter{equation}
		
		Recall now the fact that $\mathbf{M}_u(z)$ is a $\mathbb{C}_b$-bounded TFM and that the realisation $(A_M,B_M,C_M,D_M)$ is minimal, which implies that the matrix pencil $A_M-zI_{n_M}$ is $\mathbb{C}_g$-admissible. Since the realisation $(A_{22}, B_{u2},	C_2, O)$ is minimal, it follows by direct application of the PBH (Popov-Belevitch-Hautus; see, for example, Section~3.2 of \cite{zhou}) test that both of the state-space systems from \eqref{eq:comb} are $\mathbb{C}_b$-observable. Consequently, $(\widehat{A}_{11},\widehat{B}_{1},C_{t1},O)$ is a $\mathbb{C}_b$-irreducible realisation (we eliminate the $n_{h2}$ uncontrollable modes from the right-hand realisation in \eqref{eq:comb}) of $C(zI_{n_x}-A)^{-1}B_u\mathbf{M}_u(z)$. Recalling that the latter TFM is $\mathbb{C}_b$-bounded, it follows by standard state-space theory that $\widehat{A}_{11}-zI_{n_{h1}}$ is a $\mathbb{C}_g$-admissible pencil.\vspace{-3mm}
		
		In order to conclude the proof, we now apply a coordinate transformation given by the matrix $\widehat{T}_c:=\tiny\begin{bmatrix}
			I_{n_1} & O \\ O & T_c
		\end{bmatrix}$ to the realisation from \eqref{eq:G_hat_before}, which yields the identity\vspace{-3mm}
		\begin{equation}\label{eq:G_hat_after}
			\hspace{-3mm}\widehat{\mathbf{G}}_u(z) \hspace{-0.5mm}=\hspace{-0.5mm} \left[\tiny\begin{array}{ccc|c}
				A_{11}-zI_{n_1} & A_{12}           & B_{u1}C_M        & B_{u1}D_M\\
				O      & \widehat{A}_{11}-zI_{n_{h1}} & \widehat{A}_{12} & \widehat{B}_{1}\\
				O      & O                & \widehat{A}_{22}-zI_{n_{h2}} & O\\\hline
				I_{n_1}& O                & O                & O\\
				O      & \widehat{C}_1    & \widehat{C}_2    & O\\
				O      & O                & O                & O\\
			\end{array}\right]\hspace{-1mm}.\hspace{-2mm}
		\end{equation}
		By once again removing the uncontrollable modes which are associated with the matrix block $\widehat{A}_{22}$ from \eqref{eq:G_hat_after}, we get that $\widehat{\mathbf{G}}_u(z)$ can also be expressed as follows\vspace{-3mm}
		\begin{equation}\label{eq:G_hat_final}
			\widehat{\mathbf{G}}_u(z) = \left[\tiny\begin{array}{cc|c}
				A_{11}-zI_{n_1} & A_{12}           & B_{u1}D_M\\
				O      & \widehat{A}_{11}-zI_{n_{h1}} & \widehat{B}_{1}\\\hline
				I_{n_1}& O                & O\\
				O      & \widehat{C}_1    & O\\
				O      & O                & O\\
			\end{array}\right].\vspace{-3mm}
		\end{equation}
		Recalling that, by Assumption~A1), the realisation from \eqref{eq:G_trans} is $\mathbb{C}_b$-observable, we get that the pencil $A_{11}-zI_{n_1}$ must be $\mathbb{C}_g$-admissible. Since the same property holds for $\widehat{A}_{11}-zI_{n_{h1}}$, then the pole-pencil of the realisation expressed in \eqref{eq:G_hat_final} is also $\mathbb{C}_g$-admissible. We thus conclude that both $\widehat{\mathbf{G}}_u(z)$ and $\widetilde{\mathbf{G}}_u(z)$ are $\mathbb{C}_b$-bounded TFMs.\vspace{-6mm}
	\end{pf}

	\begin{figure*}
		\begin{equation}\label{eq:sens}\tag{B.15}
			(I_{n_u}-\mathbf{\Phi}(z)-\mathbf{\Gamma}(z)\mathbf{G}_u(z))^{-1}=\mathbf{M}(z)\left(\widetilde{\mathbf{Y}}_{\mathbf{Q}}(z)\mathbf{M}(z)-\widetilde{\mathbf{X}}_{\mathbf{Q}}(z)\mathbf{N}(z)\right)^{-1}\widetilde{\mathbf{Y}}_{\mathbf{Q}}^{\mathrm{diag}}(z)=\mathbf{M}(z)\widetilde{\mathbf{Y}}_{\mathbf{Q}}^{\mathrm{diag}}(z).
		\end{equation}\hrulefill\vspace{-1mm}
		\begin{subequations}
			\begin{align}\label{eq:Gd_1}\tag{B.18a}
				(\mathbf{N}(z)\widetilde{\mathbf{X}}_{\mathbf{Q}}(z)+I_n)\mathbf{G}_d(z)= (\mathbf{Y}(z)+\mathbf{N}(z)\mathbf{Q}(z))\widetilde{\mathbf{M}}(z)\mathbf{G}_d(z)= (\mathbf{Y}(z)+\mathbf{N}(z)\mathbf{Q}(z))\widetilde{\mathbf{M}}(z)(zI_{n_x}-A)^{-1}B_d,&\\
				\mathbf{M}(z)\widetilde{\mathbf{X}}_{\mathbf{Q}}(z)\mathbf{G}_d(z)=(\mathbf{X}(z)+\mathbf{M}(z)\mathbf{Q}(z))\widetilde{\mathbf{M}}(z)(zI_{n_x}-A)^{-1}B_d.&\label{eq:Gd_2}\tag{B.18b}
			\end{align}
		\end{subequations}
		\begin{equation*}
			\vspace{-9mm}
		\end{equation*}
		\hrulefill
	\end{figure*}
	
	By proceeding to leverage Lemma~\ref{lem:pole_pencil}, we are now able to prove the main result of our paper.\vspace{-3mm}
	
	\textbf{Proof of Theorem~\ref{thm:NRF_state}}\vspace{-3mm}
	
	The proof of this result reduces to a series of straightforward, yet laborious, algebraic manipulations.\vspace{-3mm}
	
	We begin by writing down the state dynamics of the network and the first layer, in terms of their $\mathcal{Z}$-transforms, while taking into account the closed-loop interconnections depicted in Figure~\ref{fig:NRF_implem}. Following this, we rearrange the terms of the resulting algebraic expressions, and we retrieve the sought-after identities by employing the notions related to DCFs and NRF pairs from Appendix~\ref{app:aux}.\vspace{-3mm}
	
	In order to express the closed-loop dynamics from Figure~\ref{fig:NRF_implem}, we first employ the realisations given in \eqref{eq:area_NRF} to denote $B_w:=\tiny\begin{bmatrix}
		B_{w1}^\top & \dots & B_{wN}^\top 
	\end{bmatrix}^\top$ and $D_w:=\tiny\begin{bmatrix}
		D_{w1}^\top & \dots & D_{wN}^\top 
	\end{bmatrix}^\top$, along with $w[k]:=\tiny\begin{bmatrix}
		w_1^\top[k] & \dots & w_N^\top[k] 
	\end{bmatrix}^\top$, where the $w_i[k]$ vectors denote the state variables of the NRF subcontrollers from \eqref{eq:area_NRF}. Moving on, we denote the $\mathcal{Z}$-transform of all the signals and state variables from Figure~\ref{fig:NRF_implem} and, by adopting the notation given in \eqref{eq:Z_not} and located at the top of this page, we are able to express the dynamics of the closed-loop interconnection via the identities from \eqref{eq:cl_dyn_freq_a}-\eqref{eq:cl_dyn_freq_c}, also located at the top of this page.\vspace{-3mm}
	
	By first embedding \eqref{eq:cl_dyn_freq_a} along with \eqref{eq:cl_dyn_freq_b} into \eqref{eq:cl_dyn_freq_c}, and then recalling the realisations of $\mathbf{G}_u(z)$, $\mathbf{G}_d(z)$ and $\mathbf{K}_{\mathbf{D}i}(z)$, for all $i\in\{1:N\}$, standard algebraic substitutions in \eqref{eq:cl_dyn_freq_c} yield the identity from \eqref{eq:U_f_cl}, located at the top of this page. Notice, however, that after left-multiplying \eqref{eq:U_f_cl} with the term $(I_{n_u}-\mathbf{\Phi}(z)-\mathbf{\Gamma}(z)\mathbf{G}_u(z))^{-1}$, the resulting expressions do not immediately resemble the entries of either $\tiny\begin{bmatrix}
		O & I_{n_u}
	\end{bmatrix}\mathbf{F}_{\mathbf{Q}}(z)$ or $\tiny\begin{bmatrix}
		O & I_{n_u}
	\end{bmatrix}\mathbf{I}_{\mathbf{Q}}(z)$. In order to retrieve these TFMs, we must first rewrite $(I_{n_u}-\mathbf{\Phi}(z)-\mathbf{\Gamma}(z)\mathbf{G}_u(z))^{-1}$ by recalling \eqref{eq:aux_NRF} and \eqref{eq:NRF_def}, along with the fact that $\mathbf{G}_u(z)=\mathbf{N}(z)\mathbf{M}^{-1}(z)$. Using all of these facts and then performing straightforward substitutions, we obtain the identity from \eqref{eq:sens}, located at the top of the next page.\vspace{-3mm}
	
	By left-multiplying in \eqref{eq:U_f_cl} with the rightmost TFM from \eqref{eq:sens} and by using the definitions given in \eqref{eq:NRF_def}, we retrieve all of the entries of $\tiny\begin{bmatrix}
		O & I_{n_u}
	\end{bmatrix}\mathbf{F}_{\mathbf{Q}}(z)$ in a straightforward manner, except for the block corresponding to the signal vector $\beta_{u}$. The aforementioned block is retrieved by employing \eqref{eq:Bezout}-\eqref{eq:aux_NRF}, in order to get that\vspace{-3mm}
	\begin{equation*}
		\mathbf{M}(z)\widetilde{\mathbf{Y}}_{\mathbf{Q}}^{\mathrm{diag}}(z)\mathbf{\Gamma}(z)\mathbf{G}_u(z)=\mathbf{M}(z)\widetilde{\mathbf{Y}}_{\mathbf{Q}}(z)-I_{n_u}.\vspace{-3mm}
	\end{equation*}
	Additionally, note that left-multiplying \eqref{eq:U_f_cl} with the TFM $\mathbf{M}(z)\widetilde{\mathbf{Y}}_{\mathbf{Q}}^{\mathrm{diag}}(z)$ also yields the terms which make up $\tiny\begin{bmatrix}
		O & I_{n_u}
	\end{bmatrix}\mathbf{I}_{\mathbf{Q}}(z)\tiny\begin{bmatrix}
		O & I_{n_w}
	\end{bmatrix}^\top$, while enabling us to state that\vspace{-3mm}
	\scriptsize\begin{equation*}
		\tiny\begin{bmatrix}
			O & I_{n_u}
		\end{bmatrix}\mathbf{I}_{\mathbf{Q}}(z)\tiny\begin{bmatrix}
			I_{n_x} \\ O
		\end{bmatrix}=(\mathbf{X}(z)+\mathbf{M}(z)\mathbf{Q}(z))\widetilde{\mathbf{M}}(z)(zI_{n_x}-A)^{-1}z,\vspace{-3mm}
	\end{equation*}\normalsize
	which once again follows by employing \eqref{eq:Bezout}-\eqref{eq:aux_NRF}, as done previously. By performing now the aforementioned left-multiplication in \eqref{eq:U_f_cl} and then taking the inverse $\mathcal{Z}$-transform of the resulting expression, we recover the last $n_u$ rows of the identity from \eqref{eq:cl_dyn}. To obtain the first $n_x$ rows from \eqref{eq:cl_dyn}, it suffices to plug the expression of $\mathbf{U}_f(z)$ obtained from \eqref{eq:U_f_cl} into \eqref{eq:cl_dyn_freq_a}. Performing all of the required substitutions and taking the inverse $\mathcal{Z}$-transform of the result yields the desired expression.\vspace{-3mm}
	
	At this point in the proof, we have obtained the closed-loop dynamics expressed in \eqref{eq:cl_dyn}, while also validating the statements made in points i)-iii) of the result. All that remains is to show that point iv) holds, as well.\vspace{-3mm}
	
	The fact that $\mathbf{F}_{\mathbf{Q}}(z)$ in \eqref{eq:F_Q} is proper follows directly from its (block-)entries being obtained via the multiplication and addition of proper TFMs. Notice, moreover, that we may rewrite the following expression\stepcounter{equation}\stepcounter{equation}\stepcounter{equation}\vspace{-3mm}
	\begin{equation}\label{eq:z_mult_1}
		(zI_{n_x}-A)^{-1}z=I_{n_x}+ (zI_{n_x}-A)^{-1}A,\vspace{-3mm}
	\end{equation}
	along with its counterpart\vspace{-3mm}
	\begin{equation}\label{eq:z_mult_2}
		(zI_{n_w}-A_w)^{-1}z=I_{n_w} + (zI_{n_w}-A_w)^{-1}A_w.\vspace{-3mm}
	\end{equation}
	The result of these computations is a pair of proper TFMs, which enables us to employ the same arguments with respect to $\mathbf{J}_1(z)$ and $\mathbf{J}_2(z)$, along with $\mathbf{I}_{\mathbf{Q}}(z)$, in order to conclude that all three TFMs in \eqref{eq:I_Q} are proper.\vspace{-3mm}
	
	We now address the property of $\mathbb{C}_b$-boundedness. Direct inspection of \eqref{eq:F_Q} yields the fact that the TFM $\mathbf{F}_{\mathbf{Q}}(z)\tiny\begin{bmatrix}
		I_{(n_x+2n_u)}&O
	\end{bmatrix}^\top$ is $\mathbb{C}_b$-bounded due to it being obtained via the multiplication and the addition of $\mathbb{C}_b$-bounded TFMs. With respect to the blocks corresponding to the signal vector $d$, it suffices to employ the identities from \eqref{eq:Bezout} in order to rewrite the expressions shown in \eqref{eq:Gd_1}-\eqref{eq:Gd_2} and located at the top of this page.\vspace{-3mm}
	
	It is now straightforward to apply Lemma~\ref{lem:pole_pencil} for the system represented by the realisation $(A^\top,I_{n_x},B_u^\top,O)$ and for the pair of proper, $\mathbb{C}_b$-bounded TFMs $\widetilde{\mathbf{N}}^\top(z)$ and $\widetilde{\mathbf{M}}^\top(z)$. We may do so since the aforementioned realisation is, as per Assumption~A1) in the result's statement, $\mathbb{C}_b$-irreducible and since the following identity holds\vspace{-4mm}
	\begin{equation*}
		\mathbf{G}_u^\top(z):=B_u^\top\left(zI_{n_x}-A^\top\right)^{-1}=\widetilde{\mathbf{N}}^\top(z)\left(\widetilde{\mathbf{M}}^\top(z)\right)^{-1}.\vspace{-4mm}
	\end{equation*}
	By applying Lemma~\ref{lem:pole_pencil}, we get that $\widetilde{\mathbf{M}}(z)(zI_{n_x}-A)^{-1}$ along with its transpose are $\mathbb{C}_b$-bounded TFMs, akin to $(\mathbf{X}(z)+\mathbf{M}(z)\mathbf{Q}(z))$ and to $(\mathbf{Y}(z)+\mathbf{N}(z)\mathbf{Q}(z))$. Therefore, $(\mathbf{N}(z)\widetilde{\mathbf{X}}_{\mathbf{Q}}(z)+I_{n_x})\mathbf{G}_d(z)$ and $\mathbf{M}(z)\widetilde{\mathbf{X}}_{\mathbf{Q}}(z)\mathbf{G}_d(z)$ are also $\mathbb{C}_b$-bounded, since they are obtained by multiplying and adding $\mathbb{C}_b$-bounded TFMs.\vspace{-3mm}
	
	Finally, we turn our attention to the TFMs defined in \eqref{eq:I_Q}. Recalling \eqref{eq:z_mult_1}, we rewrite $\mathbf{J}_1(z)$ as \stepcounter{equation}\vspace{-3mm}
	\begin{equation}\label{eq:J1_aux}
		\mathbf{J}_1(z) = \widetilde{\mathbf{M}}(z)+ \widetilde{\mathbf{M}}(z)(zI_{n_x}-A)^{-1}A,\vspace{-3mm}
	\end{equation}
	and we point out that we have previously shown the fact that $\widetilde{\mathbf{M}}(z)(zI_{n_x}-A)^{-1}$ is a $\mathbb{C}_b$-bounded TFM. Since all of the sums and multiplications performed in \eqref{eq:J1_aux} involve $\mathbb{C}_b$-bounded TFMs, it follows that the same property extends to $\mathbf{J}_1(z)$. The only remaining obstacle is to show that the same property holds for $\mathbf{J}_2(z)$. In order to do so, notice first that the TFMs defined as $\mathbf{N}_{\mathbf{K}}(z):=\scriptsize\begin{bmatrix}
		\widetilde{\mathbf{Y}}_{\mathbf{Q}}^{\mathrm{diag}}(z)-\widetilde{\mathbf{Y}}_{\mathbf{Q}} & \widetilde{\mathbf{X}}_{\mathbf{Q}}(z)
	\end{bmatrix}^\top$ along with $\mathbf{M}_{\mathbf{K}}(z):=\left(\widetilde{\mathbf{Y}}_{\mathbf{Q}}^{\mathrm{diag}}(z)\right)^\top$ are proper and $\mathbb{C}_b$-bounded, and that the following identity holds\vspace{-3mm}
	\begin{equation*}
		\mathbf{N}_{\mathbf{K}}^{}(z)\mathbf{M}_{\mathbf{K}}^{-1}(z)=\mathbf{K}_{\mathbf{D}}^\top(z)=B_w^\top(zI_{n_w}-A_w)^{-1}C_w^\top+D_w^\top.\vspace{-3mm}
	\end{equation*}
	Recall now that each of the realisations from \eqref{eq:Kd_implem} is minimal, as per point ii) of Proposition~\ref{prop:ss_implem}. Then, by employing the properties of NRF-based distributed controllers, it is possible to show (see the proof of Theorem~III.6 in \cite{NRF}) that the realisation $(A_w,B_w,C_w,D_w)$ is $\mathbb{C}_b$-irreducible. Moreover, it is straightforward to show, by employing standard PBH tests, that the latter property holds for the realisation $(A_w^\top,C_w^\top,B_w^\top,D_w^\top)$, as well.\vspace{-3mm}
	
	By applying now Lemma~\ref{lem:pole_pencil}, we get that the TFM $(zI_{n_w}-A_w^\top)^{-1}C_w^\top\mathbf{M}_{\mathbf{K}}(z)$ is $\mathbb{C}_b$-bounded (along with its transpose) and, by recalling \eqref{eq:z_mult_2}, we rewrite $\mathbf{J}_2(z)$ as\vspace{-3mm}
	\begin{equation*}
		\mathbf{J}_2(z)=\widetilde{\mathbf{Y}}_{\mathbf{Q}}^{\mathrm{diag}}(z)C_w+\widetilde{\mathbf{Y}}_{\mathbf{Q}}^{\mathrm{diag}}(z)C_w(zI_{n_w}-A_w)^{-1}A_w,\vspace{-3mm}
	\end{equation*}
	which is a $\mathbb{C}_b$-bounded TFM, by the same arguments as those employed when investigating the same property for $\mathbf{J}_1(z)$. In conclusion, since the TFMs $\mathbf{J}_1(z)$ and $\mathbf{J}_2(z)$, along with the TFM pair $(\mathbf{X}(z)+\mathbf{M}(z)\mathbf{Q}(z))$ and $(\mathbf{Y}(z)+\mathbf{N}(z)\mathbf{Q}(z))$, are all $\mathbb{C}_b$-bounded, we once again employ the same set of arguments in order to deduce the fact that $\mathbf{I}_{\mathbf{Q}}(z)$ is $\mathbb{C}_b$-bounded, as well.
	
\end{document}